\documentclass[twocolumn,floatfix,prl,aps,showpacs]{revtex4-1}

\usepackage{graphicx,amsmath,amssymb,color}
\usepackage{nicefrac}
\usepackage[titletoc,title]{appendix}

\usepackage{subfigure}

\usepackage{psfrag}
\usepackage{bm,color}
\usepackage{array,multirow}
\usepackage{booktabs}

\newcommand{\be}{\begin{equation}}
\newcommand{\ee}{\end{equation}}

\newcommand{\ba}{\begin{eqnarray}}
\newcommand{\ea}{\end{eqnarray}}

\renewcommand{\vec}[1]{{\textbf{\textit{#1}}}}

\begin{document}

\title{Landau-Level Mixing and Particle-Hole Symmetry Breaking for Spin Transitions in the Fractional Quantum Hall Effect}

\author{Yuhe Zhang,$^1$ A. W\'ojs,$^2$ and J. K. Jain$^{1,3}$}

\affiliation{$^1$Department of Physics, The Pennsylvania State University, University Park, Pennsylvania, 16802, USA}
\affiliation{$^{2}$Department of Theoretical Physics, Wroclaw University of Technology, Wybrzeze Wyspianskiego 27, 50-370 Wroclaw, Poland}
\affiliation{$^3$Department of Physics, Indian Institute of Science, Bengaluru, 560012, India}

\date{\today}

\begin{abstract}
The spin transitions in the fractional quantum Hall effect provide a direct measure of the tiny energy differences between differently spin-polarized states, and thereby serve as an extremely sensitive test of the quantitative accuracy of the theory of the fractional quantum Hall effect, and, in particular, of the role of Landau-level mixing in lifting the particle-hole symmetry. We report on an accurate quantitative study of this physics, evaluating the effect of Landau-level mixing in a nonperturbative manner using a fixed-phase diffusion Monte Carlo method. We find excellent agreement between our calculated critical Zeeman energies and the experimentally measured values. In particular, we find, as also do experiments, that the critical Zeeman energies for fractional quantum Hall states at filling factors $\nu=2-n/(2n\pm 1)$ are significantly higher than those for $\nu=n/(2n\pm 1)$, a quantitative signature of the lifting of particle-hole symmetry due to Landau-level mixing. 
\end{abstract}

\maketitle

The role of particle-hole symmetry in the lowest Landau level (LLL) as well as its breaking due to Landau-level (LL) mixing has come into renewed focus in the contexts of the competition between the Pfaffian and the anti-Pfaffian wave functions for the $\nu=5/2$ fractional quantum Hall (FQH) effect \cite{Moore91,Levin07,Lee07,Bishara09,Wojs10, Sodemann13, Simon13,Peterson13,Peterson14} and of the nature of the composite-fermion (CF) Fermi sea at $\nu=1/2$ \cite{Halperin93,Barkeshli15,Son15,Wang15b,Metlitski15,Wang15,Geraedts15,Potter16,Murthy15,Metlitski15a,Mross16,Kachru15,Mulligan16,Levin16,Balram16b}.  LL mixing also affects various observable quantities in the FQH effect, and a lack of its quantitative understanding has been one of the major impediments to the goal of an accurate comparison between theory and experiment.  The effect of LL mixing has been treated in a perturbative approach~\cite{Bishara09,Wojs10, Sodemann13, Simon13,Peterson13,Peterson14}, but the extent of its validity for typical experiments has remained unclear because the relevant parameter controlling the strength of LL mixing, namely the ratio of the Coulomb interaction to the cyclotron energy $\kappa=(e^2/\epsilon \ell)/\hbar\omega_c$, is typically $\sim 1$ and sometimes as high as $\sim 2$. (Here, $\ell=\sqrt{\hbar c/eB}$ is the magnetic length, $\epsilon$ is the dielectric constant of the background material, and $\omega_c=eB/m_{b}c$ is the cyclotron frequency.)

We study in this work the effect of LL mixing through the nonperturbative method of fixed-phase diffusion Monte Carlo calculations \cite{Ortiz93,Melik-Alaverdian97, Melik-Alaverdian01}. We focus here on the phase transitions between differently spin-polarized FQH states as a function of the Zeeman energy, which are an ideal testing ground for the role of LL mixing, both because a wealth of experimental information exists for the critical energies where such transitions occur \cite{Eisenstein89, Eisenstein90, Engel92, Du95,Kang97, Kukushkin99, Yeh99,Kukushkin00, Melinte00,Freytag01, Tiemann12,Feldman13,Liu14}, and because they depend sensitively on LL mixing  \cite{Liu14,Balram15a}. The critical Zeeman energy $E_{\rm Z}^{\rm crit}$, quoted below in terms of the dimensionless ratio $\alpha_{\rm Z}^{\rm crit}=E_{\rm Z}^{\rm crit}/(e^{2}/\epsilon \ell)$, is a direct measure of the tiny energy differences between differently spin polarized states, and thus serves as an extremely sensitive test of the quantitative accuracy of the theory.  In particular, a long-standing puzzle has been that the observed values of $\alpha_{\rm Z}^{\rm crit}$ for spin transitions at the filling factor $\nu=2-n/(2n\pm 1)$ are significantly higher than those at $\nu=n/(2n\pm 1)$. Because  particle-hole symmetry  in a system confined to the LLL guarantees that the transitions at $\nu$ and $2-\nu$ occur at the same $\alpha_{\rm Z}^{\rm crit}$, it is clear that LL mixing, which breaks particle-hole symmetry, is responsible for the effect. Surprisingly, for heterojunction samples, $\alpha_{\rm Z}^{\rm crit}$ for spin transitions at the filling factor $\nu=2-n/(2n\pm 1)$ are higher even than the theoretical values for systems with zero width and zero LL mixing, which is counterintuitive because the corrections due to finite width and finite LL mixing are both expected to weaken the interaction and, thus, reduce $\alpha_{\rm Z}^{\rm crit}$.  

If the fixed-phase Diffusion Monte Carlo (DMC) method can be demonstrated to provide a quantitative account of these experiments, it will not only reveal the role of Landau-level mixing in a quantitative fashion but, in principle, also enable an investigation of the effect of LL mixing on various other issues, including the 5/2 Pfaffian/anti-Pfaffian state and the 1/2 CF Fermi sea, in a nonperturbative approach.

The DMC method~\cite{Reynolds82,Foulkes01} 
solves the many-body Schr\"{o}dinger equation by noting that its imaginary time ($t\rightarrow i t$) version can be interpreted as a diffusion equation. The wave function $\Phi$ of interest plays the role of the density of diffusing particles, which is valid when $\Phi$ is always real and non-negative, such as for Bose systems in their ground states. In order to treat Fermi statistics, a fixed-node approximation is used which does not allow diffusion through the nodal surface. The fixed-node DMC method, suitable for real wave function, cannot be applied directly to FQH systems, which, due to the broken time-reversal symmetry, produce complex valued eigenfunctions for interacting fermions. For such systems, a fixed-phase approximation was introduced by Ortiz, Ceperley and Martin~\cite{Ortiz93} who express the wave function as $\Phi(\mathcal{R})=|\Phi(\mathcal{R})|e^{i\varphi_T(\mathcal{R})}$ and solve the appropriate Schr\"odinger equation for the real non-negative wave function $|\Phi(\mathcal{R})|$ by the DMC method. Here, $\mathcal{R}=({\bf r}_{1}, {\bf r}_{2}, ... ,{\bf r}_{N})$ denotes the coordinates collectively, and the phase $\varphi_T(\mathcal{R})$ is fixed with the help of an initial ``trial" or ``guiding" wave function $\psi_{T} (\mathcal{R})= |\psi_{T}(\mathcal{R})| e^{i\varphi_{T}(\mathcal{R})}$.  The DMC algorithm gives the lowest energy consistent with the prescribed trial phase $\varphi_{T}(\mathcal{R})$ and the accuracy of the results depends on the choice of $\varphi_{T}(\mathcal{R})$. It was found by G\"u\c{c}l\"u and Umrigar\cite{Guclu05} that the Coulomb eigenstate of the LLL subspace is an excellent choice for $\psi_T$, i.e., LL mixing does not significantly alter the phase. We will, therefore, choose for our fixed-phase DMC calculation the phases of the wave functions of the CF theory, which are known to accurately represent the actual Coulomb eigenstates~\cite{Jain89, Jain07}. 

We follow the method presented by Melik-Alaveridan, Bonesteel and Ortiz \cite{Melik-Alaverdian97, Melik-Alaverdian01}, who have generalized the fixed-phase DMC method to the spherical manifold~\cite{ZhangSM1}. The electrons are confined to the surface of a sphere~\cite{Haldane83} of radius $R_{0}$ with a magnetic monopole of strength $Q$ at the center, producing a total flux of $2Q\phi_{0}$. In order to simulate the diffusion process conveniently, a stereographic projection is employed to represent the electrons' positions by planar coordinates ${\bf r}=(x,y)=(\cos{\phi},\sin{\phi})\cot(\theta/2)$, where $\theta$ and $\phi$ are the usual spherical angles. The Hamiltonian is then written as
\begin{equation}
H=\frac{1}{2m_{\rm b}}\sum_{i}{D({\bf r}_{i})[-i\hbar\nabla_{i}+e{\bf A}({\bf r}_{i})]^{2} + V(\mathcal{R})},
\label{Hamiltonian}
\end{equation}
where $D({\bf r}_{i})=(1+r_{i}^{2})^{2}/4R_{0}^{2}$. The vector potential ${\bf A}=-\frac{\hbar c Q}{eR_{0}} \cot \theta \bm{\hat{\phi}}$ produces a radial magnetic field $B=2Q\phi_{0}/4\pi R_{0}^{2}$ in the Haldane gauge. 
At filling factor $\nu=n/(2pn\pm1)$, for trial function $\psi_{T} (\mathcal{R})$
we choose the wave functions of the CF theory (suppressing the spin part) \cite{Jain89,Jain07} 
\begin{equation}
\begin{aligned}
\Psi_{n/(2pn\pm 1)}  = \mathcal{P}_{\text{LLL}} \Phi_{\pm n_{\uparrow}} \Phi_{\pm n_{\downarrow}} \Phi_1^{2p} 
\label{CFWF}
\end{aligned}
\end{equation}
Here $\Phi_n$ is the wave function for $n$ filled Landau levels, $\Phi_{-n}\equiv [\Phi_n]^*$, and $\mathcal{P}_{\text{LLL}}$ denotes LLL projection, performed below using the method in Refs.~\cite{Jain97,Jain97b,Jain07,Davenport12}. The state of spinfull composite fermions with $n_{\uparrow}$ spin-up and $n_{\downarrow}$ spin-down filled $\Lambda$ levels (CF LLs) is denoted as $(n_{\uparrow},n_{\downarrow})$, with $n=n_{\uparrow}+n_{\downarrow}$.

Our goal is to compute the critical Zeeman energy where a FQH system undergoes a transition from a fully spin-polarized (FP) state into either a partially spin-polarized (PP) or a spin-singlet (SS) state. We first obtain the per particle interaction energies $E_{(n_{\uparrow}, n_{\downarrow})}$ of the states $(n_{\uparrow}, n_{\downarrow})$. The dimensionless critical Zeeman energy $\alpha_{\rm Z}^{\rm crit}$ for the transition between two successive states $(n_{\uparrow}, n_{\downarrow})$ and $(n_{\uparrow}-1, n_{\downarrow}+1)$ is given by
\begin{equation}
\alpha_{\rm Z}^{\rm crit}  = (n_{\uparrow}+n_{\downarrow})\left [ \frac{ E_{(n_{\uparrow}, n_{\downarrow})} - E_{(n_{\uparrow}-1, n_{\downarrow}+1)} }{e^{2}/\epsilon \ell} \right] .
\label{Ec}
\end{equation}
Many previous studies~\cite{Park98, Park99, Davenport12, Balram15a} have used variational Monte Carlo (VMC) calculations to evaluate $\alpha_{\rm Z}^{\rm crit}$ using the LLL wave functions of Eq. (\ref{CFWF}). (For other approaches, see Refs.~\onlinecite{Mandal96,Lopez01,Murthy03,Murthy07}.) To study the effect of LL mixing, we perform a DMC calculation as a function of $\kappa$, which, for parameters appropriate for electron-doped GaAs ($\epsilon=12.5$, electron band mass $m_{\rm b}=0.067m_{e}$), is given by $\kappa\approx 2.6/\sqrt{B [T]} \approx 1.28\sqrt{\nu/(\rho/10^{11}\text{cm}^{-2})}$, where $\rho$ is the areal density.  The DMC result reduces to a VMC result in the limit of $\kappa=0$.

The nonzero transverse width of GaAs-${\mathrm{Al}}_{\mathrm{x}}$${\mathrm{Ga}}_{1\mathrm{-}\mathrm{x}}$As heterojunctions and quantum wells also has a quantitative effect, producing an effective two-dimensional interaction dependent on the transverse wave function $\xi(z)$:
\begin{equation}
V^{\text{eff}}(r) = \frac{e^{2}}{\epsilon} \int dz_{1} \int dz_{2} \frac{|\xi(z_{1})|^{2} |\xi(z_{2})|^{2}}{[r^{2} + (z_{1} - z_{2})^{2}]^{1/2}},
\end{equation}
where $z_{1}$ and $z_{2}$ denote the coordinates perpendicular to the 2D plane, and $r =\sqrt{ (x_{1} - x_{2})^{2}+ (y_{1} - y_{2})^{2}}$. $V^{\text{eff}}(r)$ is less repulsive than the ideal 2D interaction $e^{2}/\epsilon r$ at short distances. In this work, we calculate the critical Zeeman energy using $V^{\text{eff}}(r)$ to include the effect of the finite transverse width. A realistic $\xi(z)$ for each density and geometry is obtained by solving the Schr\"odinger and Poisson equations self-consistently through the local density approximation~\cite{Ortalano97}. Note that the finite-width correction in the VMC results depends on the density through $\xi(z)$.

\begin{figure*}
\resizebox{0.334\textwidth}{!}{\includegraphics{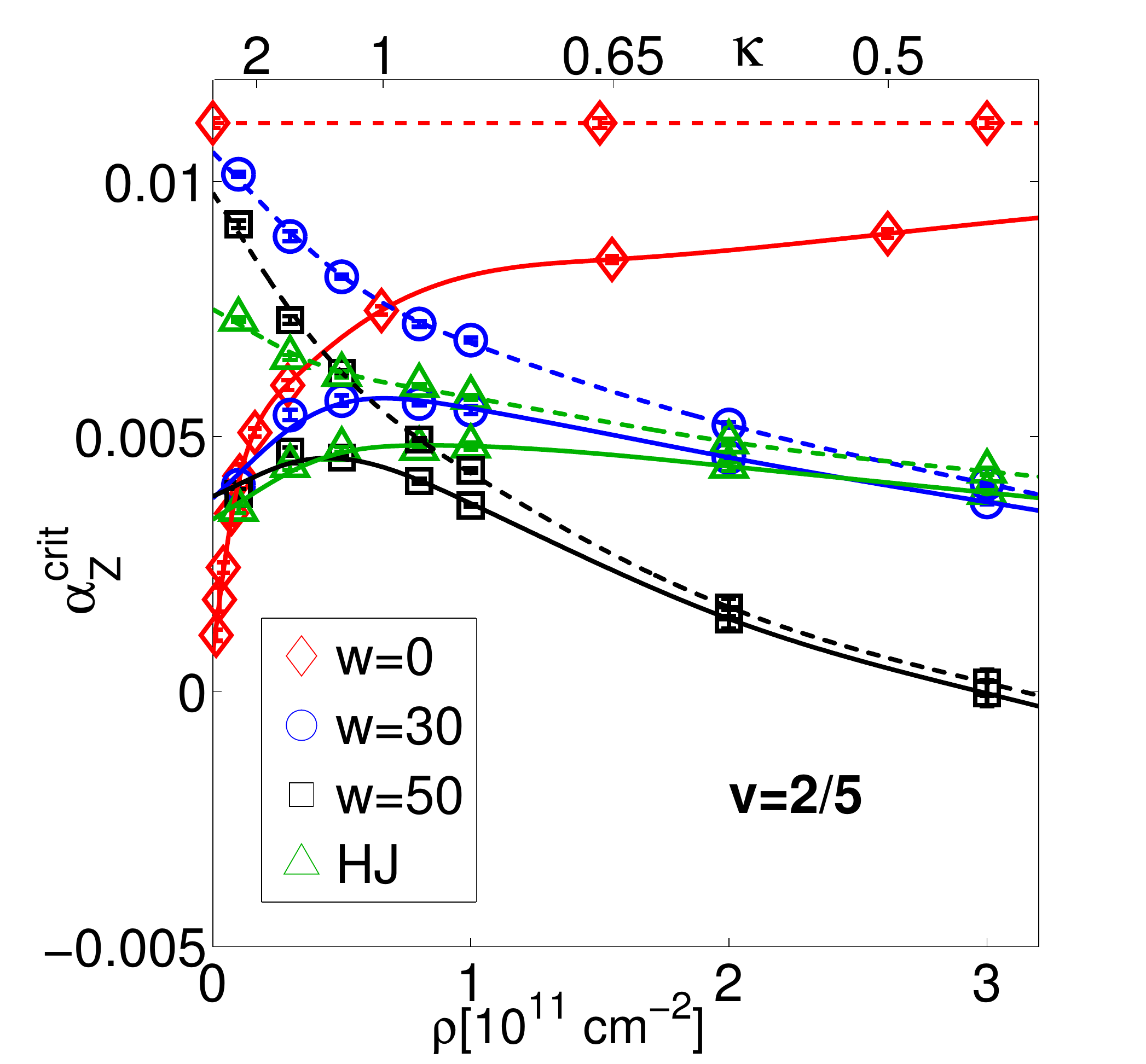}}
\hspace{-3mm}
\resizebox{0.31\textwidth}{!}{\includegraphics{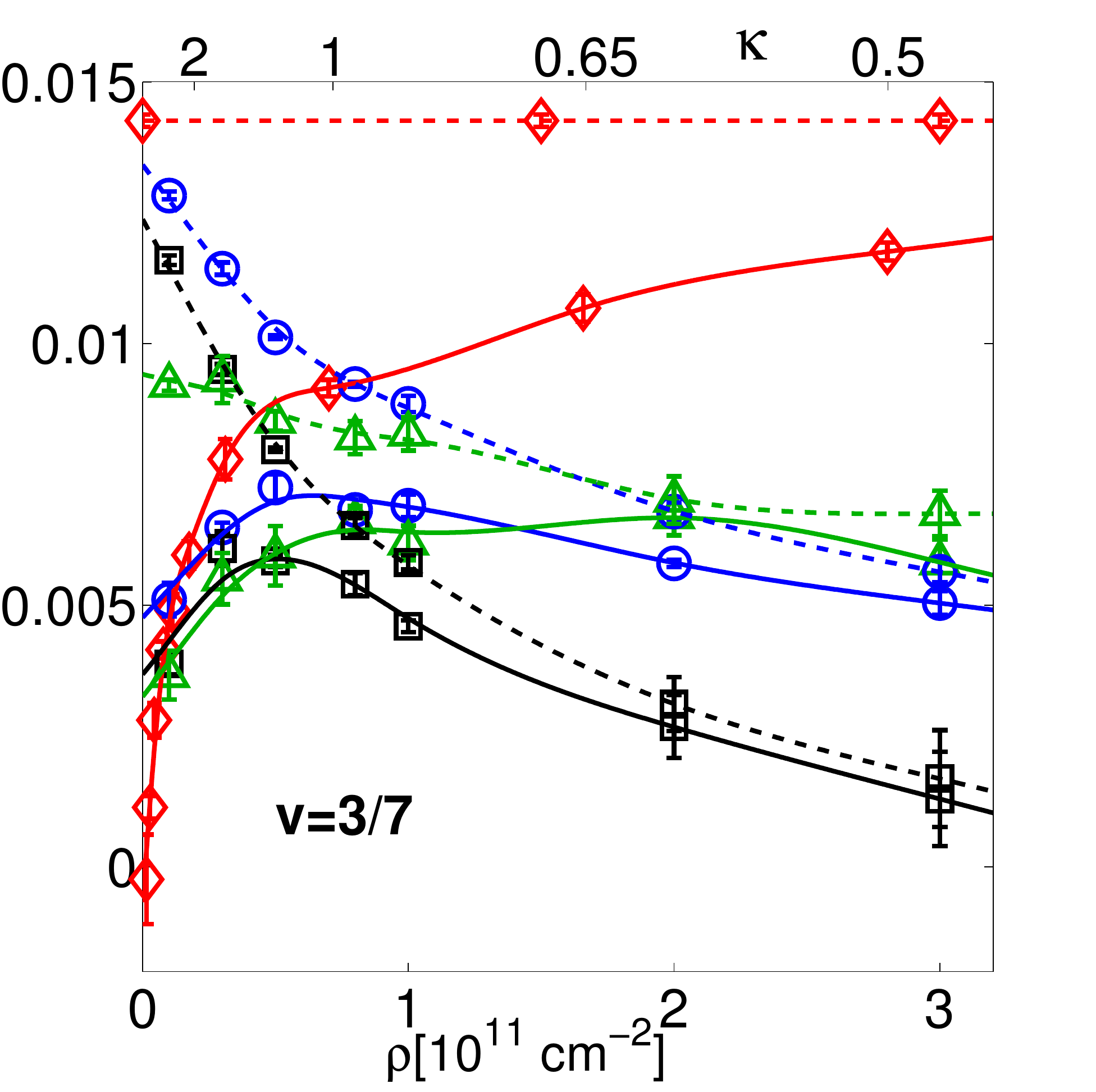}}
\hspace{-3mm}
\resizebox{0.31\textwidth}{!}{\includegraphics{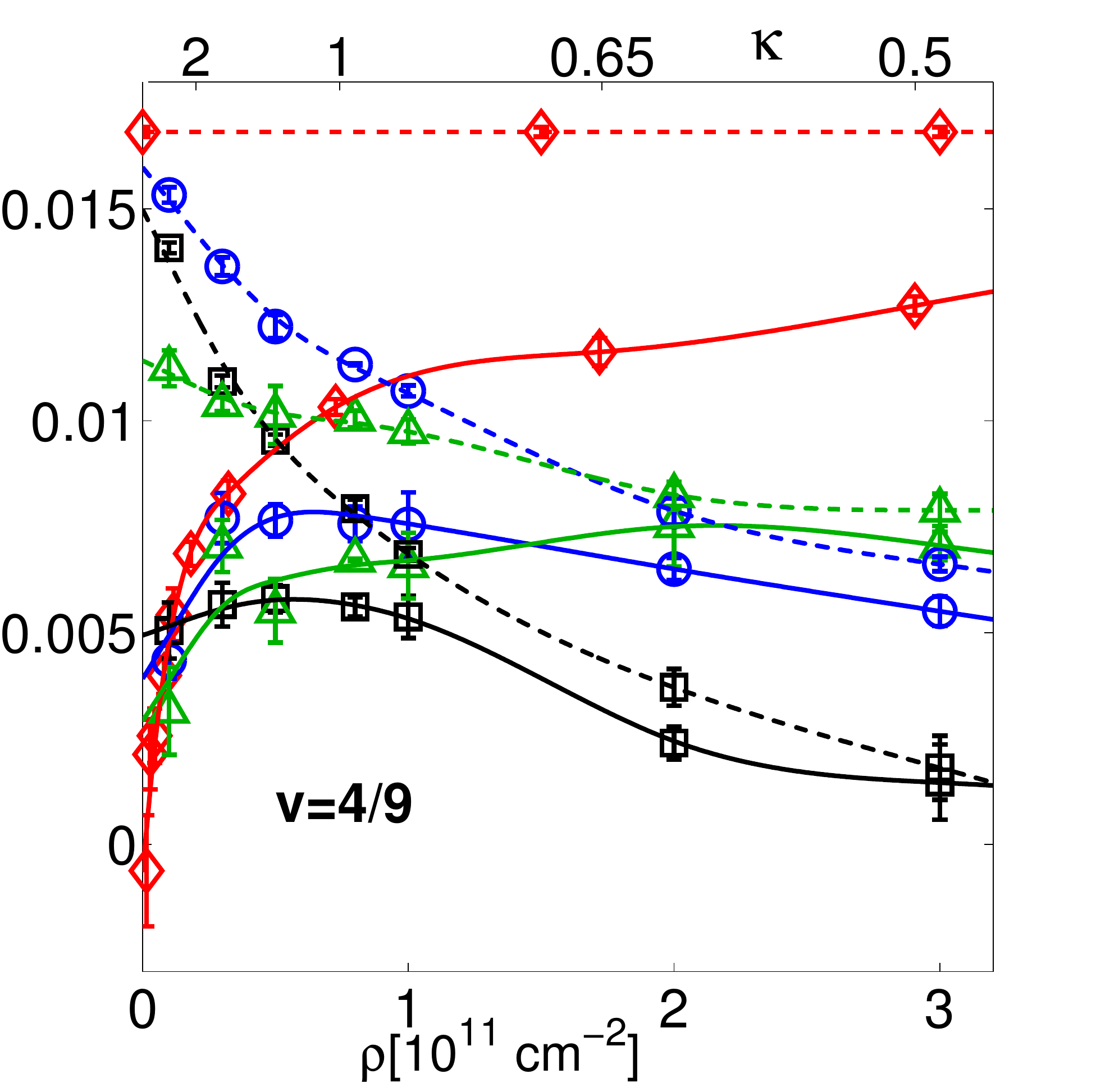}}\\
\vspace{-3mm}
\caption{The theoretical critical Zeeman energies $\alpha_{\rm Z}^{\rm crit}=E_{\rm Z}^{\rm crit}/(e^{2}/\epsilon \ell)$ for $\nu=2/5$ (left), $3/7$ (middle), $4/9$ (right) are shown with empty symbols with error bars, calculated from both DMC (solid line) and VMC (dashed line) methods, for an ideal 2D system ($w=0$), quantum wells with widths $w=30$nm and $50$nm, and heterojunction (HJ). The symbol $\rho$ denotes the electron density. The solid lines for quantum wells and heterojunction display a ``hill'' shape, where, roughly speaking, LL mixing correction dominates on the left of the ``hill'' (at small $\rho$) and finite width correction on the right  (at large $\rho$).
}
\label{v25}
\end{figure*}

In the following, we show our numerical results for $\alpha_{\rm Z}^{\rm crit}$ in the thermodynamic limit and compare them with those obtained from transport experiments. We have used two methods to perform extrapolation to $N \rightarrow \infty$. In method I, we extrapolate the energy difference to the thermodynamic limit. For this purpose, we correct for the finite-size deviation of the density from its asymptotic value by multiplying the finite-size energy with a factor $(2Q\nu/N)^{1/2}$~\cite{Morf86b}, and, if needed, also interpolate the energy to the appropriate particle number. In method II, we extrapolate the density-corrected per particle energies of SS, FP or PP states to the thermodynamic limit separately, and then obtain $\alpha_{\rm Z}^{\rm crit}$ according to Eq. (\ref{Ec}). The results quoted below are obtained from method I unless specified otherwise~\cite{ZhangSM1}. The errors shown below arise primarily from the extrapolation; the statistical error from the Monte Carlo sampling is comparatively negligible.

We first study the FQH states with fillings $\nu=n/(2n+1)$. The critical Zeeman energies $\alpha_{\rm Z}^{\rm crit}$ for $\nu=2/5,3/7$ and 4/9 are shown in Fig. \ref{v25} for an ideal 2D system with width $w=0$, for GaAs-${\mathrm{Al}}_{\mathrm{x}}$${\mathrm{Ga}}_{1\mathrm{-}\mathrm{x}}$As quantum wells with widths $w=30$ and 50 nm, and also for a GaAs-${\mathrm{Al}}_{\mathrm{x}}$${\mathrm{Ga}}_{1\mathrm{-}\mathrm{x}}$As heterojunction (HJ). $\alpha_{\rm Z}^{\rm crit}$ calculated from the DMC and VMC methods are plotted as a function of density $\rho$ with solid and dashed lines, respectively. The value of $\kappa$ is shown at the upper $x$ axis. For quantum wells, $\alpha_{\rm Z}^{\rm crit}$ from the VMC calculation (no LL mixing) decreases with increasing $w$ or $\rho$. The behavior of the $\alpha_{\rm Z}^{\rm crit}$ from the DMC calculation, which includes the correction due to LL mixing, is more complicated. At large $\rho$ (small $\kappa$), the DMC results are close to the VMC results for each width. On the other hand, with decreasing $\rho$ (increasing $\kappa$), the DMC results are increasingly lower than the VMC results. For $\kappa \gtrsim 2$ the DMC results are largely insensitive to $w$, implying that the dominating correction here is due to LL mixing. We note that we have not included in our calculations any physics relating to an instability of the FQH effect into a Wigner crystal at large $\kappa$ \cite{Price93}.

One of the main messages of our calculation is that LL mixing and finite-width corrections significantly reduce the critical Zeeman energy at $\nu=n/(2n+1)$, by a factor of 2 or more for the experimental systems. This is consistent with the fact that, in typical experiments, the FQH states at $\nu<1/2$ are fully spin polarized even with zero tilt of magnetic field. The transitions at $\nu=2/5, 3/7$ have been seen by Kang {\em et al.}~\cite{Kang97} in transport experiments only by significantly decreasing the Land\'e factor $g_{0}$ with the application of hydrostatic pressure. 

For FQH states at $\nu=n/(2n-1)$, where the composite fermions are in a negative effective magnetic field, the wave functions of nonfully spin-polarized states in Eq. (\ref{CFWF}), evaluated with the projection method in Refs.~\cite{Jain97,Davenport12} are not as accurate as those for $n/(2n+1)$ and are known to produce, for $w=0$ and $\kappa=0$, values of $\alpha_{\rm Z}^{\rm crit}$ that are off by up to a factor of 2 relative to the exact results~\cite{Balram15a}. For example, for $\nu=2/3$, the value of $\alpha_{\rm Z}^{\rm crit}=0.0082 (1)$ obtained from the wave functions in Eq. (\ref{CFWF}) is much lower than the value 0.0183(5) obtained from exact diagonalization (ED) for $\kappa=0$ at $w=0$. The reason is because our projection method\cite{Jain97,Davenport12} slightly overestimates the probability of spatial coincidence of electrons in the nonfully polarized states, and thereby overestimates their energies. (The ``hard-core" projection of Ref.~\cite{Wu93} produces very accurate wave functions, but is not amenable to numerical evaluations.) Fortunately, we find that for $\kappa \gtrsim 2$ the results are insensitive to slight differences in the initial trial wave function $\psi_{T}$ because of the relatively large modification due to LL mixing. Taking again the example of $\nu=2/3$, for $\kappa=1.91$, both the exact wave function and the wave function in Eq. (\ref{CFWF}) produce $\alpha_{\rm Z}^{\rm crit} \approx 0.0090$ (see Fig.~\ref{v23v43} and Fig.~S1
in the Supplemental Material~\cite{ZhangSM1}). 

 \begin{figure}
\resizebox{0.46\textwidth}{!}{\includegraphics{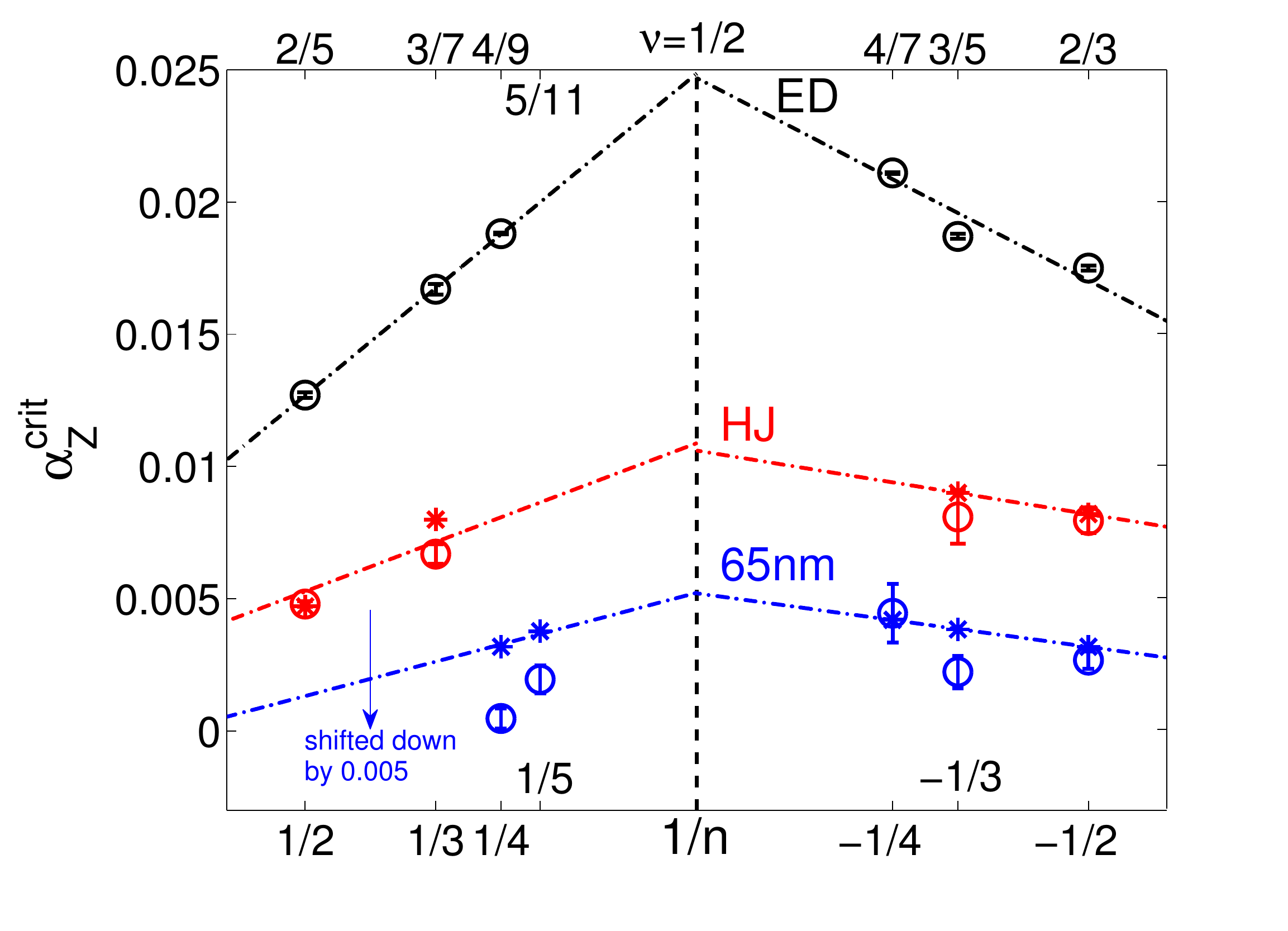}}\\  
\vspace{-4mm}
\caption{
Comparison between experimental values (stars) and theoretical DMC values (empty circles) of $\alpha_{\rm Z}^{\rm crit}=E_{\rm Z}^{\rm crit}/(e^{2}/\epsilon \ell)$ for a $w=65$nm quantum well (blue) from Liu {\em et al.} \cite{Liu14}, and heterojunctions (red) from Engel {\em et al.}\cite{Engel92} and Kang {\em et al.}\cite{Kang97}. (For the experiment of Kang {\em et al.}, we estimate the value of the Land\'e factor $g_0$ by assuming that it changes linearly and passes through zero at a pressure of roughly 18 Kbar \cite{Leadley97}.) The filling factors $\nu=n/(2n+1)$ are shown on top and $1/n$ at the bottom. The black circles show the results obtained from exact-diagonalization (ED) without including any LL mixing or finite-width corrections~\cite{Balram15a} (these do not involve the DMC calculation). The results for the 65 nm quantum well are shifted down by 0.005 for ease of depiction. The dashed lines are a guide to the eye. For the heterojunction, some other experimental values (theoretical predictions) of $\alpha_{\rm Z}^{\rm crit}$ are 0.0109 [0.0076(4)] \cite{Eisenstein90} and 0.0078 [0.0065(4)] \cite{Engel92} at $\nu=2/3$, and 0.0081[0.0080(20)] \cite{Engel92} at $\nu=3/5$; these are not shown on the figure to avoid clutter.}
\label{compare}
\end{figure}

Figure \ref{compare} shows the comparison between experimental data (stars) and  theoretical results (circles) for $\alpha_{\rm Z}^{\rm crit}$ for many states at $\nu=n/(2n\pm 1)$. The theoretical results (red and blue circles) are obtained with DMC calculations for the specific experimental parameters ($\rho$, $w$). The black empty circles show the $\alpha_{\rm Z}^{\rm crit}$ obtained from ED with $\kappa=0$ and $w=0$, taken from Ref.~\cite{Balram15a}.  The experimental values for $\alpha_{\rm Z}^{\rm crit}$ are significantly lower than the ED values, but in reasonably good agreement with our DMC results.

\begin{figure}
\resizebox{0.42\textwidth}{!}{\includegraphics{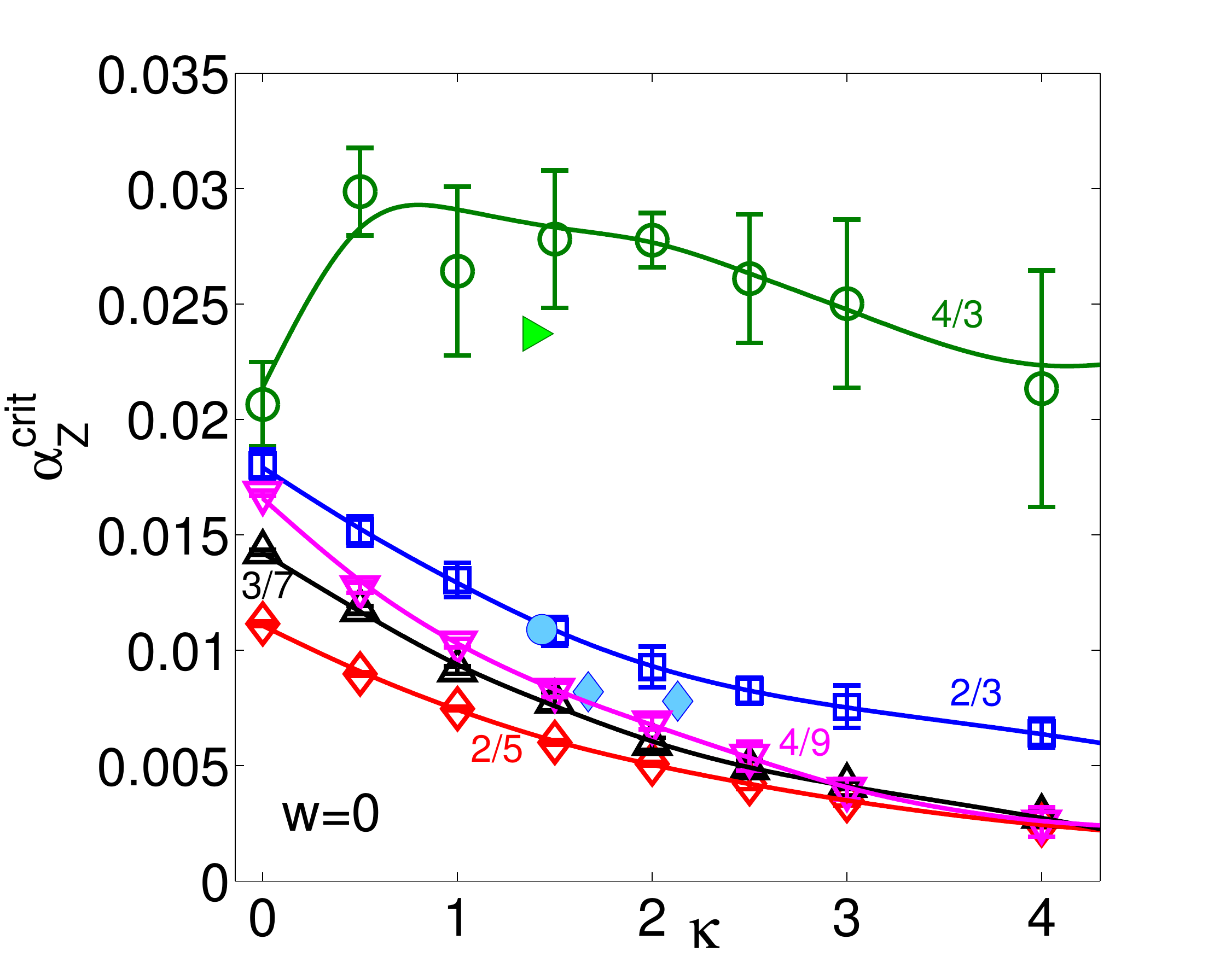}}\\  
\vspace{-3mm}
\caption{Theoretical critical Zeeman energies for the $w=0$ model as a function of the LL mixing parameter $\kappa$ obtained from the DMC method for $\nu=4/3$ (green circle), $2/3$ (blue square), $4/9$ (magenta downward triangle), $3/7$ (black upward triangle), and $2/5$ (red diamond). For the fractions $n/(2n+1)$, the wave functions of Eq. (\ref{CFWF}) are used to fix the phase. For $\nu=2/3$ and $\nu=4/3$ the exact Coulomb state in the LLL is used to fix the phase of the wave function. The solid lines are an approximate guide to the eye. The filled symbols indicate the experimental data from heterojunction samples at $\nu=2/3$ (light blue) and $4/3$ (green) taken from Eisenstein {\em et al.} \cite{Eisenstein90} (circle), Engel {\em et al.} \cite{Engel92} (diamond), and Du {\em et al.}\cite{Du95} (rightward triangle). }
\label{v23v43}
\end{figure}

The corrections due to LL mixing enter in a more dramatic manner when one compares the spin transitions between the filling factor regions $0<\nu<1$ and $1<\nu<2$. Experiments have found (see Fig.~3 of Ref.~\onlinecite{Liu14}) that the $\alpha_{\rm Z}^{\rm crit}$'s for the latter are significantly higher than those for the former.   As noted above, the difference arises from, and, thus, is a measure of, the breaking of the particle-hole symmetry by LL mixing. To address this issue, we find it most convenient (for reasons of computational cost) to compare the spin transitions at $\nu=2/3$ and $\nu=4/3$. To obtain accurate results, we use for our $\psi_{T}$ the exact $\kappa=0$ Coulomb wave functions for the SS states at $2/3$ and $4/3$, Eq. (\ref{CFWF}) for the $2/3$ FP state, and $\Phi_{1\uparrow} \Psi_{1/3 \downarrow}$ for the $4/3$ PP state. For the SS states, we can only calculate for small systems as the exact states contain a large number of Slater determinants. Figure \ref{v23v43} shows the $\alpha_{\rm Z}^{\rm crit}$ for $\nu=4/3$ (green circle) and $\nu=2/3$ (blue square) obtained from the extrapolation method II. The value of $\alpha_{\rm Z}^{\rm crit}$ at $\kappa=0$ is approximately consistent with the exact value $0.0175$~\cite{Balram15a}, giving us confidence in our calculated $\alpha_{\rm Z}^{\rm crit}$ with relatively small system sizes. The main message of Fig. \ref{v23v43} is that the $\alpha_{\rm Z}^{\rm crit}$ at 4/3 is substantially higher than that at 2/3 for the typical experimental value of $\kappa\approx 1-2$.  Note that we only show the zero-width results, because the extrapolation of finite-width results to thermodynamic limit has a poor statistics for such small systems~\cite{ZhangSM1}. We also show in Fig. \ref{v23v43} the experimental data from GaAs-${\mathrm{Al}}_{\mathrm{x}}$${\mathrm{Ga}}_{1\mathrm{-}\mathrm{x}}$As heterojunction samples, because these have the smallest effective width, with solid symbols for $\nu=2/3$ (light blue) and $\nu=4/3$ (green). The agreement with the $w=0$ results is very good, which is not surprising because we know from Fig. \ref{v25} that at relatively large $\kappa$ ($\gtrsim 2$), $\alpha_{\rm Z}^{\rm crit}$ is not very sensitive to the width $w$.

\begin{table}
\begin{tabular}{|c|c|c|}
\hline
\multirow{2}{*}{\quad$\nu$ \quad} & \multicolumn{2}{c|}{$d(\alpha_{\rm Z}^{\rm crit})/d\kappa$}\\
\cline{2-3}  
 & Perturbative & Nonperturbative (DMC)\\
\hline
\quad2/5 \quad & -0.0023 & -0.0043 \\
\hline
\quad3/7 \quad &  -0.0025 &  -0.0050\\
\hline
\quad2/3 \quad &  -0.0135 & -0.0057 \\
\hline
\quad4/3 \quad  & 0.0339 & 0.0184\\
\hline
\end{tabular}
\caption{This table compares the values of $d(\alpha_{\rm Z}^{\rm crit})/d\kappa$ at $\kappa=0$ obtained from the perturbative and the nonperturbative DMC calculations.}
\label{perturbative}
\end{table}

It is natural to ask how well our results agree with those obtained from the perturbative approach in which the effect of LL mixing is incorporated within the LLL theory through an effective interaction, which contains perturbative corrections to the two-body interaction, and, minimally, also a three-body interaction (because the two-body interaction does not break particle-hole symmetry). We discuss this issue for $w=0$. As seen in Fig. \ref{v23v43}, the perturbation theory is {\em in principle} valid for up to $\kappa \approx 1$ for the states $n/(2n\pm 1)$, and up to $\kappa\approx 0.5$ for the states at $2-n/(2n\pm 1)$. In practice, one cannot keep all two-body, three-body and $n$-body terms in the calculation. We have evaluated $\alpha_{\rm Z}^{\rm crit}$~\cite{ZhangSM1} using the interaction given by Peterson and Nayak\cite{Peterson13}, including corrections to the two-body pseudopotentials $V^{(2)}_m$ for $m\leq 5$ and three-body pseudopotentials $V^{(3)}_m$ for $m\leq 3$. Table \ref{perturbative} compares the perturbative $d(\alpha_{\rm Z}^{\rm crit})/d\kappa$ with that deduced from Fig. \ref{v23v43} at small $\kappa$. The two results are substantially different. For example, if the perturbative result is applied to $\kappa=1.5$, it would produce $\alpha_{\rm Z}^{\rm crit}\sim 0.068$ and $-0.003$ for $\nu=4/3$ and 2/3, respectively, to be compared to the DMC values of $\alpha_{\rm Z}^{\rm crit}\sim0.027$ and 0.012.  An exhaustive study of the quantitative importance of the terms left out in the perturbative study 
 is outside the scope of the current study.

To conclude, we 
find that LL mixing substantially suppresses the critical Zeeman energies for the $\nu=n/(2n \pm 1)$ FQH states, and brings theory into satisfactory agreement with experiment. We also find that LL mixing causes an enhancement of the critical Zeeman energy for $\nu=2-n/(2n \pm 1)$, as also seen experimentally. In addition to providing an accurate quantitative comparison between FQH  theory and experiment, our work shows how the quantitative study of the spin physics can shed fundamental light on the role of LL mixing in breaking the particle-hole symmetry of the lowest LL.

\underline{Acknowledgments:}  We are grateful to A. C. Balram, J. Shabani, C. T\"oke, Y.-H. Wu and especially M. Shayegan for very useful discussions. We acknowledge financial support from the DOE Grant No. DE-SC0005042 (Y.Z. and J.K.J), Polish NCN Grant No. 2014/14/A/ST3/00654 (A.W.), and thank Research Computing and CyberInfrastructure at the Pennsylvania State University.

\bibliography{biblio_fqhe.bib}

\begin{thebibliography}{62}%
\makeatletter
\providecommand \@ifxundefined [1]{%
 \@ifx{#1\undefined}
}%
\providecommand \@ifnum [1]{%
 \ifnum #1\expandafter \@firstoftwo
 \else \expandafter \@secondoftwo
 \fi
}%
\providecommand \@ifx [1]{%
 \ifx #1\expandafter \@firstoftwo
 \else \expandafter \@secondoftwo
 \fi
}%
\providecommand \natexlab [1]{#1}%
\providecommand \enquote  [1]{``#1''}%
\providecommand \bibnamefont  [1]{#1}%
\providecommand \bibfnamefont [1]{#1}%
\providecommand \citenamefont [1]{#1}%
\providecommand \href@noop [0]{\@secondoftwo}%
\providecommand \href [0]{\begingroup \@sanitize@url \@href}%
\providecommand \@href[1]{\@@startlink{#1}\@@href}%
\providecommand \@@href[1]{\endgroup#1\@@endlink}%
\providecommand \@sanitize@url [0]{\catcode `\\12\catcode `\$12\catcode
  `\&12\catcode `\#12\catcode `\^12\catcode `\_12\catcode `\%12\relax}%
\providecommand \@@startlink[1]{}%
\providecommand \@@endlink[0]{}%
\providecommand \url  [0]{\begingroup\@sanitize@url \@url }%
\providecommand \@url [1]{\endgroup\@href {#1}{\urlprefix }}%
\providecommand \urlprefix  [0]{URL }%
\providecommand \Eprint [0]{\href }%
\providecommand \doibase [0]{http://dx.doi.org/}%
\providecommand \selectlanguage [0]{\@gobble}%
\providecommand \bibinfo  [0]{\@secondoftwo}%
\providecommand \bibfield  [0]{\@secondoftwo}%
\providecommand \translation [1]{[#1]}%
\providecommand \BibitemOpen [0]{}%
\providecommand \bibitemStop [0]{}%
\providecommand \bibitemNoStop [0]{.\EOS\space}%
\providecommand \EOS [0]{\spacefactor3000\relax}%
\providecommand \BibitemShut  [1]{\csname bibitem#1\endcsname}%
\let\auto@bib@innerbib\@empty
\bibitem [{\citenamefont {Moore}\ and\ \citenamefont {Read}(1991)}]{Moore91}%
  \BibitemOpen
  \bibfield  {author} {\bibinfo {author} {\bibfnamefont {G.}~\bibnamefont
  {Moore}}\ and\ \bibinfo {author} {\bibfnamefont {N.}~\bibnamefont {Read}},\
  }\href {\doibase 10.1016/0550-3213(91)90407-O} {\bibfield  {journal}
  {\bibinfo  {journal} {Nucl. Phys. B}\ }\textbf {\bibinfo {volume} {360}},\
  \bibinfo {pages} {362 } (\bibinfo {year} {1991})}\BibitemShut {NoStop}%
\bibitem [{\citenamefont {Levin}\ \emph {et~al.}(2007)\citenamefont {Levin},
  \citenamefont {Halperin},\ and\ \citenamefont {Rosenow}}]{Levin07}%
  \BibitemOpen
  \bibfield  {author} {\bibinfo {author} {\bibfnamefont {M.}~\bibnamefont
  {Levin}}, \bibinfo {author} {\bibfnamefont {B.~I.}\ \bibnamefont {Halperin}},
  \ and\ \bibinfo {author} {\bibfnamefont {B.}~\bibnamefont {Rosenow}},\ }\href
  {\doibase 10.1103/PhysRevLett.99.236806} {\bibfield  {journal} {\bibinfo
  {journal} {Phys. Rev. Lett.}\ }\textbf {\bibinfo {volume} {99}},\ \bibinfo
  {pages} {236806} (\bibinfo {year} {2007})}\BibitemShut {NoStop}%
\bibitem [{\citenamefont {Lee}\ \emph {et~al.}(2007)\citenamefont {Lee},
  \citenamefont {Ryu}, \citenamefont {Nayak},\ and\ \citenamefont
  {Fisher}}]{Lee07}%
  \BibitemOpen
  \bibfield  {author} {\bibinfo {author} {\bibfnamefont {S.-S.}\ \bibnamefont
  {Lee}}, \bibinfo {author} {\bibfnamefont {S.}~\bibnamefont {Ryu}}, \bibinfo
  {author} {\bibfnamefont {C.}~\bibnamefont {Nayak}}, \ and\ \bibinfo {author}
  {\bibfnamefont {M.~P.~A.}\ \bibnamefont {Fisher}},\ }\href {\doibase
  10.1103/PhysRevLett.99.236807} {\bibfield  {journal} {\bibinfo  {journal}
  {Phys. Rev. Lett.}\ }\textbf {\bibinfo {volume} {99}},\ \bibinfo {pages}
  {236807} (\bibinfo {year} {2007})}\BibitemShut {NoStop}%
\bibitem [{\citenamefont {Bishara}\ and\ \citenamefont
  {Nayak}(2009)}]{Bishara09}%
  \BibitemOpen
  \bibfield  {author} {\bibinfo {author} {\bibfnamefont {W.}~\bibnamefont
  {Bishara}}\ and\ \bibinfo {author} {\bibfnamefont {C.}~\bibnamefont
  {Nayak}},\ }\href {\doibase 10.1103/PhysRevB.80.121302} {\bibfield  {journal}
  {\bibinfo  {journal} {Phys. Rev. B}\ }\textbf {\bibinfo {volume} {80}},\
  \bibinfo {pages} {121302} (\bibinfo {year} {2009})}\BibitemShut {NoStop}%
\bibitem [{\citenamefont {W\'ojs}\ \emph {et~al.}(2010)\citenamefont {W\'ojs},
  \citenamefont {T\H{o}ke},\ and\ \citenamefont {Jain}}]{Wojs10}%
  \BibitemOpen
  \bibfield  {author} {\bibinfo {author} {\bibfnamefont {A.}~\bibnamefont
  {W\'ojs}}, \bibinfo {author} {\bibfnamefont {C.}~\bibnamefont {T\H{o}ke}}, \
  and\ \bibinfo {author} {\bibfnamefont {J.~K.}\ \bibnamefont {Jain}},\ }\href
  {\doibase 10.1103/PhysRevLett.105.096802} {\bibfield  {journal} {\bibinfo
  {journal} {Phys. Rev. Lett.}\ }\textbf {\bibinfo {volume} {105}},\ \bibinfo
  {pages} {096802} (\bibinfo {year} {2010})}\BibitemShut {NoStop}%
\bibitem [{\citenamefont {Sodemann}\ and\ \citenamefont
  {MacDonald}(2013)}]{Sodemann13}%
  \BibitemOpen
  \bibfield  {author} {\bibinfo {author} {\bibfnamefont {I.}~\bibnamefont
  {Sodemann}}\ and\ \bibinfo {author} {\bibfnamefont {A.~H.}\ \bibnamefont
  {MacDonald}},\ }\href {\doibase 10.1103/PhysRevB.87.245425} {\bibfield
  {journal} {\bibinfo  {journal} {Phys. Rev. B}\ }\textbf {\bibinfo {volume}
  {87}},\ \bibinfo {pages} {245425} (\bibinfo {year} {2013})}\BibitemShut
  {NoStop}%
\bibitem [{\citenamefont {Simon}\ and\ \citenamefont {Rezayi}(2013)}]{Simon13}%
  \BibitemOpen
  \bibfield  {author} {\bibinfo {author} {\bibfnamefont {S.~H.}\ \bibnamefont
  {Simon}}\ and\ \bibinfo {author} {\bibfnamefont {E.~H.}\ \bibnamefont
  {Rezayi}},\ }\href {\doibase 10.1103/PhysRevB.87.155426} {\bibfield
  {journal} {\bibinfo  {journal} {Phys. Rev. B}\ }\textbf {\bibinfo {volume}
  {87}},\ \bibinfo {pages} {155426} (\bibinfo {year} {2013})}\BibitemShut
  {NoStop}%
\bibitem [{\citenamefont {Peterson}\ and\ \citenamefont
  {Nayak}(2013)}]{Peterson13}%
  \BibitemOpen
  \bibfield  {author} {\bibinfo {author} {\bibfnamefont {M.~R.}\ \bibnamefont
  {Peterson}}\ and\ \bibinfo {author} {\bibfnamefont {C.}~\bibnamefont
  {Nayak}},\ }\href {\doibase 10.1103/PhysRevB.87.245129} {\bibfield  {journal}
  {\bibinfo  {journal} {Phys. Rev. B}\ }\textbf {\bibinfo {volume} {87}},\
  \bibinfo {pages} {245129} (\bibinfo {year} {2013})}\BibitemShut {NoStop}%
\bibitem [{\citenamefont {Peterson}\ and\ \citenamefont
  {Nayak}(2014)}]{Peterson14}%
  \BibitemOpen
  \bibfield  {author} {\bibinfo {author} {\bibfnamefont {M.~R.}\ \bibnamefont
  {Peterson}}\ and\ \bibinfo {author} {\bibfnamefont {C.}~\bibnamefont
  {Nayak}},\ }\href {\doibase 10.1103/PhysRevLett.113.086401} {\bibfield
  {journal} {\bibinfo  {journal} {Phys. Rev. Lett.}\ }\textbf {\bibinfo
  {volume} {113}},\ \bibinfo {pages} {086401} (\bibinfo {year}
  {2014})}\BibitemShut {NoStop}%
\bibitem [{\citenamefont {Halperin}\ \emph {et~al.}(1993)\citenamefont
  {Halperin}, \citenamefont {Lee},\ and\ \citenamefont {Read}}]{Halperin93}%
  \BibitemOpen
  \bibfield  {author} {\bibinfo {author} {\bibfnamefont {B.~I.}\ \bibnamefont
  {Halperin}}, \bibinfo {author} {\bibfnamefont {P.~A.}\ \bibnamefont {Lee}}, \
  and\ \bibinfo {author} {\bibfnamefont {N.}~\bibnamefont {Read}},\ }\href
  {\doibase 10.1103/PhysRevB.47.7312} {\bibfield  {journal} {\bibinfo
  {journal} {Phys. Rev. B}\ }\textbf {\bibinfo {volume} {47}},\ \bibinfo
  {pages} {7312} (\bibinfo {year} {1993})}\BibitemShut {NoStop}%
\bibitem [{\citenamefont {Barkeshli}\ \emph {et~al.}(2015)\citenamefont
  {Barkeshli}, \citenamefont {Mulligan},\ and\ \citenamefont
  {Fisher}}]{Barkeshli15}%
  \BibitemOpen
  \bibfield  {author} {\bibinfo {author} {\bibfnamefont {M.}~\bibnamefont
  {Barkeshli}}, \bibinfo {author} {\bibfnamefont {M.}~\bibnamefont {Mulligan}},
  \ and\ \bibinfo {author} {\bibfnamefont {M.~P.~A.}\ \bibnamefont {Fisher}},\
  }\href {\doibase 10.1103/PhysRevB.92.165125} {\bibfield  {journal} {\bibinfo
  {journal} {Phys. Rev. B}\ }\textbf {\bibinfo {volume} {92}},\ \bibinfo
  {pages} {165125} (\bibinfo {year} {2015})}\BibitemShut {NoStop}%
\bibitem [{\citenamefont {Son}(2015)}]{Son15}%
  \BibitemOpen
  \bibfield  {author} {\bibinfo {author} {\bibfnamefont {D.~T.}\ \bibnamefont
  {Son}},\ }\href {\doibase 10.1103/PhysRevX.5.031027} {\bibfield  {journal}
  {\bibinfo  {journal} {Phys. Rev. X}\ }\textbf {\bibinfo {volume} {5}},\
  \bibinfo {pages} {031027} (\bibinfo {year} {2015})}\BibitemShut {NoStop}%
\bibitem [{\citenamefont {Wang}\ and\ \citenamefont {Senthil}(2015)}]{Wang15b}%
  \BibitemOpen
  \bibfield  {author} {\bibinfo {author} {\bibfnamefont {C.}~\bibnamefont
  {Wang}}\ and\ \bibinfo {author} {\bibfnamefont {T.}~\bibnamefont {Senthil}},\
  }\href {\doibase 10.1103/PhysRevX.5.041031} {\bibfield  {journal} {\bibinfo
  {journal} {Phys. Rev. X}\ }\textbf {\bibinfo {volume} {5}},\ \bibinfo {pages}
  {041031} (\bibinfo {year} {2015})}\BibitemShut {NoStop}%
\bibitem [{\citenamefont {Metlitski}\ \emph {et~al.}(2015)\citenamefont
  {Metlitski}, \citenamefont {Mross}, \citenamefont {Sachdev},\ and\
  \citenamefont {Senthil}}]{Metlitski15}%
  \BibitemOpen
  \bibfield  {author} {\bibinfo {author} {\bibfnamefont {M.~A.}\ \bibnamefont
  {Metlitski}}, \bibinfo {author} {\bibfnamefont {D.~F.}\ \bibnamefont
  {Mross}}, \bibinfo {author} {\bibfnamefont {S.}~\bibnamefont {Sachdev}}, \
  and\ \bibinfo {author} {\bibfnamefont {T.}~\bibnamefont {Senthil}},\ }\href
  {\doibase 10.1103/PhysRevB.91.115111} {\bibfield  {journal} {\bibinfo
  {journal} {Phys. Rev. B}\ }\textbf {\bibinfo {volume} {91}},\ \bibinfo
  {pages} {115111} (\bibinfo {year} {2015})}\BibitemShut {NoStop}%
\bibitem [{\citenamefont {Wang}\ and\ \citenamefont {Senthil}(2016)}]{Wang15}%
  \BibitemOpen
  \bibfield  {author} {\bibinfo {author} {\bibfnamefont {C.}~\bibnamefont
  {Wang}}\ and\ \bibinfo {author} {\bibfnamefont {T.}~\bibnamefont {Senthil}},\
  }\href {\doibase 10.1103/PhysRevB.93.085110} {\bibfield  {journal} {\bibinfo
  {journal} {Phys. Rev. B}\ }\textbf {\bibinfo {volume} {93}},\ \bibinfo
  {pages} {085110} (\bibinfo {year} {2016})}\BibitemShut {NoStop}%
\bibitem [{\citenamefont {Geraedts}\ \emph {et~al.}(2016)\citenamefont
  {Geraedts}, \citenamefont {Zaletel}, \citenamefont {Mong}, \citenamefont
  {Metlitski}, \citenamefont {Vishwanath},\ and\ \citenamefont
  {Motrunich}}]{Geraedts15}%
  \BibitemOpen
  \bibfield  {author} {\bibinfo {author} {\bibfnamefont {S.~D.}\ \bibnamefont
  {Geraedts}}, \bibinfo {author} {\bibfnamefont {M.~P.}\ \bibnamefont
  {Zaletel}}, \bibinfo {author} {\bibfnamefont {R.~S.~K.}\ \bibnamefont
  {Mong}}, \bibinfo {author} {\bibfnamefont {M.~A.}\ \bibnamefont {Metlitski}},
  \bibinfo {author} {\bibfnamefont {A.}~\bibnamefont {Vishwanath}}, \ and\
  \bibinfo {author} {\bibfnamefont {O.~I.}\ \bibnamefont {Motrunich}},\ }\href
  {\doibase 10.1126/science.aad4302} {\bibfield  {journal} {\bibinfo  {journal}
  {Science}\ }\textbf {\bibinfo {volume} {352}},\ \bibinfo {pages} {197}
  (\bibinfo {year} {2016})},\ \Eprint
  {http://arxiv.org/abs/http://science.sciencemag.org/content/352/6282/197.full.pdf}
  {http://science.sciencemag.org/content/352/6282/197.full.pdf} \BibitemShut
  {NoStop}%
\bibitem [{\citenamefont {Potter}\ \emph {et~al.}(2016)\citenamefont {Potter},
  \citenamefont {Serbyn},\ and\ \citenamefont {Vishwanath}}]{Potter16}%
  \BibitemOpen
  \bibfield  {author} {\bibinfo {author} {\bibfnamefont {A.~C.}\ \bibnamefont
  {Potter}}, \bibinfo {author} {\bibfnamefont {M.}~\bibnamefont {Serbyn}}, \
  and\ \bibinfo {author} {\bibfnamefont {A.}~\bibnamefont {Vishwanath}},\
  }\href {\doibase 10.1103/PhysRevX.6.031026} {\bibfield  {journal} {\bibinfo
  {journal} {Phys. Rev. X}\ }\textbf {\bibinfo {volume} {6}},\ \bibinfo {pages}
  {031026} (\bibinfo {year} {2016})}\BibitemShut {NoStop}%
\bibitem [{\citenamefont {Murthy}\ and\ \citenamefont
  {Shankar}(2016)}]{Murthy15}%
  \BibitemOpen
  \bibfield  {author} {\bibinfo {author} {\bibfnamefont {G.}~\bibnamefont
  {Murthy}}\ and\ \bibinfo {author} {\bibfnamefont {R.}~\bibnamefont
  {Shankar}},\ }\href {\doibase 10.1103/PhysRevB.93.085405} {\bibfield
  {journal} {\bibinfo  {journal} {Phys. Rev. B}\ }\textbf {\bibinfo {volume}
  {93}},\ \bibinfo {pages} {085405} (\bibinfo {year} {2016})}\BibitemShut
  {NoStop}%
\bibitem [{\citenamefont {{Metlitski}}(2015)}]{Metlitski15a}%
  \BibitemOpen
  \bibfield  {author} {\bibinfo {author} {\bibfnamefont {M.~A.}\ \bibnamefont
  {{Metlitski}}},\ }\href@noop {} {\bibfield  {journal} {\bibinfo  {journal}
  {ArXiv e-prints}\ } (\bibinfo {year} {2015})},\ \Eprint
  {http://arxiv.org/abs/1510.05663} {arXiv:1510.05663 [hep-th]} \BibitemShut
  {NoStop}%
\bibitem [{\citenamefont {Mross}\ \emph {et~al.}(2016)\citenamefont {Mross},
  \citenamefont {Alicea},\ and\ \citenamefont {Motrunich}}]{Mross16}%
  \BibitemOpen
  \bibfield  {author} {\bibinfo {author} {\bibfnamefont {D.~F.}\ \bibnamefont
  {Mross}}, \bibinfo {author} {\bibfnamefont {J.}~\bibnamefont {Alicea}}, \
  and\ \bibinfo {author} {\bibfnamefont {O.~I.}\ \bibnamefont {Motrunich}},\
  }\href {\doibase 10.1103/PhysRevLett.117.016802} {\bibfield  {journal}
  {\bibinfo  {journal} {Phys. Rev. Lett.}\ }\textbf {\bibinfo {volume} {117}},\
  \bibinfo {pages} {016802} (\bibinfo {year} {2016})}\BibitemShut {NoStop}%
\bibitem [{\citenamefont {Kachru}\ \emph {et~al.}(2015)\citenamefont {Kachru},
  \citenamefont {Mulligan}, \citenamefont {Torroba},\ and\ \citenamefont
  {Wang}}]{Kachru15}%
  \BibitemOpen
  \bibfield  {author} {\bibinfo {author} {\bibfnamefont {S.}~\bibnamefont
  {Kachru}}, \bibinfo {author} {\bibfnamefont {M.}~\bibnamefont {Mulligan}},
  \bibinfo {author} {\bibfnamefont {G.}~\bibnamefont {Torroba}}, \ and\
  \bibinfo {author} {\bibfnamefont {H.}~\bibnamefont {Wang}},\ }\href {\doibase
  10.1103/PhysRevB.92.235105} {\bibfield  {journal} {\bibinfo  {journal} {Phys.
  Rev. B}\ }\textbf {\bibinfo {volume} {92}},\ \bibinfo {pages} {235105}
  (\bibinfo {year} {2015})}\BibitemShut {NoStop}%
\bibitem [{\citenamefont {Mulligan}\ \emph {et~al.}(2016)\citenamefont
  {Mulligan}, \citenamefont {Raghu},\ and\ \citenamefont
  {Fisher}}]{Mulligan16}%
  \BibitemOpen
  \bibfield  {author} {\bibinfo {author} {\bibfnamefont {M.}~\bibnamefont
  {Mulligan}}, \bibinfo {author} {\bibfnamefont {S.}~\bibnamefont {Raghu}}, \
  and\ \bibinfo {author} {\bibfnamefont {M.~P.~A.}\ \bibnamefont {Fisher}},\
  }\href {\doibase 10.1103/PhysRevB.94.075101} {\bibfield  {journal} {\bibinfo
  {journal} {Phys. Rev. B}\ }\textbf {\bibinfo {volume} {94}},\ \bibinfo
  {pages} {075101} (\bibinfo {year} {2016})}\BibitemShut {NoStop}%
\bibitem [{\citenamefont {Levin}\ and\ \citenamefont {Son}()}]{Levin16}%
  \BibitemOpen
  \bibfield  {author} {\bibinfo {author} {\bibfnamefont {M.}~\bibnamefont
  {Levin}}\ and\ \bibinfo {author} {\bibfnamefont {D.~T.}\ \bibnamefont
  {Son}},\ }\href@noop {} {\enquote {\bibinfo {title} {Particle-hole symmetry
  and the nature of the composite fermion},}\ }\bibinfo {note}
  {Unpublished}\BibitemShut {NoStop}%
\bibitem [{\citenamefont {Balram}\ and\ \citenamefont
  {Jain}(2016)}]{Balram16b}%
  \BibitemOpen
  \bibfield  {author} {\bibinfo {author} {\bibfnamefont {A.~C.}\ \bibnamefont
  {Balram}}\ and\ \bibinfo {author} {\bibfnamefont {J.~K.}\ \bibnamefont
  {Jain}},\ }\href {\doibase 10.1103/PhysRevB.93.235152} {\bibfield  {journal}
  {\bibinfo  {journal} {Phys. Rev. B}\ }\textbf {\bibinfo {volume} {93}},\
  \bibinfo {pages} {235152} (\bibinfo {year} {2016})}\BibitemShut {NoStop}%
\bibitem [{\citenamefont {Ortiz}\ \emph {et~al.}(1993)\citenamefont {Ortiz},
  \citenamefont {Ceperley},\ and\ \citenamefont {Martin}}]{Ortiz93}%
  \BibitemOpen
  \bibfield  {author} {\bibinfo {author} {\bibfnamefont {G.}~\bibnamefont
  {Ortiz}}, \bibinfo {author} {\bibfnamefont {D.~M.}\ \bibnamefont {Ceperley}},
  \ and\ \bibinfo {author} {\bibfnamefont {R.~M.}\ \bibnamefont {Martin}},\
  }\href {\doibase 10.1103/PhysRevLett.71.2777} {\bibfield  {journal} {\bibinfo
   {journal} {Phys. Rev. Lett.}\ }\textbf {\bibinfo {volume} {71}},\ \bibinfo
  {pages} {2777} (\bibinfo {year} {1993})}\BibitemShut {NoStop}%
\bibitem [{\citenamefont {Melik-Alaverdian}\ \emph {et~al.}(1997)\citenamefont
  {Melik-Alaverdian}, \citenamefont {Bonesteel},\ and\ \citenamefont
  {Ortiz}}]{Melik-Alaverdian97}%
  \BibitemOpen
  \bibfield  {author} {\bibinfo {author} {\bibfnamefont {V.}~\bibnamefont
  {Melik-Alaverdian}}, \bibinfo {author} {\bibfnamefont {N.~E.}\ \bibnamefont
  {Bonesteel}}, \ and\ \bibinfo {author} {\bibfnamefont {G.}~\bibnamefont
  {Ortiz}},\ }\href {\doibase 10.1103/PhysRevLett.79.5286} {\bibfield
  {journal} {\bibinfo  {journal} {Phys. Rev. Lett.}\ }\textbf {\bibinfo
  {volume} {79}},\ \bibinfo {pages} {5286} (\bibinfo {year}
  {1997})}\BibitemShut {NoStop}%
\bibitem [{\citenamefont {Melik-Alaverdian}\ \emph {et~al.}(2001)\citenamefont
  {Melik-Alaverdian}, \citenamefont {Ortiz},\ and\ \citenamefont
  {Bonesteel}}]{Melik-Alaverdian01}%
  \BibitemOpen
  \bibfield  {author} {\bibinfo {author} {\bibfnamefont {V.}~\bibnamefont
  {Melik-Alaverdian}}, \bibinfo {author} {\bibfnamefont {G.}~\bibnamefont
  {Ortiz}}, \ and\ \bibinfo {author} {\bibfnamefont {N.}~\bibnamefont
  {Bonesteel}},\ }\href {\doibase 10.1023/A:1010326231389} {\bibfield
  {journal} {\bibinfo  {journal} {Journal of Statistical Physics}\ }\textbf
  {\bibinfo {volume} {104}},\ \bibinfo {pages} {449} (\bibinfo {year}
  {2001})}\BibitemShut {NoStop}%
\bibitem [{\citenamefont {Eisenstein}\ \emph {et~al.}(1989)\citenamefont
  {Eisenstein}, \citenamefont {Stormer}, \citenamefont {Pfeiffer},\ and\
  \citenamefont {West}}]{Eisenstein89}%
  \BibitemOpen
  \bibfield  {author} {\bibinfo {author} {\bibfnamefont {J.~P.}\ \bibnamefont
  {Eisenstein}}, \bibinfo {author} {\bibfnamefont {H.~L.}\ \bibnamefont
  {Stormer}}, \bibinfo {author} {\bibfnamefont {L.}~\bibnamefont {Pfeiffer}}, \
  and\ \bibinfo {author} {\bibfnamefont {K.~W.}\ \bibnamefont {West}},\ }\href
  {\doibase 10.1103/PhysRevLett.62.1540} {\bibfield  {journal} {\bibinfo
  {journal} {Phys. Rev. Lett.}\ }\textbf {\bibinfo {volume} {62}},\ \bibinfo
  {pages} {1540} (\bibinfo {year} {1989})}\BibitemShut {NoStop}%
\bibitem [{\citenamefont {Eisenstein}\ \emph {et~al.}(1990)\citenamefont
  {Eisenstein}, \citenamefont {Stormer}, \citenamefont {Pfeiffer},\ and\
  \citenamefont {West}}]{Eisenstein90}%
  \BibitemOpen
  \bibfield  {author} {\bibinfo {author} {\bibfnamefont {J.~P.}\ \bibnamefont
  {Eisenstein}}, \bibinfo {author} {\bibfnamefont {H.~L.}\ \bibnamefont
  {Stormer}}, \bibinfo {author} {\bibfnamefont {L.~N.}\ \bibnamefont
  {Pfeiffer}}, \ and\ \bibinfo {author} {\bibfnamefont {K.~W.}\ \bibnamefont
  {West}},\ }\href {\doibase 10.1103/PhysRevB.41.7910} {\bibfield  {journal}
  {\bibinfo  {journal} {Phys. Rev. B}\ }\textbf {\bibinfo {volume} {41}},\
  \bibinfo {pages} {7910} (\bibinfo {year} {1990})}\BibitemShut {NoStop}%
\bibitem [{\citenamefont {Engel}\ \emph {et~al.}(1992)\citenamefont {Engel},
  \citenamefont {Hwang}, \citenamefont {Sajoto}, \citenamefont {Tsui},\ and\
  \citenamefont {Shayegan}}]{Engel92}%
  \BibitemOpen
  \bibfield  {author} {\bibinfo {author} {\bibfnamefont {L.~W.}\ \bibnamefont
  {Engel}}, \bibinfo {author} {\bibfnamefont {S.~W.}\ \bibnamefont {Hwang}},
  \bibinfo {author} {\bibfnamefont {T.}~\bibnamefont {Sajoto}}, \bibinfo
  {author} {\bibfnamefont {D.~C.}\ \bibnamefont {Tsui}}, \ and\ \bibinfo
  {author} {\bibfnamefont {M.}~\bibnamefont {Shayegan}},\ }\href {\doibase
  10.1103/PhysRevB.45.3418} {\bibfield  {journal} {\bibinfo  {journal} {Phys.
  Rev. B}\ }\textbf {\bibinfo {volume} {45}},\ \bibinfo {pages} {3418}
  (\bibinfo {year} {1992})}\BibitemShut {NoStop}%
\bibitem [{\citenamefont {Du}\ \emph {et~al.}(1995)\citenamefont {Du},
  \citenamefont {Yeh}, \citenamefont {Stormer}, \citenamefont {Tsui},
  \citenamefont {Pfeiffer},\ and\ \citenamefont {West}}]{Du95}%
  \BibitemOpen
  \bibfield  {author} {\bibinfo {author} {\bibfnamefont {R.~R.}\ \bibnamefont
  {Du}}, \bibinfo {author} {\bibfnamefont {A.~S.}\ \bibnamefont {Yeh}},
  \bibinfo {author} {\bibfnamefont {H.~L.}\ \bibnamefont {Stormer}}, \bibinfo
  {author} {\bibfnamefont {D.~C.}\ \bibnamefont {Tsui}}, \bibinfo {author}
  {\bibfnamefont {L.~N.}\ \bibnamefont {Pfeiffer}}, \ and\ \bibinfo {author}
  {\bibfnamefont {K.~W.}\ \bibnamefont {West}},\ }\href {\doibase
  10.1103/PhysRevLett.75.3926} {\bibfield  {journal} {\bibinfo  {journal}
  {Phys. Rev. Lett.}\ }\textbf {\bibinfo {volume} {75}},\ \bibinfo {pages}
  {3926} (\bibinfo {year} {1995})}\BibitemShut {NoStop}%
\bibitem [{\citenamefont {Kang}\ \emph {et~al.}(1997)\citenamefont {Kang},
  \citenamefont {Young}, \citenamefont {Hannahs}, \citenamefont {Palm},
  \citenamefont {Campman},\ and\ \citenamefont {Gossard}}]{Kang97}%
  \BibitemOpen
  \bibfield  {author} {\bibinfo {author} {\bibfnamefont {W.}~\bibnamefont
  {Kang}}, \bibinfo {author} {\bibfnamefont {J.~B.}\ \bibnamefont {Young}},
  \bibinfo {author} {\bibfnamefont {S.~T.}\ \bibnamefont {Hannahs}}, \bibinfo
  {author} {\bibfnamefont {E.}~\bibnamefont {Palm}}, \bibinfo {author}
  {\bibfnamefont {K.~L.}\ \bibnamefont {Campman}}, \ and\ \bibinfo {author}
  {\bibfnamefont {A.~C.}\ \bibnamefont {Gossard}},\ }\href {\doibase
  10.1103/PhysRevB.56.R12776} {\bibfield  {journal} {\bibinfo  {journal} {Phys.
  Rev. B}\ }\textbf {\bibinfo {volume} {56}},\ \bibinfo {pages} {R12776}
  (\bibinfo {year} {1997})}\BibitemShut {NoStop}%
\bibitem [{\citenamefont {Kukushkin}\ \emph {et~al.}(1999)\citenamefont
  {Kukushkin}, \citenamefont {v.~Klitzing},\ and\ \citenamefont
  {Eberl}}]{Kukushkin99}%
  \BibitemOpen
  \bibfield  {author} {\bibinfo {author} {\bibfnamefont {I.~V.}\ \bibnamefont
  {Kukushkin}}, \bibinfo {author} {\bibfnamefont {K.}~\bibnamefont
  {v.~Klitzing}}, \ and\ \bibinfo {author} {\bibfnamefont {K.}~\bibnamefont
  {Eberl}},\ }\href {\doibase 10.1103/PhysRevLett.82.3665} {\bibfield
  {journal} {\bibinfo  {journal} {Phys. Rev. Lett.}\ }\textbf {\bibinfo
  {volume} {82}},\ \bibinfo {pages} {3665} (\bibinfo {year}
  {1999})}\BibitemShut {NoStop}%
\bibitem [{\citenamefont {Yeh}\ \emph {et~al.}(1999)\citenamefont {Yeh},
  \citenamefont {Stormer}, \citenamefont {Tsui}, \citenamefont {Pfeiffer},
  \citenamefont {Baldwin},\ and\ \citenamefont {West}}]{Yeh99}%
  \BibitemOpen
  \bibfield  {author} {\bibinfo {author} {\bibfnamefont {A.~S.}\ \bibnamefont
  {Yeh}}, \bibinfo {author} {\bibfnamefont {H.~L.}\ \bibnamefont {Stormer}},
  \bibinfo {author} {\bibfnamefont {D.~C.}\ \bibnamefont {Tsui}}, \bibinfo
  {author} {\bibfnamefont {L.~N.}\ \bibnamefont {Pfeiffer}}, \bibinfo {author}
  {\bibfnamefont {K.~W.}\ \bibnamefont {Baldwin}}, \ and\ \bibinfo {author}
  {\bibfnamefont {K.~W.}\ \bibnamefont {West}},\ }\href {\doibase
  10.1103/PhysRevLett.82.592} {\bibfield  {journal} {\bibinfo  {journal} {Phys.
  Rev. Lett.}\ }\textbf {\bibinfo {volume} {82}},\ \bibinfo {pages} {592}
  (\bibinfo {year} {1999})}\BibitemShut {NoStop}%
\bibitem [{\citenamefont {Kukushkin}\ \emph {et~al.}(2000)\citenamefont
  {Kukushkin}, \citenamefont {Smet}, \citenamefont {von Klitzing},\ and\
  \citenamefont {Eberl}}]{Kukushkin00}%
  \BibitemOpen
  \bibfield  {author} {\bibinfo {author} {\bibfnamefont {I.~V.}\ \bibnamefont
  {Kukushkin}}, \bibinfo {author} {\bibfnamefont {J.~H.}\ \bibnamefont {Smet}},
  \bibinfo {author} {\bibfnamefont {K.}~\bibnamefont {von Klitzing}}, \ and\
  \bibinfo {author} {\bibfnamefont {K.}~\bibnamefont {Eberl}},\ }\href
  {\doibase 10.1103/PhysRevLett.85.3688} {\bibfield  {journal} {\bibinfo
  {journal} {Phys. Rev. Lett.}\ }\textbf {\bibinfo {volume} {85}},\ \bibinfo
  {pages} {3688} (\bibinfo {year} {2000})}\BibitemShut {NoStop}%
\bibitem [{\citenamefont {Melinte}\ \emph {et~al.}(2000)\citenamefont
  {Melinte}, \citenamefont {Freytag}, \citenamefont {Horvatic}, \citenamefont
  {Berthier}, \citenamefont {L\'evy}, \citenamefont {Bayot},\ and\
  \citenamefont {Shayegan}}]{Melinte00}%
  \BibitemOpen
  \bibfield  {author} {\bibinfo {author} {\bibfnamefont {S.}~\bibnamefont
  {Melinte}}, \bibinfo {author} {\bibfnamefont {N.}~\bibnamefont {Freytag}},
  \bibinfo {author} {\bibfnamefont {M.}~\bibnamefont {Horvatic}}, \bibinfo
  {author} {\bibfnamefont {C.}~\bibnamefont {Berthier}}, \bibinfo {author}
  {\bibfnamefont {L.~P.}\ \bibnamefont {L\'evy}}, \bibinfo {author}
  {\bibfnamefont {V.}~\bibnamefont {Bayot}}, \ and\ \bibinfo {author}
  {\bibfnamefont {M.}~\bibnamefont {Shayegan}},\ }\href {\doibase
  10.1103/PhysRevLett.84.354} {\bibfield  {journal} {\bibinfo  {journal} {Phys.
  Rev. Lett.}\ }\textbf {\bibinfo {volume} {84}},\ \bibinfo {pages} {354}
  (\bibinfo {year} {2000})}\BibitemShut {NoStop}%
\bibitem [{\citenamefont {Freytag}\ \emph {et~al.}(2001)\citenamefont
  {Freytag}, \citenamefont {Tokunaga}, \citenamefont {Horvati\'{c}},
  \citenamefont {Berthier}, \citenamefont {Shayegan},\ and\ \citenamefont
  {L\'evy}}]{Freytag01}%
  \BibitemOpen
  \bibfield  {author} {\bibinfo {author} {\bibfnamefont {N.}~\bibnamefont
  {Freytag}}, \bibinfo {author} {\bibfnamefont {Y.}~\bibnamefont {Tokunaga}},
  \bibinfo {author} {\bibfnamefont {M.}~\bibnamefont {Horvati\'{c}}}, \bibinfo
  {author} {\bibfnamefont {C.}~\bibnamefont {Berthier}}, \bibinfo {author}
  {\bibfnamefont {M.}~\bibnamefont {Shayegan}}, \ and\ \bibinfo {author}
  {\bibfnamefont {L.~P.}\ \bibnamefont {L\'evy}},\ }\href {\doibase
  10.1103/PhysRevLett.87.136801} {\bibfield  {journal} {\bibinfo  {journal}
  {Phys. Rev. Lett.}\ }\textbf {\bibinfo {volume} {87}},\ \bibinfo {pages}
  {136801} (\bibinfo {year} {2001})}\BibitemShut {NoStop}%
\bibitem [{\citenamefont {Tiemann}\ \emph {et~al.}(2012)\citenamefont
  {Tiemann}, \citenamefont {Gamez}, \citenamefont {Kumada},\ and\ \citenamefont
  {Muraki}}]{Tiemann12}%
  \BibitemOpen
  \bibfield  {author} {\bibinfo {author} {\bibfnamefont {L.}~\bibnamefont
  {Tiemann}}, \bibinfo {author} {\bibfnamefont {G.}~\bibnamefont {Gamez}},
  \bibinfo {author} {\bibfnamefont {N.}~\bibnamefont {Kumada}}, \ and\ \bibinfo
  {author} {\bibfnamefont {K.}~\bibnamefont {Muraki}},\ }\href {\doibase
  10.1126/science.1216697} {\bibfield  {journal} {\bibinfo  {journal}
  {Science}\ }\textbf {\bibinfo {volume} {335}},\ \bibinfo {pages} {828}
  (\bibinfo {year} {2012})},\ \Eprint
  {http://arxiv.org/abs/http://www.sciencemag.org/content/335/6070/828.full.pdf}
  {http://www.sciencemag.org/content/335/6070/828.full.pdf} \BibitemShut
  {NoStop}%
\bibitem [{\citenamefont {Feldman}\ \emph {et~al.}(2013)\citenamefont
  {Feldman}, \citenamefont {Levin}, \citenamefont {Krauss}, \citenamefont
  {Abanin}, \citenamefont {Halperin}, \citenamefont {Smet},\ and\ \citenamefont
  {Yacoby}}]{Feldman13}%
  \BibitemOpen
  \bibfield  {author} {\bibinfo {author} {\bibfnamefont {B.~E.}\ \bibnamefont
  {Feldman}}, \bibinfo {author} {\bibfnamefont {A.~J.}\ \bibnamefont {Levin}},
  \bibinfo {author} {\bibfnamefont {B.}~\bibnamefont {Krauss}}, \bibinfo
  {author} {\bibfnamefont {D.~A.}\ \bibnamefont {Abanin}}, \bibinfo {author}
  {\bibfnamefont {B.~I.}\ \bibnamefont {Halperin}}, \bibinfo {author}
  {\bibfnamefont {J.~H.}\ \bibnamefont {Smet}}, \ and\ \bibinfo {author}
  {\bibfnamefont {A.}~\bibnamefont {Yacoby}},\ }\href {\doibase
  10.1103/PhysRevLett.111.076802} {\bibfield  {journal} {\bibinfo  {journal}
  {Phys. Rev. Lett.}\ }\textbf {\bibinfo {volume} {111}},\ \bibinfo {pages}
  {076802} (\bibinfo {year} {2013})}\BibitemShut {NoStop}%
\bibitem [{\citenamefont {Liu}\ \emph {et~al.}(2014)\citenamefont {Liu},
  \citenamefont {Hasdemir}, \citenamefont {W\'ojs}, \citenamefont {Jain},
  \citenamefont {Pfeiffer}, \citenamefont {West}, \citenamefont {Baldwin},\
  and\ \citenamefont {Shayegan}}]{Liu14}%
  \BibitemOpen
  \bibfield  {author} {\bibinfo {author} {\bibfnamefont {Y.}~\bibnamefont
  {Liu}}, \bibinfo {author} {\bibfnamefont {S.}~\bibnamefont {Hasdemir}},
  \bibinfo {author} {\bibfnamefont {A.}~\bibnamefont {W\'ojs}}, \bibinfo
  {author} {\bibfnamefont {J.~K.}\ \bibnamefont {Jain}}, \bibinfo {author}
  {\bibfnamefont {L.~N.}\ \bibnamefont {Pfeiffer}}, \bibinfo {author}
  {\bibfnamefont {K.~W.}\ \bibnamefont {West}}, \bibinfo {author}
  {\bibfnamefont {K.~W.}\ \bibnamefont {Baldwin}}, \ and\ \bibinfo {author}
  {\bibfnamefont {M.}~\bibnamefont {Shayegan}},\ }\href {\doibase
  10.1103/PhysRevB.90.085301} {\bibfield  {journal} {\bibinfo  {journal} {Phys.
  Rev. B}\ }\textbf {\bibinfo {volume} {90}},\ \bibinfo {pages} {085301}
  (\bibinfo {year} {2014})}\BibitemShut {NoStop}%
\bibitem [{\citenamefont {Balram}\ \emph {et~al.}(2015)\citenamefont {Balram},
  \citenamefont {T\"oke}, \citenamefont {W\'ojs},\ and\ \citenamefont
  {Jain}}]{Balram15a}%
  \BibitemOpen
  \bibfield  {author} {\bibinfo {author} {\bibfnamefont {A.~C.}\ \bibnamefont
  {Balram}}, \bibinfo {author} {\bibfnamefont {C.}~\bibnamefont {T\"oke}},
  \bibinfo {author} {\bibfnamefont {A.}~\bibnamefont {W\'ojs}}, \ and\ \bibinfo
  {author} {\bibfnamefont {J.~K.}\ \bibnamefont {Jain}},\ }\href {\doibase
  10.1103/PhysRevB.92.075410} {\bibfield  {journal} {\bibinfo  {journal} {Phys.
  Rev. B}\ }\textbf {\bibinfo {volume} {92}},\ \bibinfo {pages} {075410}
  (\bibinfo {year} {2015})}\BibitemShut {NoStop}%
\bibitem [{\citenamefont {Reynolds}\ \emph {et~al.}(1982)\citenamefont
  {Reynolds}, \citenamefont {Ceperley}, \citenamefont {Alder},\ and\
  \citenamefont {Lester~Jr.}}]{Reynolds82}%
  \BibitemOpen
  \bibfield  {author} {\bibinfo {author} {\bibfnamefont {P.~J.}\ \bibnamefont
  {Reynolds}}, \bibinfo {author} {\bibfnamefont {D.~M.}\ \bibnamefont
  {Ceperley}}, \bibinfo {author} {\bibfnamefont {B.~J.}\ \bibnamefont {Alder}},
  \ and\ \bibinfo {author} {\bibfnamefont {W.~A.}\ \bibnamefont {Lester~Jr.}},\
  }\href {\doibase http://dx.doi.org/10.1063/1.443766} {\bibfield  {journal}
  {\bibinfo  {journal} {J. Chem. Phys.}\ }\textbf {\bibinfo {volume} {77}},\
  \bibinfo {pages} {5593} (\bibinfo {year} {1982})}\BibitemShut {NoStop}%
\bibitem [{\citenamefont {Foulkes}\ \emph {et~al.}(2001)\citenamefont
  {Foulkes}, \citenamefont {Mitas}, \citenamefont {Needs},\ and\ \citenamefont
  {Rajagopal}}]{Foulkes01}%
  \BibitemOpen
  \bibfield  {author} {\bibinfo {author} {\bibfnamefont {W.~M.~C.}\
  \bibnamefont {Foulkes}}, \bibinfo {author} {\bibfnamefont {L.}~\bibnamefont
  {Mitas}}, \bibinfo {author} {\bibfnamefont {R.~J.}\ \bibnamefont {Needs}}, \
  and\ \bibinfo {author} {\bibfnamefont {G.}~\bibnamefont {Rajagopal}},\ }\href
  {\doibase 10.1103/RevModPhys.73.33} {\bibfield  {journal} {\bibinfo
  {journal} {Rev. Mod. Phys.}\ }\textbf {\bibinfo {volume} {73}},\ \bibinfo
  {pages} {33} (\bibinfo {year} {2001})}\BibitemShut {NoStop}%
\bibitem [{\citenamefont {G\"u\c{c}l\"u}\ and\ \citenamefont
  {Umrigar}(2005)}]{Guclu05}%
  \BibitemOpen
  \bibfield  {author} {\bibinfo {author} {\bibfnamefont {A.~D.}\ \bibnamefont
  {G\"u\c{c}l\"u}}\ and\ \bibinfo {author} {\bibfnamefont {C.~J.}\ \bibnamefont
  {Umrigar}},\ }\href {\doibase 10.1103/PhysRevB.72.045309} {\bibfield
  {journal} {\bibinfo  {journal} {Phys. Rev. B}\ }\textbf {\bibinfo {volume}
  {72}},\ \bibinfo {pages} {045309} (\bibinfo {year} {2005})}\BibitemShut
  {NoStop}%
\bibitem [{\citenamefont {Jain}(1989)}]{Jain89}%
  \BibitemOpen
  \bibfield  {author} {\bibinfo {author} {\bibfnamefont {J.~K.}\ \bibnamefont
  {Jain}},\ }\href {\doibase 10.1103/PhysRevLett.63.199} {\bibfield  {journal}
  {\bibinfo  {journal} {Phys. Rev. Lett.}\ }\textbf {\bibinfo {volume} {63}},\
  \bibinfo {pages} {199} (\bibinfo {year} {1989})}\BibitemShut {NoStop}%
\bibitem [{\citenamefont {Jain}(2007)}]{Jain07}%
  \BibitemOpen
  \bibfield  {author} {\bibinfo {author} {\bibfnamefont {J.~K.}\ \bibnamefont
  {Jain}},\ }\href@noop {} {\emph {\bibinfo {title} {Composite Fermions}}}\
  (\bibinfo  {publisher} {Cambridge University Press, New York, US (Cambridge
  Books Online)},\ \bibinfo {year} {2007})\BibitemShut {NoStop}%
\bibitem [{Zha()}]{ZhangSM1}%
  \BibitemOpen
  \href@noop {} {}\bibinfo {note} {See Supplemental Material which includes
  details for evaluation of drift velocity and local energy in the fixed-phase
  DMC calculation, the critical Zeeman energy $\alpha_{\rm Z}^{\rm crit}$ for
  reverse-flux-attached states, extrapolation to the thermodynamic limit,
  change in the energy as a function of $\kappa$ for various FQH states and the
  CF Fermi sea, and perturbative calculation of $\alpha_{\rm Z}^{\rm crit}$
  with LL mixing.}\BibitemShut {Stop}%
\bibitem [{\citenamefont {Haldane}(1983)}]{Haldane83}%
  \BibitemOpen
  \bibfield  {author} {\bibinfo {author} {\bibfnamefont {F.~D.~M.}\
  \bibnamefont {Haldane}},\ }\href {\doibase 10.1103/PhysRevLett.51.605}
  {\bibfield  {journal} {\bibinfo  {journal} {Phys. Rev. Lett.}\ }\textbf
  {\bibinfo {volume} {51}},\ \bibinfo {pages} {605} (\bibinfo {year}
  {1983})}\BibitemShut {NoStop}%
\bibitem [{\citenamefont {Jain}\ and\ \citenamefont
  {Kamilla}(1997{\natexlab{a}})}]{Jain97}%
  \BibitemOpen
  \bibfield  {author} {\bibinfo {author} {\bibfnamefont {J.~K.}\ \bibnamefont
  {Jain}}\ and\ \bibinfo {author} {\bibfnamefont {R.~K.}\ \bibnamefont
  {Kamilla}},\ }\href {\doibase 10.1142/S0217979297001301} {\bibfield
  {journal} {\bibinfo  {journal} {Int. J. Mod. Phys. B}\ }\textbf {\bibinfo
  {volume} {11}},\ \bibinfo {pages} {2621} (\bibinfo {year}
  {1997}{\natexlab{a}})}\BibitemShut {NoStop}%
\bibitem [{\citenamefont {Jain}\ and\ \citenamefont
  {Kamilla}(1997{\natexlab{b}})}]{Jain97b}%
  \BibitemOpen
  \bibfield  {author} {\bibinfo {author} {\bibfnamefont {J.~K.}\ \bibnamefont
  {Jain}}\ and\ \bibinfo {author} {\bibfnamefont {R.~K.}\ \bibnamefont
  {Kamilla}},\ }\href {\doibase 10.1103/PhysRevB.55.R4895} {\bibfield
  {journal} {\bibinfo  {journal} {Phys. Rev. B}\ }\textbf {\bibinfo {volume}
  {55}},\ \bibinfo {pages} {R4895} (\bibinfo {year}
  {1997}{\natexlab{b}})}\BibitemShut {NoStop}%
\bibitem [{\citenamefont {Davenport}\ and\ \citenamefont
  {Simon}(2012)}]{Davenport12}%
  \BibitemOpen
  \bibfield  {author} {\bibinfo {author} {\bibfnamefont {S.~C.}\ \bibnamefont
  {Davenport}}\ and\ \bibinfo {author} {\bibfnamefont {S.~H.}\ \bibnamefont
  {Simon}},\ }\href {\doibase 10.1103/PhysRevB.85.245303} {\bibfield  {journal}
  {\bibinfo  {journal} {Phys. Rev. B}\ }\textbf {\bibinfo {volume} {85}},\
  \bibinfo {pages} {245303} (\bibinfo {year} {2012})}\BibitemShut {NoStop}%
\bibitem [{\citenamefont {Park}\ and\ \citenamefont {Jain}(1998)}]{Park98}%
  \BibitemOpen
  \bibfield  {author} {\bibinfo {author} {\bibfnamefont {K.}~\bibnamefont
  {Park}}\ and\ \bibinfo {author} {\bibfnamefont {J.~K.}\ \bibnamefont
  {Jain}},\ }\href {\doibase 10.1103/PhysRevLett.80.4237} {\bibfield  {journal}
  {\bibinfo  {journal} {Phys. Rev. Lett.}\ }\textbf {\bibinfo {volume} {80}},\
  \bibinfo {pages} {4237} (\bibinfo {year} {1998})}\BibitemShut {NoStop}%
\bibitem [{\citenamefont {Park}\ and\ \citenamefont {Jain}(1999)}]{Park99}%
  \BibitemOpen
  \bibfield  {author} {\bibinfo {author} {\bibfnamefont {K.}~\bibnamefont
  {Park}}\ and\ \bibinfo {author} {\bibfnamefont {J.~K.}\ \bibnamefont
  {Jain}},\ }\href {\doibase 10.1103/PhysRevLett.83.5543} {\bibfield  {journal}
  {\bibinfo  {journal} {Phys. Rev. Lett.}\ }\textbf {\bibinfo {volume} {83}},\
  \bibinfo {pages} {5543} (\bibinfo {year} {1999})}\BibitemShut {NoStop}%
\bibitem [{\citenamefont {Mandal}\ and\ \citenamefont
  {Ravishankar}(1996)}]{Mandal96}%
  \BibitemOpen
  \bibfield  {author} {\bibinfo {author} {\bibfnamefont {S.~S.}\ \bibnamefont
  {Mandal}}\ and\ \bibinfo {author} {\bibfnamefont {V.}~\bibnamefont
  {Ravishankar}},\ }\href {\doibase 10.1103/PhysRevB.54.8699} {\bibfield
  {journal} {\bibinfo  {journal} {Phys. Rev. B}\ }\textbf {\bibinfo {volume}
  {54}},\ \bibinfo {pages} {8699} (\bibinfo {year} {1996})}\BibitemShut
  {NoStop}%
\bibitem [{\citenamefont {Lopez}\ and\ \citenamefont
  {Fradkin}(2001)}]{Lopez01}%
  \BibitemOpen
  \bibfield  {author} {\bibinfo {author} {\bibfnamefont {A.}~\bibnamefont
  {Lopez}}\ and\ \bibinfo {author} {\bibfnamefont {E.}~\bibnamefont
  {Fradkin}},\ }\href {\doibase 10.1103/PhysRevB.63.085306} {\bibfield
  {journal} {\bibinfo  {journal} {Phys. Rev. B}\ }\textbf {\bibinfo {volume}
  {63}},\ \bibinfo {pages} {085306} (\bibinfo {year} {2001})}\BibitemShut
  {NoStop}%
\bibitem [{\citenamefont {Murthy}\ and\ \citenamefont
  {Shankar}(2003)}]{Murthy03}%
  \BibitemOpen
  \bibfield  {author} {\bibinfo {author} {\bibfnamefont {G.}~\bibnamefont
  {Murthy}}\ and\ \bibinfo {author} {\bibfnamefont {R.}~\bibnamefont
  {Shankar}},\ }\href {\doibase 10.1103/RevModPhys.75.1101} {\bibfield
  {journal} {\bibinfo  {journal} {Rev. Mod. Phys.}\ }\textbf {\bibinfo {volume}
  {75}},\ \bibinfo {pages} {1101} (\bibinfo {year} {2003})}\BibitemShut
  {NoStop}%
\bibitem [{\citenamefont {Murthy}\ and\ \citenamefont
  {Shankar}(2007)}]{Murthy07}%
  \BibitemOpen
  \bibfield  {author} {\bibinfo {author} {\bibfnamefont {G.}~\bibnamefont
  {Murthy}}\ and\ \bibinfo {author} {\bibfnamefont {R.}~\bibnamefont
  {Shankar}},\ }\href {\doibase 10.1103/PhysRevB.76.075341} {\bibfield
  {journal} {\bibinfo  {journal} {Phys. Rev. B}\ }\textbf {\bibinfo {volume}
  {76}},\ \bibinfo {pages} {075341} (\bibinfo {year} {2007})}\BibitemShut
  {NoStop}%
\bibitem [{\citenamefont {Ortalano}\ \emph {et~al.}(1997)\citenamefont
  {Ortalano}, \citenamefont {He},\ and\ \citenamefont
  {Das~Sarma}}]{Ortalano97}%
  \BibitemOpen
  \bibfield  {author} {\bibinfo {author} {\bibfnamefont {M.~W.}\ \bibnamefont
  {Ortalano}}, \bibinfo {author} {\bibfnamefont {S.}~\bibnamefont {He}}, \ and\
  \bibinfo {author} {\bibfnamefont {S.}~\bibnamefont {Das~Sarma}},\ }\href
  {\doibase 10.1103/PhysRevB.55.7702} {\bibfield  {journal} {\bibinfo
  {journal} {Phys. Rev. B}\ }\textbf {\bibinfo {volume} {55}},\ \bibinfo
  {pages} {7702} (\bibinfo {year} {1997})}\BibitemShut {NoStop}%
\bibitem [{\citenamefont {Morf}\ and\ \citenamefont
  {Halperin}(1986)}]{Morf86b}%
  \BibitemOpen
  \bibfield  {author} {\bibinfo {author} {\bibfnamefont {R.}~\bibnamefont
  {Morf}}\ and\ \bibinfo {author} {\bibfnamefont {B.~I.}\ \bibnamefont
  {Halperin}},\ }\href {\doibase 10.1103/PhysRevB.33.2221} {\bibfield
  {journal} {\bibinfo  {journal} {Phys. Rev. B}\ }\textbf {\bibinfo {volume}
  {33}},\ \bibinfo {pages} {2221} (\bibinfo {year} {1986})}\BibitemShut
  {NoStop}%
\bibitem [{\citenamefont {Price}\ \emph {et~al.}(1993)\citenamefont {Price},
  \citenamefont {Platzman},\ and\ \citenamefont {He}}]{Price93}%
  \BibitemOpen
  \bibfield  {author} {\bibinfo {author} {\bibfnamefont {R.}~\bibnamefont
  {Price}}, \bibinfo {author} {\bibfnamefont {P.~M.}\ \bibnamefont {Platzman}},
  \ and\ \bibinfo {author} {\bibfnamefont {S.}~\bibnamefont {He}},\ }\href
  {\doibase 10.1103/PhysRevLett.70.339} {\bibfield  {journal} {\bibinfo
  {journal} {Phys. Rev. Lett.}\ }\textbf {\bibinfo {volume} {70}},\ \bibinfo
  {pages} {339} (\bibinfo {year} {1993})}\BibitemShut {NoStop}%
\bibitem [{\citenamefont {Wu}\ \emph {et~al.}(1993)\citenamefont {Wu},
  \citenamefont {Dev},\ and\ \citenamefont {Jain}}]{Wu93}%
  \BibitemOpen
  \bibfield  {author} {\bibinfo {author} {\bibfnamefont {X.~G.}\ \bibnamefont
  {Wu}}, \bibinfo {author} {\bibfnamefont {G.}~\bibnamefont {Dev}}, \ and\
  \bibinfo {author} {\bibfnamefont {J.~K.}\ \bibnamefont {Jain}},\ }\href
  {\doibase 10.1103/PhysRevLett.71.153} {\bibfield  {journal} {\bibinfo
  {journal} {Phys. Rev. Lett.}\ }\textbf {\bibinfo {volume} {71}},\ \bibinfo
  {pages} {153} (\bibinfo {year} {1993})}\BibitemShut {NoStop}%
\bibitem [{\citenamefont {Leadley}\ \emph {et~al.}(1997)\citenamefont
  {Leadley}, \citenamefont {Nicholas}, \citenamefont {Maude}, \citenamefont
  {Utjuzh}, \citenamefont {Portal}, \citenamefont {Harris},\ and\ \citenamefont
  {Foxon}}]{Leadley97}%
  \BibitemOpen
  \bibfield  {author} {\bibinfo {author} {\bibfnamefont {D.~R.}\ \bibnamefont
  {Leadley}}, \bibinfo {author} {\bibfnamefont {R.~J.}\ \bibnamefont
  {Nicholas}}, \bibinfo {author} {\bibfnamefont {D.~K.}\ \bibnamefont {Maude}},
  \bibinfo {author} {\bibfnamefont {A.~N.}\ \bibnamefont {Utjuzh}}, \bibinfo
  {author} {\bibfnamefont {J.~C.}\ \bibnamefont {Portal}}, \bibinfo {author}
  {\bibfnamefont {J.~J.}\ \bibnamefont {Harris}}, \ and\ \bibinfo {author}
  {\bibfnamefont {C.~T.}\ \bibnamefont {Foxon}},\ }\href {\doibase
  10.1103/PhysRevLett.79.4246} {\bibfield  {journal} {\bibinfo  {journal}
  {Phys. Rev. Lett.}\ }\textbf {\bibinfo {volume} {79}},\ \bibinfo {pages}
  {4246} (\bibinfo {year} {1997})}\BibitemShut {NoStop}%
\end{thebibliography}%

\pagebreak
\pagebreak

\section{\bf Supplemental material}{\large}

\setcounter{figure}{0}
\setcounter{equation}{0}
\renewcommand\thefigure{S\arabic{figure}}
\renewcommand\thetable{S\arabic{table}}
\renewcommand\theequation{S\arabic{equation}}

\section{I. Evaluation of drift velocity and local energy}


The details of the method of Diffusion Monte Carlo (DMC) can be found in several excellent articles \cite{Reynolds82,Foulkes01}. The fixed phase DMC used in our work was developed in Ref. \cite{Ortiz93}, and generalized to the spherical geometry in Ref. \cite{Melik-Alaverdian97}. We follow the methods developed in these articles. 

The DMC calculation requires evaluation at each step of the $3N$-dimensional drift velocity $\vec{v}_{D}(\mathcal{R})$ defined by 
\begin{equation}
\vec{v}_{D}(\mathcal{R}) = {\nabla} \ln |\psi_{T}(\mathcal{R})|
\end{equation}
We use for our trial wave function $\psi_{T}$ the wave function of Eq.~2, which involve a determinant factor. An efficient way to calculate the derivative of a determinant is as follows:
\begin{equation}
\partial_{i} \ln (\det A)=\text{Tr}[A^{-1}\partial_{i}A].
\label{derivative}
\end{equation}
Following Melik-Alaverdian, Bonesteel and Ortiz \cite{Melik-Alaverdian97} we do a stereographic projection of the spherical wave functions into the planar geometry for the purpose of DMC. 
In the following, we take $\nu=1$ and $\nu=2/5$ fully polarized states as examples to show the wave functions in stereographic coordinates as well as their derivatives.

The single-particle states are described by the ``monopole harmonics'' $Y_{Qlm}$ whose expression in the ``spinor''  coordinates ($u=\cos(\theta/2) e^{i\phi/2}$,$v=\sin(\theta/2) e^{-i\phi/2}$) can be found in the literature~\cite{Jain07}. To express $Y_{Qlm}$ in complex stereographic coordinates $z=x+iy$, we note that $z=u/v$ and $v^{2} = [(1/|z|+|z|)z]^{-1}$. The LLL single-particle wave function is then given by 
\begin{equation}
Y_{QQm} \sim v^{Q-m} u^{Q+m} = (1/|z|+|z|)^{-Q} z^{m}.
\label{monopole}
\end{equation}
The wave function of one filled Landau level can be expressed as
\begin{equation}
\Phi_{1} = \prod_{i=1}^{N}\left [  \left(\frac{1}{|z_{i}|} +|z_{i}| \right) z_{i} \right ]^{-\frac{N-1}{2}} \prod_{i<j} (z_{i} - z_{j}).
\end{equation}
and the drift velocity
\begin{equation}
\vec{v}_{k}(\mathcal{R})=\nabla_{k} \ln |\Phi_{1}|= \frac{1-N}{1+|z_{k}|^{2}}z_{k} + \sum_{j\neq k}{\frac{1}{|z_{k} - z_{j}|^{2}} (z_{k} - z_{j})}.
\end{equation}
Notice that the drift velocity on the right hand side is given as a complex number whose real part represents the x-component and the imaginary part the y-component.

The unprojected Jain wave function for $\nu=2/5$ fully polarized state is given by $\Phi_{n=2} \Phi_{1}^{2}$, where $\Phi_{n=2}$ is a Slater determinant. It turns out, as shown in Refs.~\cite{Jain97,Jain97b,Jain07,Davenport12}, that even the LLL-projected wave function can be written in this form
\begin{equation}
\Phi_{2/5} = \mathcal{P}_{\text{LLL}} \Phi_{n=2} \Phi_{1}^{2}
= \Phi_{1}^{2} \times \det A
\end{equation}
where $Q^{*}=Q-(N-1)$ and $A_{ij} = Y^{\text{CF}}_{Q^{*}l_{i}m_{i}}(z_{j})$, where $Y^{\text{CF}}_{Q^{*}l_{i}m_{i}}$ are the  ``CF monopole harmonics'' for composite fermions at effective monopole strength $Q^{*}$. The explicit expressions for the CF monopole harmonics are complicated but can be found in the literature \cite{Jain97,Jain97b,Jain07,Davenport12}.  In particular, for  $l_{i}=Q^{*}$ (i.e. composite fermions in the lowest $\Lambda$ level), $Y^{\text{CF}}_{Q^{*},l_{i}=Q^{*},m_{i}}$ is given by replacing the corresponding parameters with $Q^{*}$ and $m_{i}$ in Eq. (\ref{monopole}) apart from a coordinate-independent multiplicative factor. 
Setting $\alpha_{j} =(1/|z_{j}|+|z_{j}|)^{-Q^{*}}$, we have $Y^{\text{CF}}_{Q^{*},l_{i}= Q^{*},m_{i}}\sim z_{j}^{m_{i}} \alpha_{j}$. For $\nu=2/5$ we also have composite fermions in the second $\Lambda$ level, for which we have $l_{i} = Q^{*} + 1$.  Here, $A_{ij}=Y^{\text{CF}}_{Q^{*},l_{i}=Q^{*}+1,m_{i}} \sim z_{j}^{m_{i}} \alpha_{j} \beta_{ij}$, with $\beta_{ij}$ given by 
\begin{equation}
\beta_{ij}= -C_{i}^{(1)} \sum_{k}^{\prime} {\frac{z_{k}}{z_{j}-z_{k}}} -C_{i}^{(2)} z_{j} \sum_{k}^{\prime} \frac{1}{z_{j} - z_{k}},
\end{equation}
where $C_{i}^{(1)} = \binom {2Q^{*}+1}{Q^{*}+1-m_{i}}$ and $C_{i}^{(2)} = \binom {2Q^{*}+1}{Q^{*}-m_{i}}$. We can then calculate the derivative of each matrix element $A_{ij}$ analytically and evaluate $\partial_{i} \ln (\det A)$ according to Eq. (\ref{derivative}). The drift velocity is then given by 
\begin{equation}
\vec{v}_{k}(\mathcal{R})=\nabla_{k} \ln |\Phi_{2/5}|=2\nabla_{k} \ln |\Phi_{1}|+\text{Re}[\partial_{i} \ln (\det A)],
\end{equation}
The evaluation of wave functions and drift velocities at other filling factors can be performed analogously although the matrix elements $A_{ij}$ are more complicated and require a careful book-keeping.

The DMC also requires evaluation of the local energy 
\begin{equation}
E_{L} = \psi_{T}^{-1} H  \psi_{T}
\end{equation}
at each step. This is straightforward to evaluate because we are using a LLL projected wave function for our $\psi_{T}$, and hence the kinetic energy term of $H$ does not play a role in the evaluation of $E_L$; the interaction energy term is straightforward to evaluate.

\section{II. $\alpha_{\rm Z}^{\rm crit}$ for reverse-flux-attached states}

\begin{figure}
\resizebox{0.38\textwidth}{!}{\includegraphics{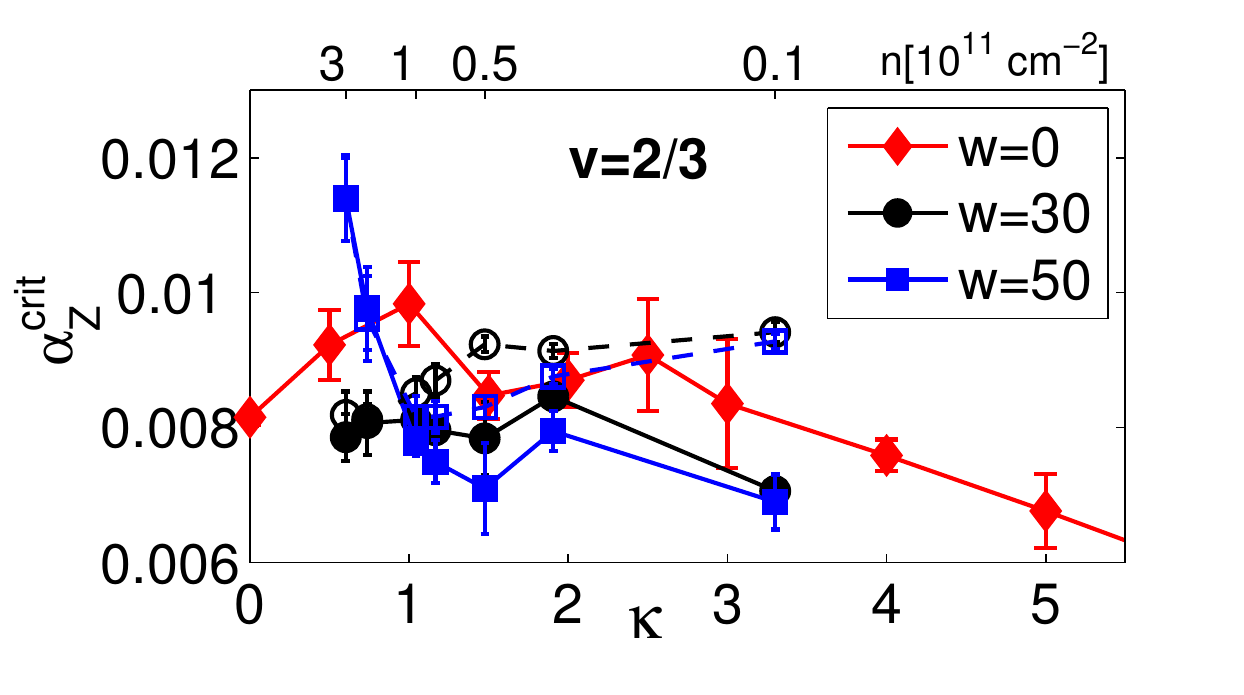}}\\
\vspace{-3.5mm}
\hspace{0.3mm}
\resizebox{0.38\textwidth}{!}{\includegraphics{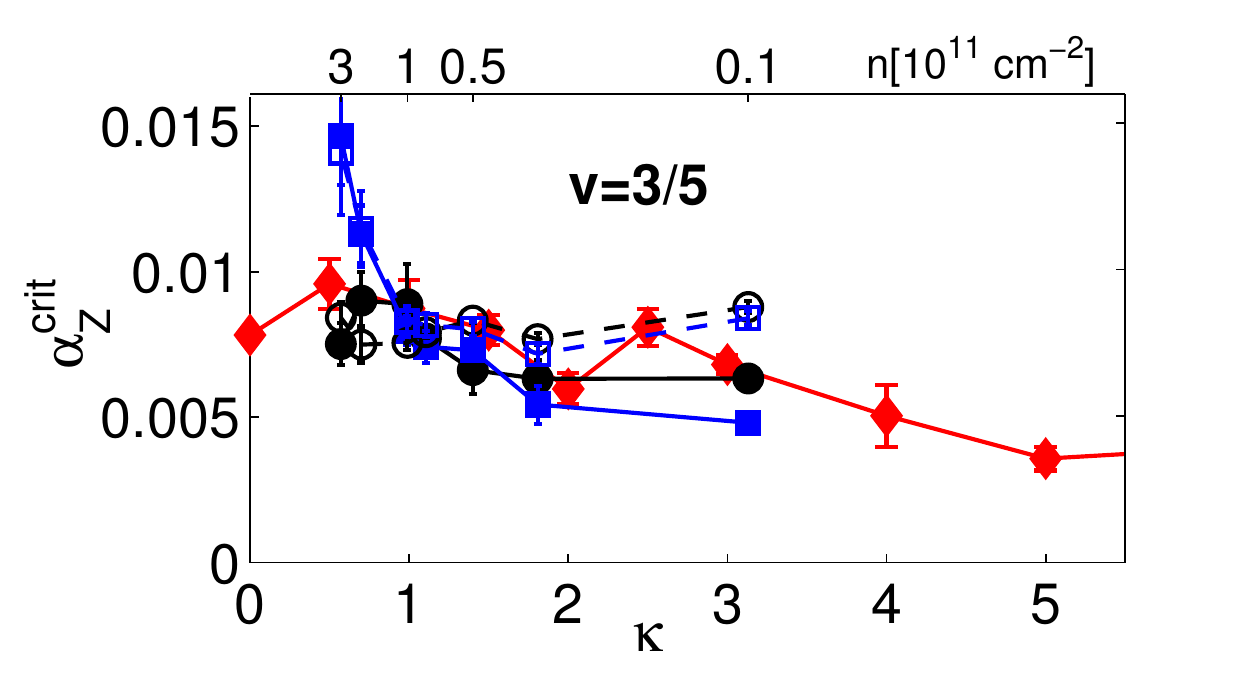}}
\vspace{-4mm}
\caption{(Color online). The critical Zeeman energy $\alpha_{\rm Z}^{\rm crit}$ for $\nu=2/3$ and $3/5$ FQH states calculated using the wave functions in Eq.~2 are plotted as a function of $\kappa$ for ideal 2D systems ($w=0$) and quantum wells with $w=30$nm and $50$nm. The DMC results are plotted with filled symbols and solid lines, while the VMC results are plotted with empty symbols and dashed lines. }
\label{v23CFWF}
\end{figure}

We have shown the $\alpha_{\rm Z}^{\rm crit}$ for fillings $\nu=n/(2n+1)$ as a function of both LL mixing and finite width in Fig.~1 in the main text. At filling factor $\nu=n/(2n-1)$, the composite fermions are in a negative effective magnetic field. As mentioned in the main text, for these states, the $\alpha_{\rm Z}^{\rm crit}$ obtained from the so-called Jain-Kamilla projection \cite{Jain97,Jain97b,Davenport12} are not as accurate as for the states at $\nu=n/(2n+1)$. For example, for $\nu=2/3$ SS state at $w=0$ and $\kappa=0$, the energy of the wave function in Eq.~2 is off by $1\%$ compared to the energy obtained by ED~\cite{Balram15a}, which leads to an error of $\sim 60\%$ in $\alpha_{\rm Z}^{\rm crit}$. The reason is that the Jain-Kamilla projection does not do a very accurate job of keeping electrons with different spins away from one another (i.e., overestimates the probability of coincidence of spin up and spin down electrons) and thus overestimates the energy of the non-fully spin polarized states, which results in a suppression of $\alpha_{\rm Z}^{\rm crit}$. This is a technical problem, which can be remedied by using the ``hard-core" projection of Ref.~\cite{Wu93}, but good methods are currently not available to evaluate the hard-core projection for large systems.

We show in this Section that for the $n/(2n-1)$ states, the Jain wave functions in Eq.~2 with the Jain-Kamilla projection are actually a satisfactory choice for $\psi_T$ for relatively large values of $\kappa$. This makes intuitive sense, because for relatively large LL mixing, one can expect the DMC energy to be less sensitive to tiny differences between the choice of the $\psi_T$, because the LL mixing itself can take care of producing good short range correlations. 

We follow the method in Ref. \onlinecite{Davenport12} to perform LLL projection, and carry out DMC calculation with the Jain wave functions for up to $N=20$ particles, from which we evaluate the thermodynamic extrapolations (next section).  Fig. \ref{v23CFWF} shows the critical Zeeman energy for $\nu=2/3$ and $3/5$ states at different quantum well widths as a function of $\kappa$. 
 Most notably, for $\kappa \sim 2$ or larger, we find that the values of $\alpha_{\rm Z}^{\rm crit}$ obtained from DMC using  the wave functions of Eq.~2 as $\psi_T$ are nicely consistent with those obtained by using exact Coulomb states as $\psi_T$ (see Fig.~3 for the latter). This justifies the use of the wave functions in Eq.~2 with Jain-Kamilla projection as the initial trial wave function $\psi_T$ of our DMC calculation.

We also note that we have restricted the comparison of our theoretical values to the critical Zeeman energies $\alpha_{\rm Z}^{\rm crit}$ obtained in transport experiments. The $\alpha_{\rm Z}^{\rm crit}$ obtained from the optical experiments \cite{Kukushkin99} are generally much higher, for reasons that are not clear to us at this moment.

\section{III. Extrapolation to the Thermodynamic Limit}

As mentioned in the main text, we have used two methods to extrapolate the finite-size results to the thermodynamic limit. For sufficiently large system sizes, these two methods produce the same results.  We find that a linear fit is satisfactory in most cases. The extrapolations for the DMC results are shown in this section. For variational Monte Carlo (VMC) results, which are not shown here, the fits are as good as or better than those for the DMC results.

\begin{figure*}
\resizebox{0.32\textwidth}{!}{\includegraphics{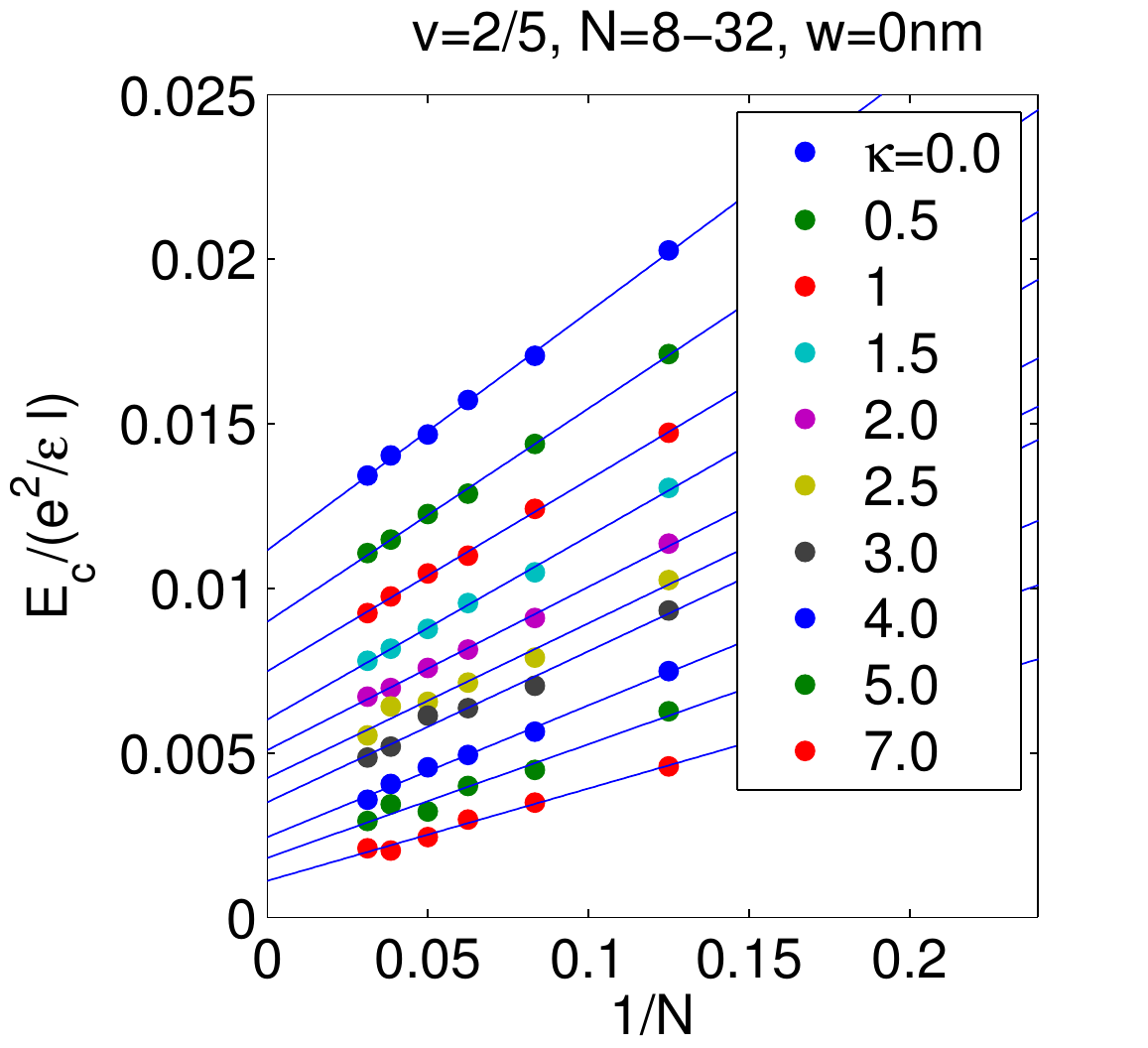}}
\hspace{-3mm}
\resizebox{0.32\textwidth}{!}{\includegraphics{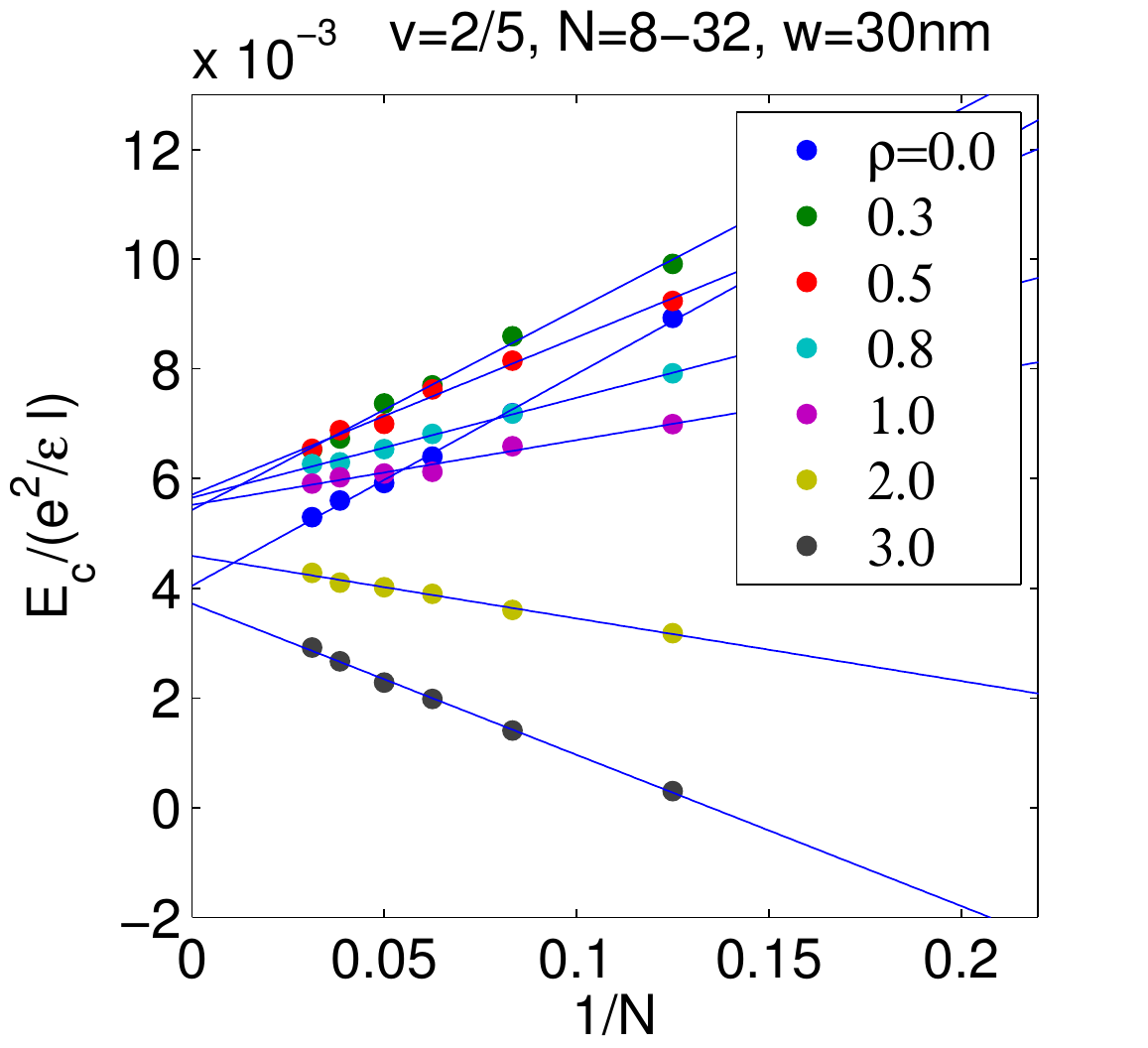}}
\hspace{-3mm}
\resizebox{0.32\textwidth}{!}{\includegraphics{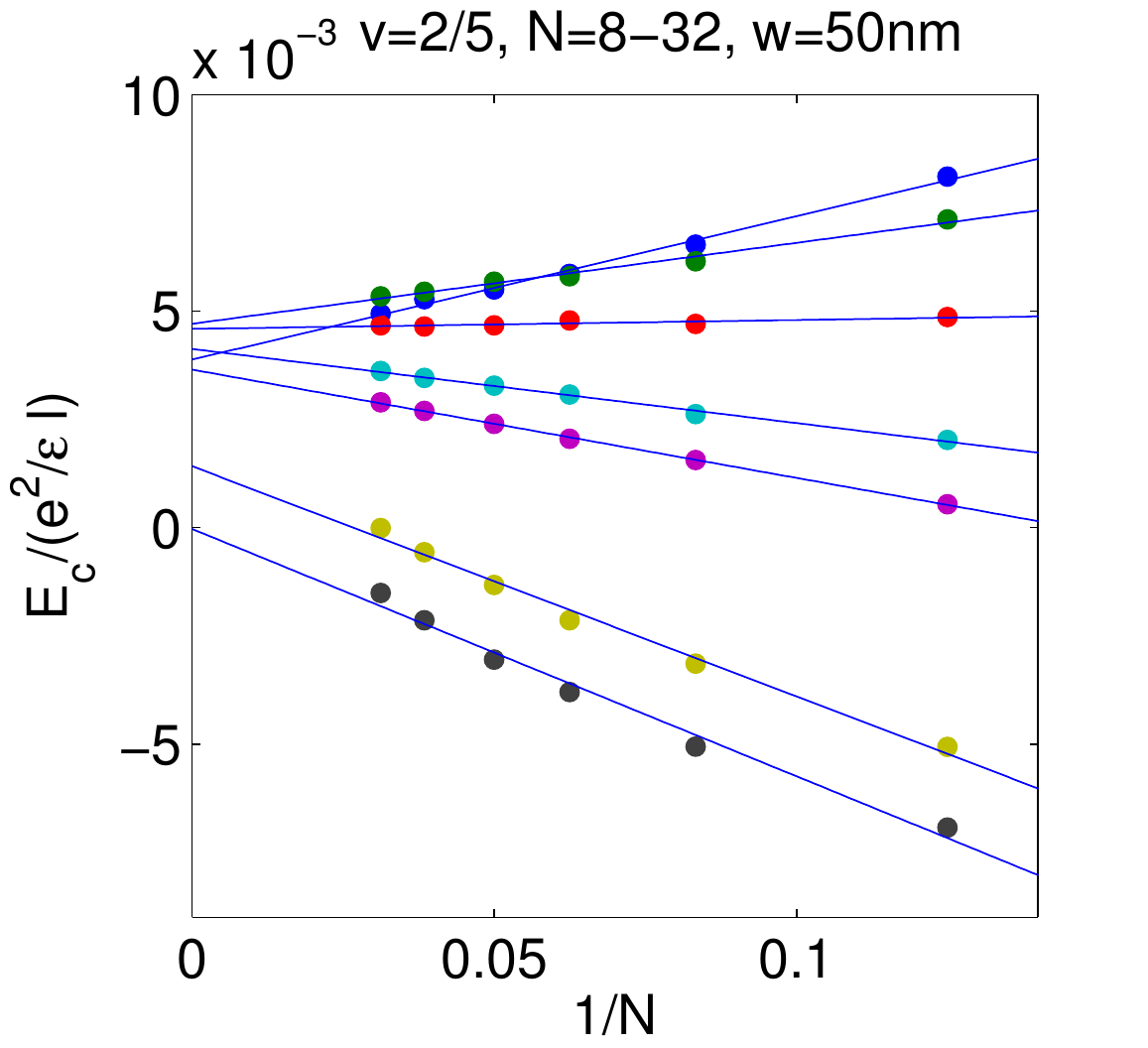}}\\
\vspace{3mm}
\resizebox{0.32\textwidth}{!}{\includegraphics{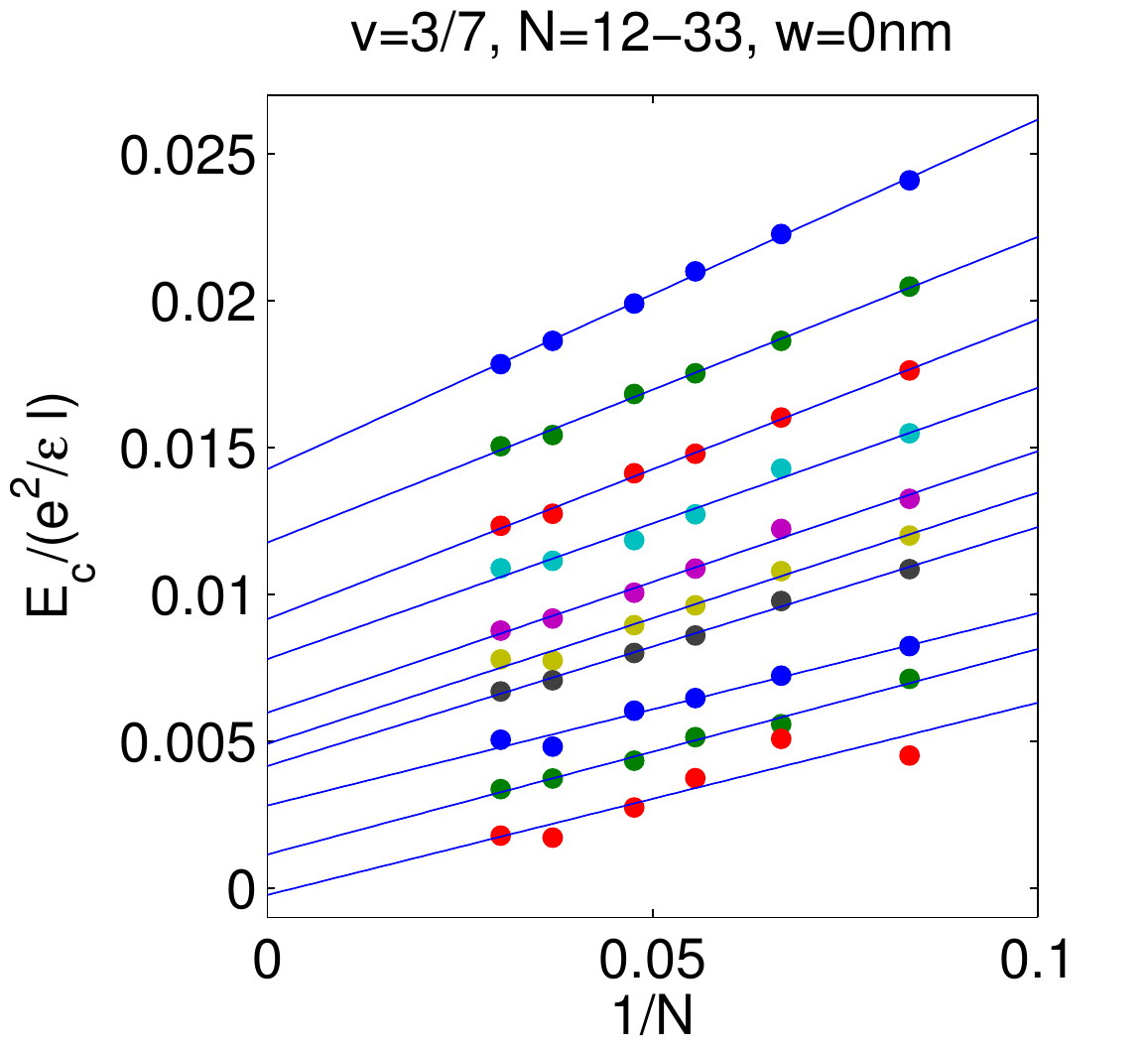}}
\hspace{-3mm}
\resizebox{0.32\textwidth}{!}{\includegraphics{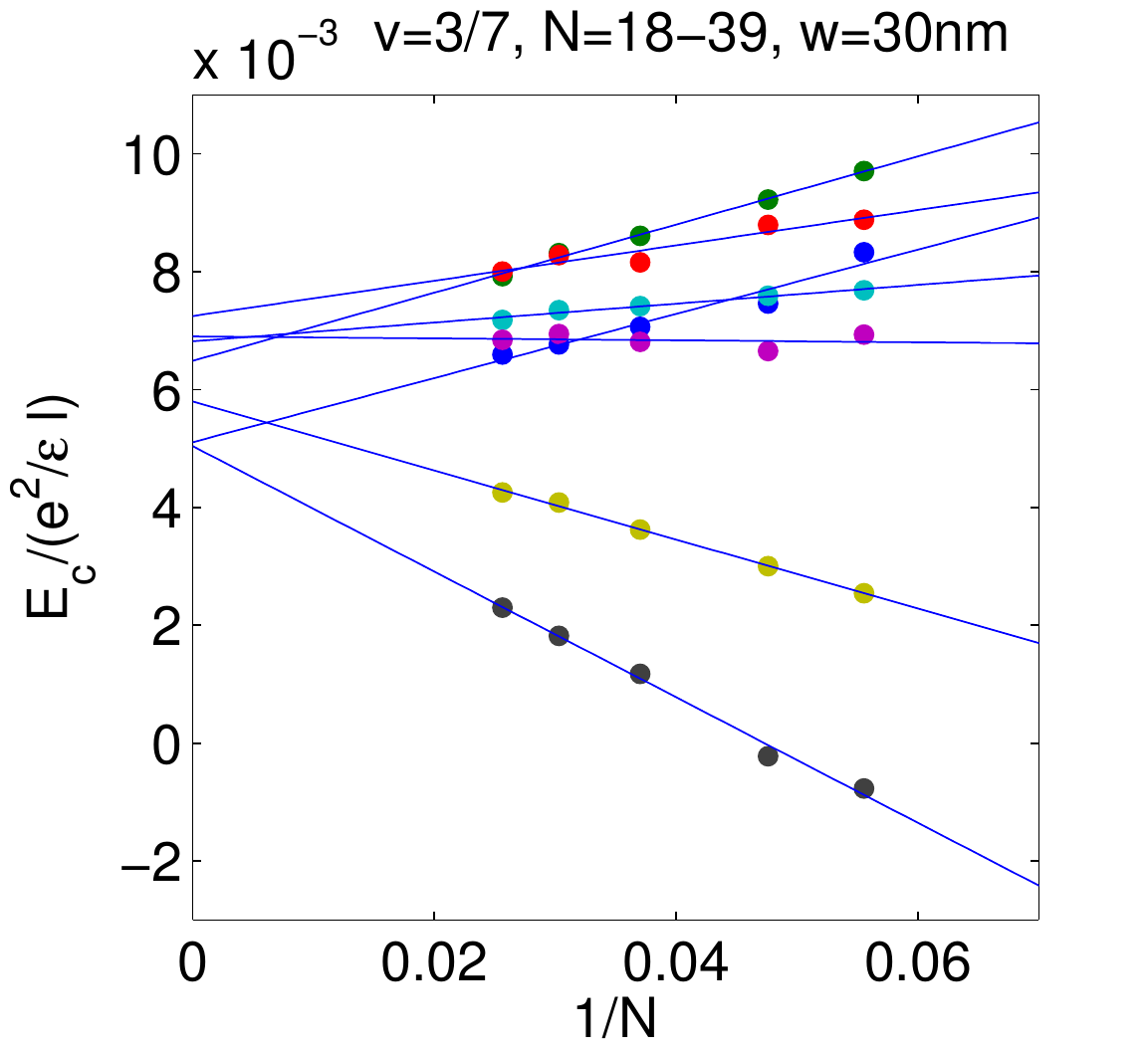}}
\hspace{-3mm}
\resizebox{0.32\textwidth}{!}{\includegraphics{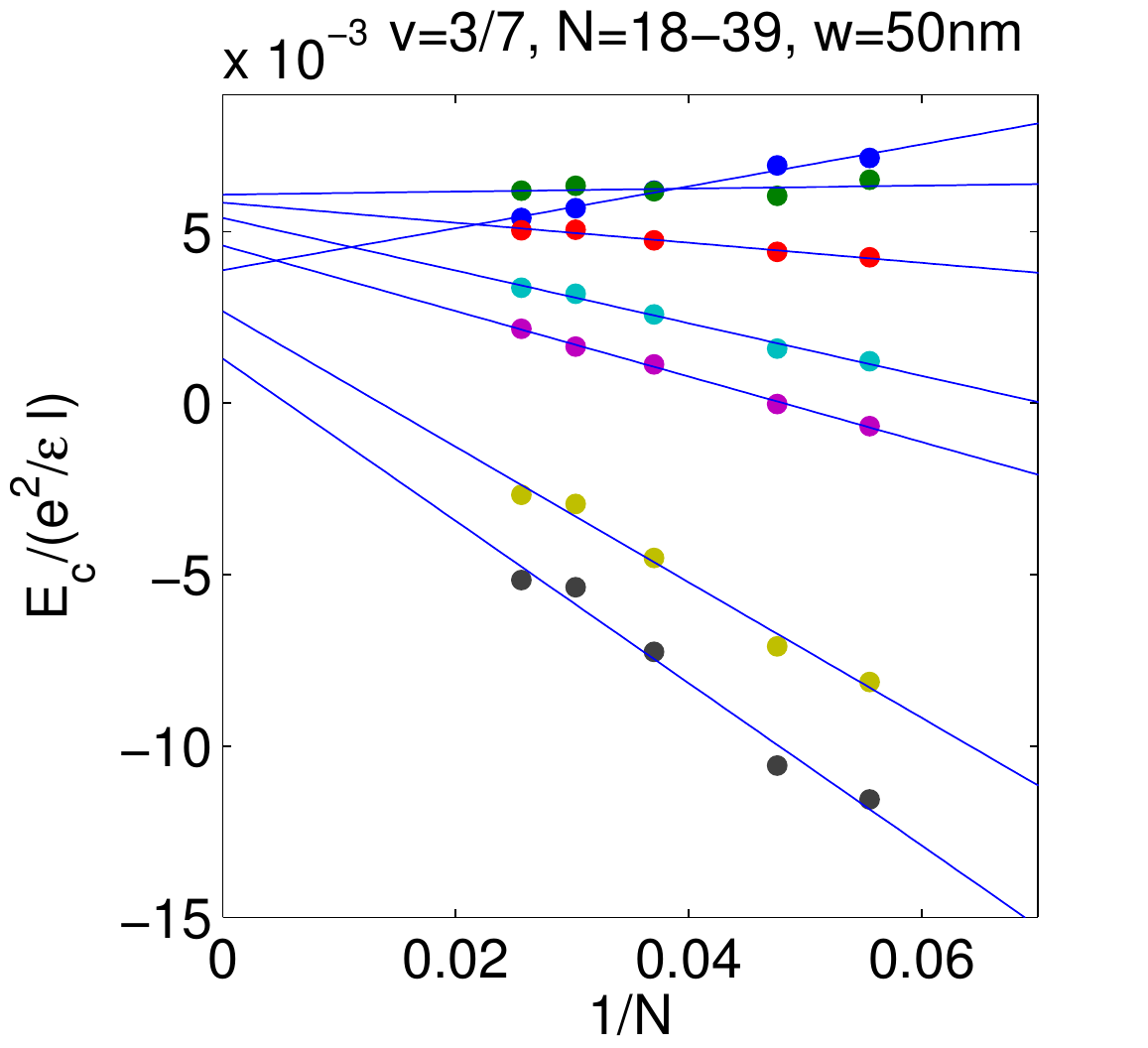}}\\
\vspace{3mm}
\resizebox{0.32\textwidth}{!}{\includegraphics{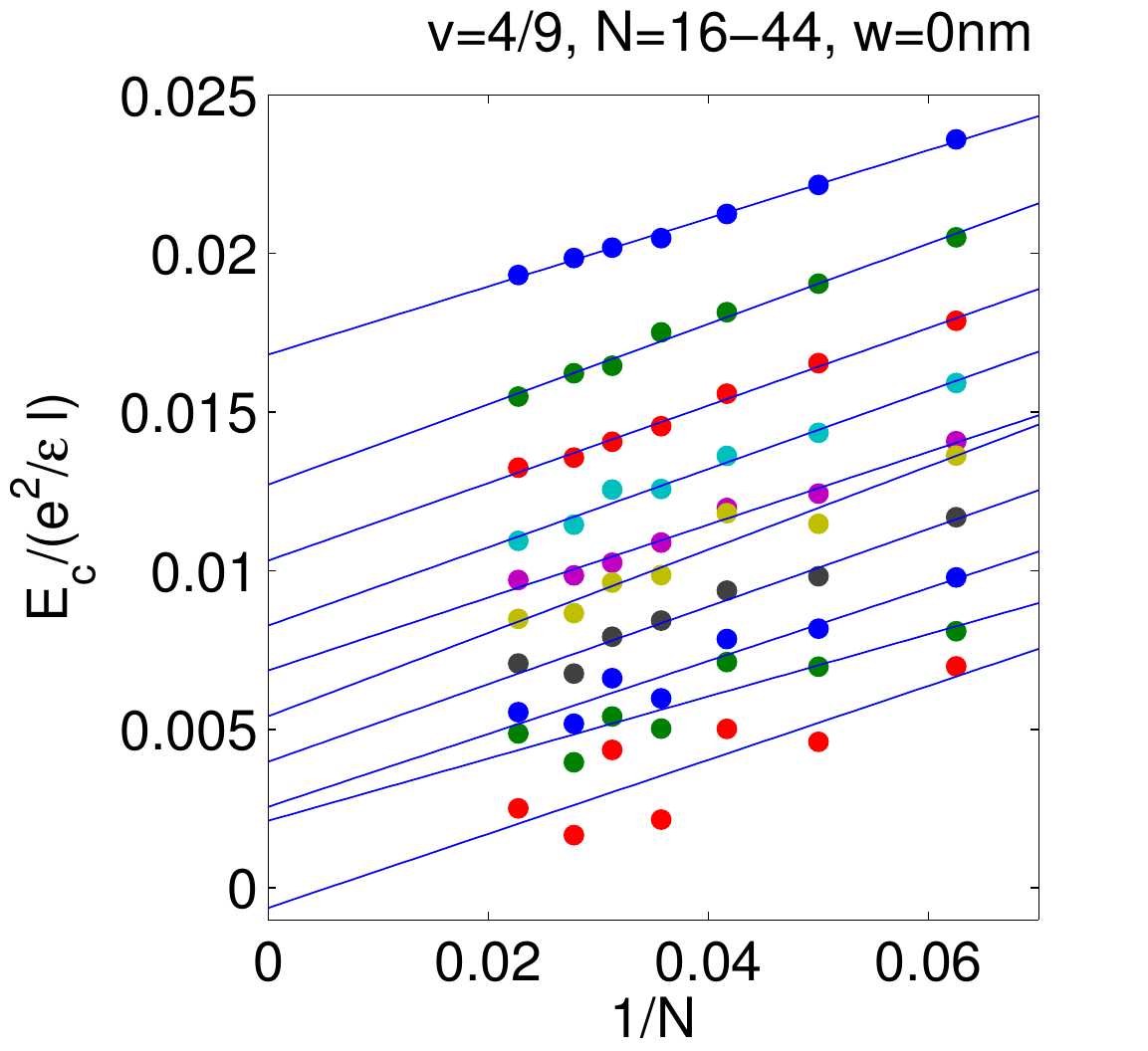}}
\hspace{-3mm}
\resizebox{0.32\textwidth}{!}{\includegraphics{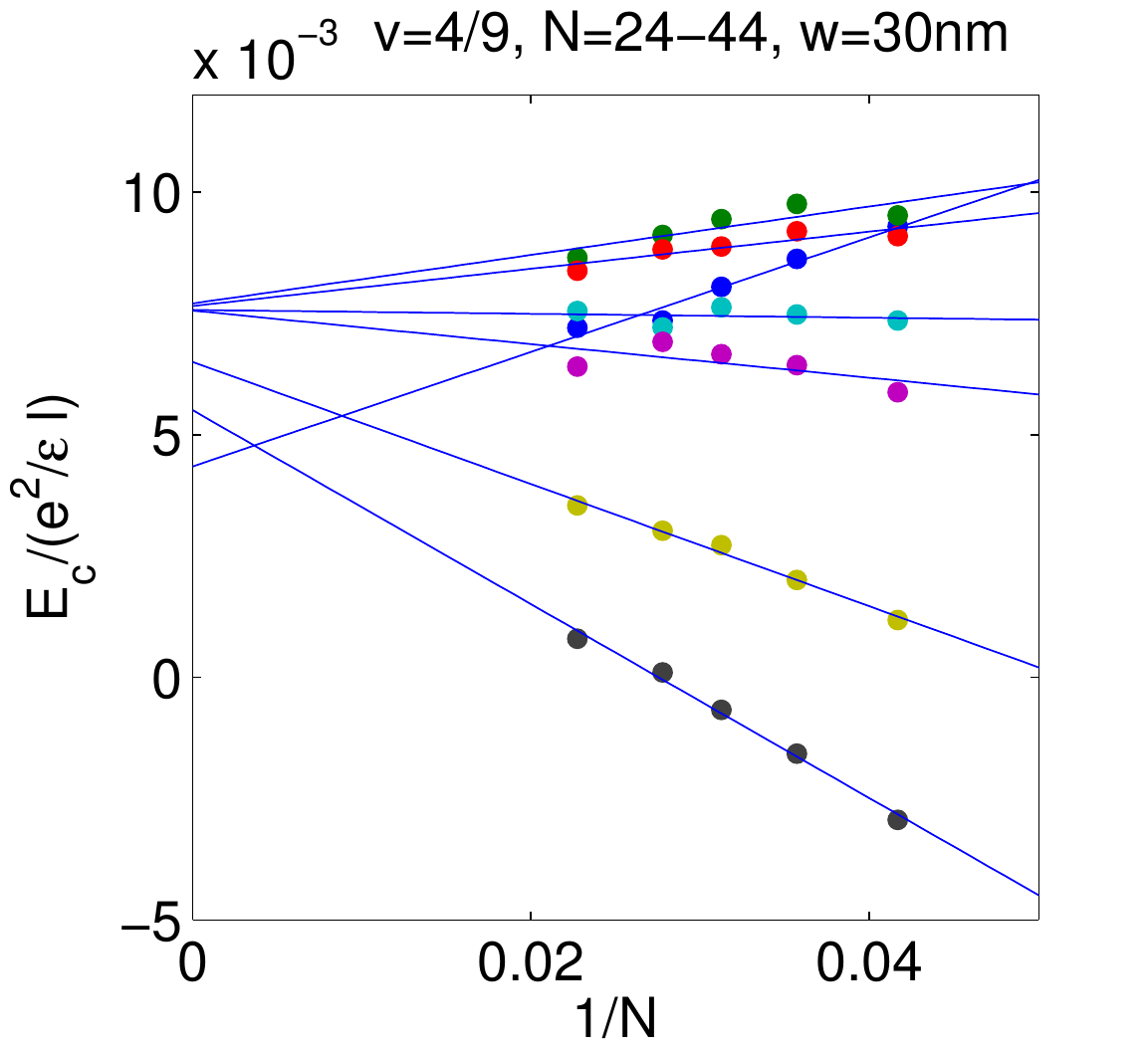}}
\hspace{-3mm}
\resizebox{0.32\textwidth}{!}{\includegraphics{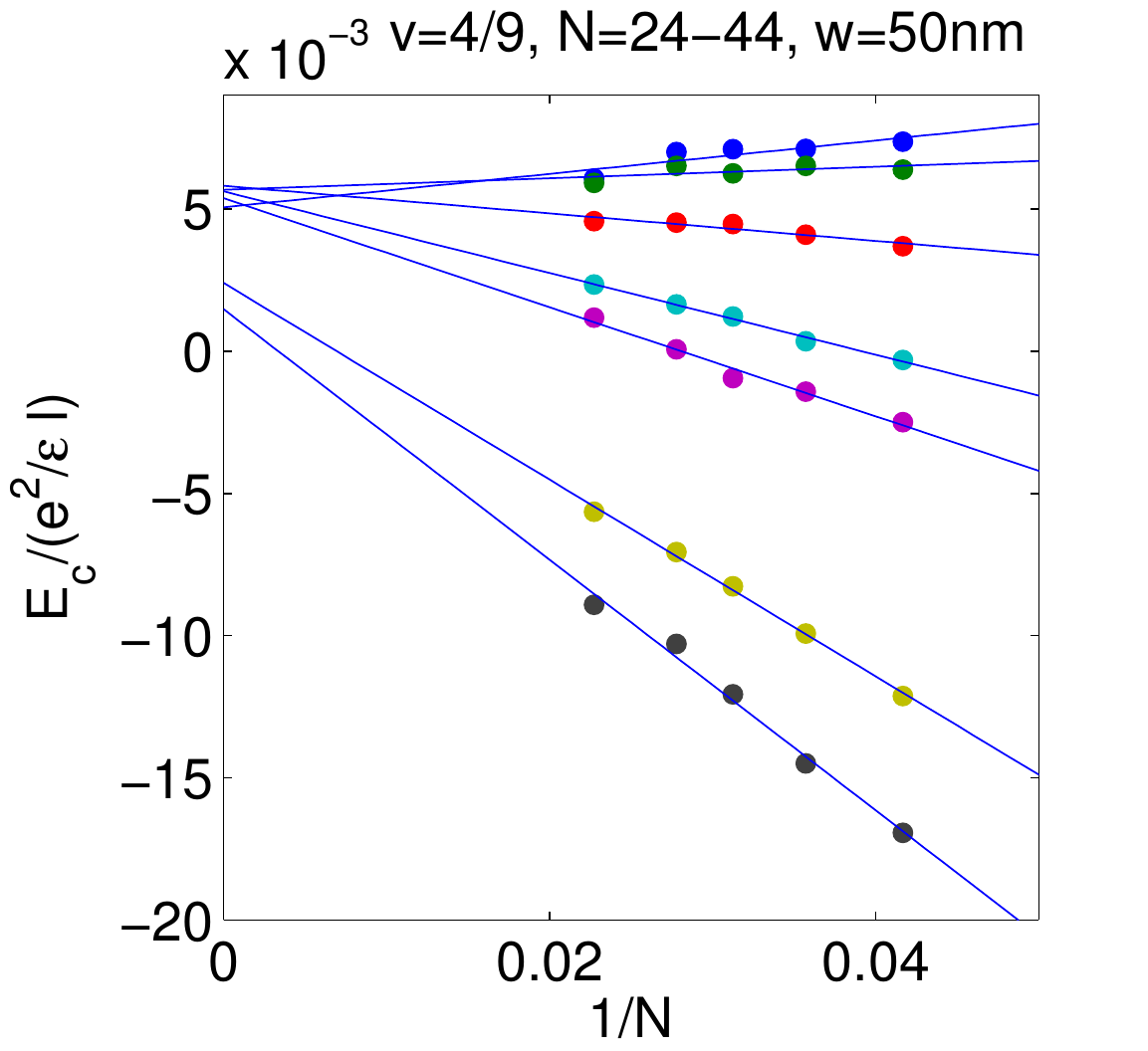}}
\caption{(Color online). The finite-size extrapolation for $\nu=2/5, 3/7, 4/9$ states using  Method I for different widths and a series of values of LL mixing parameter $\kappa$ or density $\rho[10^{11} \text{cm}^{-2}]$. The $\kappa$'s are shown in the legends of the first figure for zero-width and the $\rho$'s are in the second figure for finite-width, which are omitted in all the following figures. Linear fit is used which performs well in most cases.}
\label{fit1}
\end{figure*}

Fig. \ref{fit1} shows the finite-size extrapolation for states at $\nu=2/5, 3/7, 4/9$ using Method I, which directly extrapolates the critical Zeeman energy $E_{c}/(e^{2} /\epsilon \ell)$. We find that the accuracy of fitting is good for zero width even with relatively small particle sizes, as long as $\kappa$ is not too large. With increasing width, however, larger system sizes are needed to obtain a linear thermodynamic extrapolation. This is why we use systems with particle number $N \geq 18$ and $N \geq 24$ for the finite-width cases at $\nu=3/7$ and $4/9$ respectively.

\begin{figure*}
\resizebox{0.32\textwidth}{!}{\includegraphics{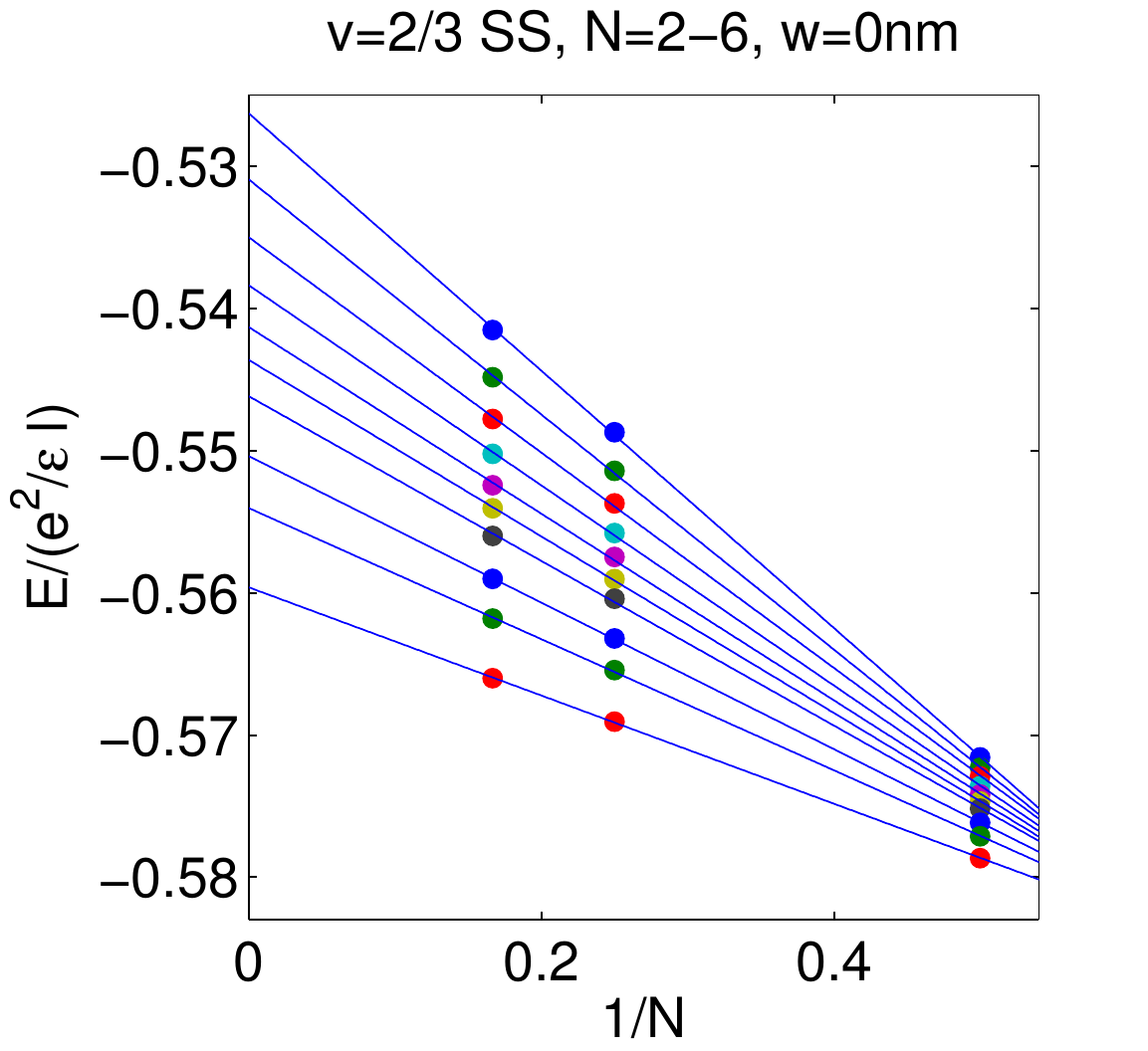}}
\hspace{-3mm}
\resizebox{0.32\textwidth}{!}{\includegraphics{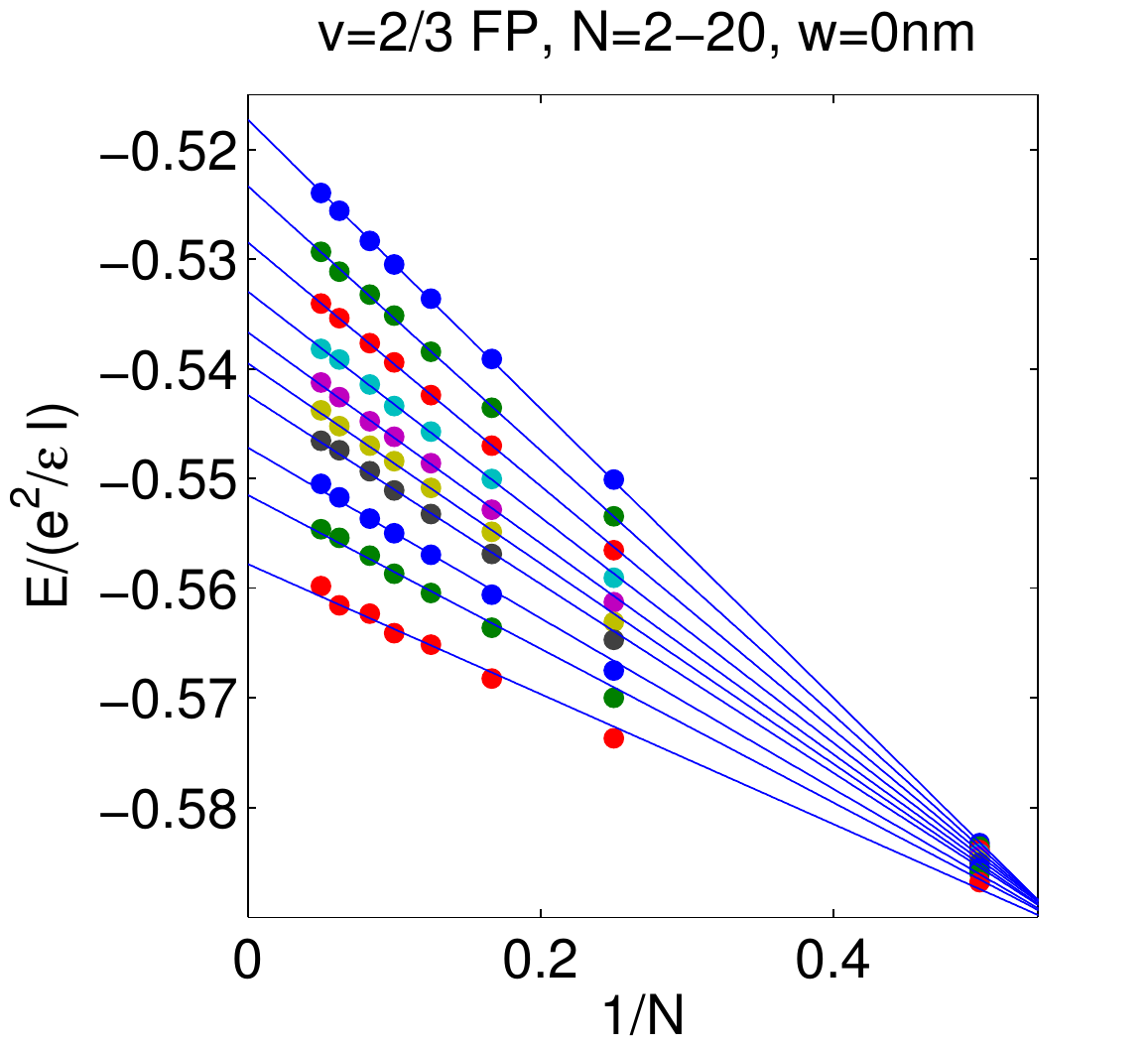}}\\
\vspace{2mm}
\resizebox{0.32\textwidth}{!}{\includegraphics{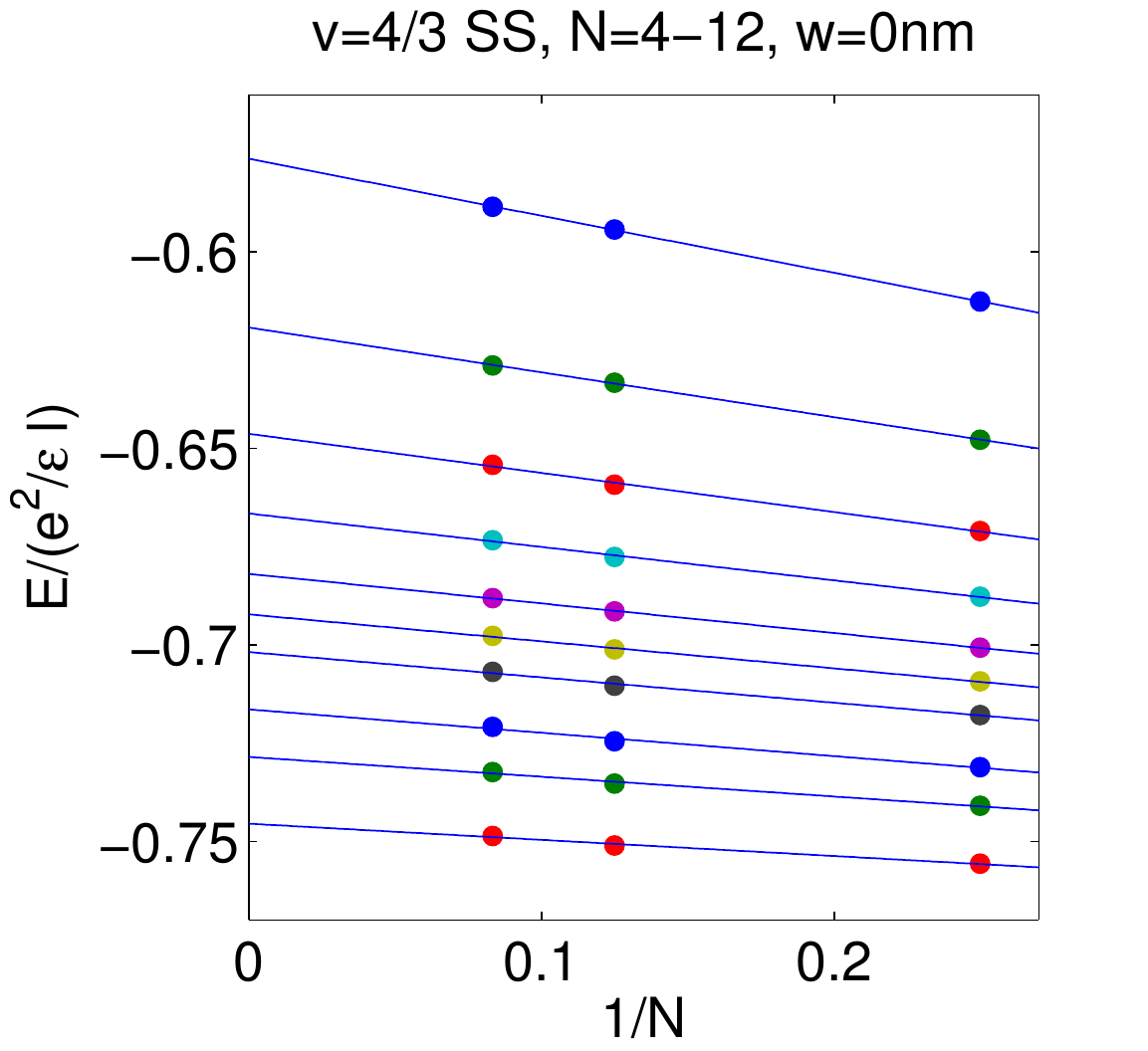}}
\hspace{-3mm}
\resizebox{0.32\textwidth}{!}{\includegraphics{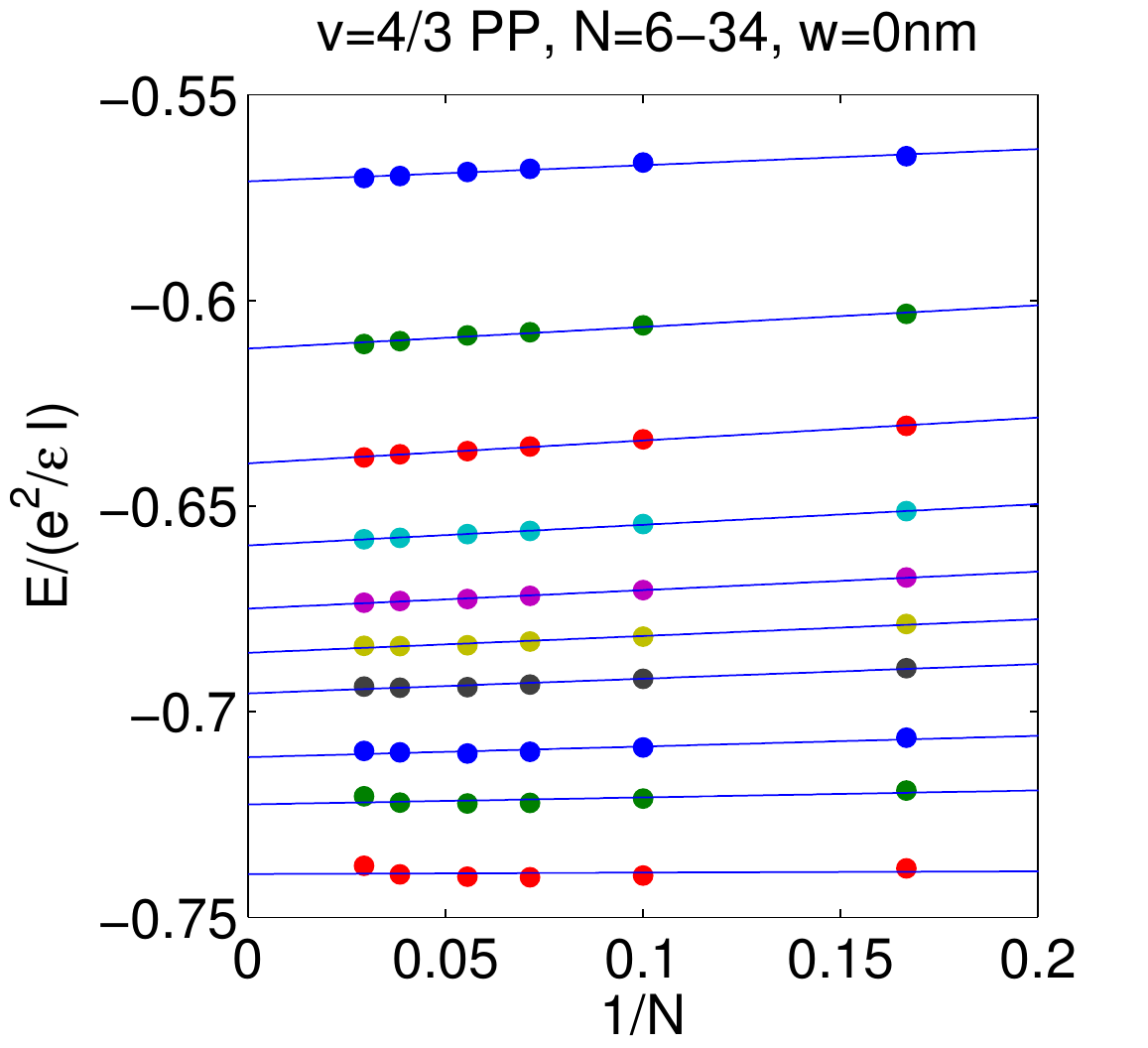}}
\vspace{-1mm}
\caption{(Color online). The finite-size extrapolation for the DMC energies at $\nu=2/3$ and $4/3$. For the spin singlet (SS) states, the exact Coulomb ground state of the LLL is used as the trial wave function $\psi_T$. The quantum well width is taken to be zero. See the top left panel of Fig. \ref{fit1} for the meaning of different colors.
}
\label{fit2}
\end{figure*}

Fig. \ref{fit2} shows the Method II extrapolation for the DMC energies obtained with exact states as the trial wave function at $\nu=2/3$ and $4/3$ at zero width. 
For the spin singlet states we use the exact LLL wave function as $\psi_T$ and can therefore only access three smallest system sizes, but the extrapolations are quite reliable. We have tested that the Method I also gives very similar results. For finite widths the extrapolations to the thermodynamic limit are not reliable and are not shown.

\begin{figure*}
\resizebox{0.3\textwidth}{!}{\includegraphics{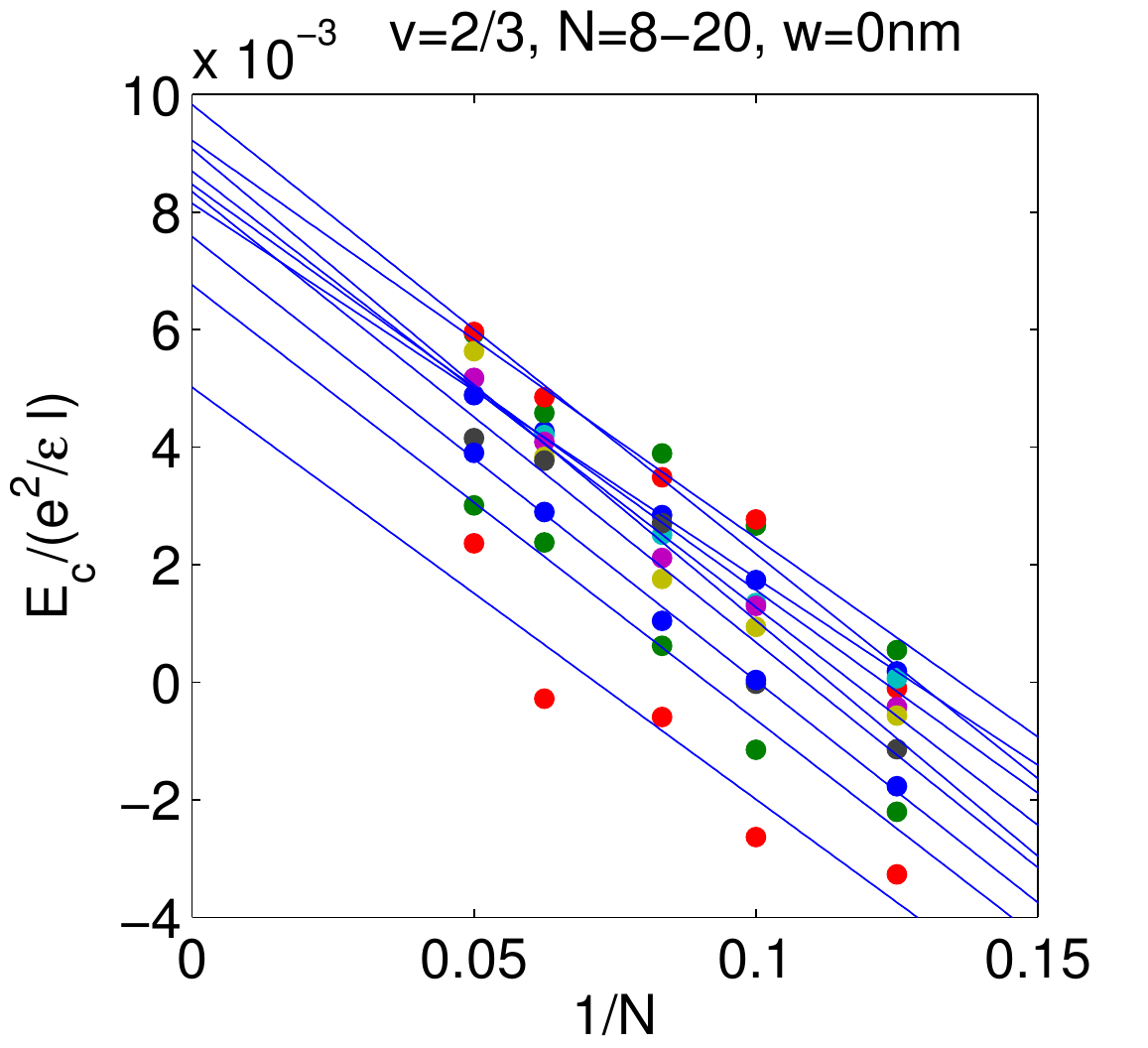}}
\hspace{-3mm}
\resizebox{0.3\textwidth}{!}{\includegraphics{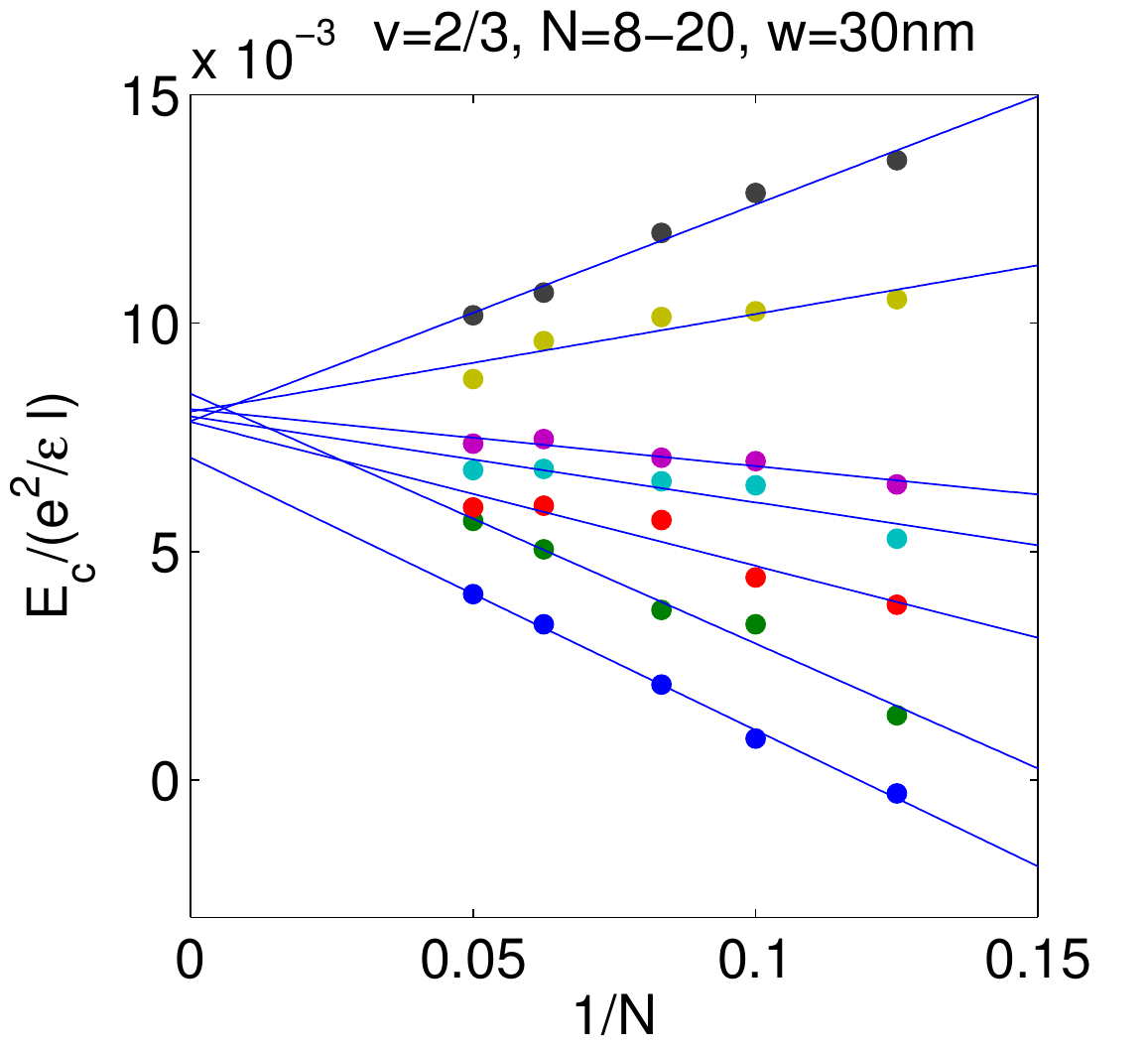}}
\hspace{-3mm}
\resizebox{0.31\textwidth}{!}{\includegraphics{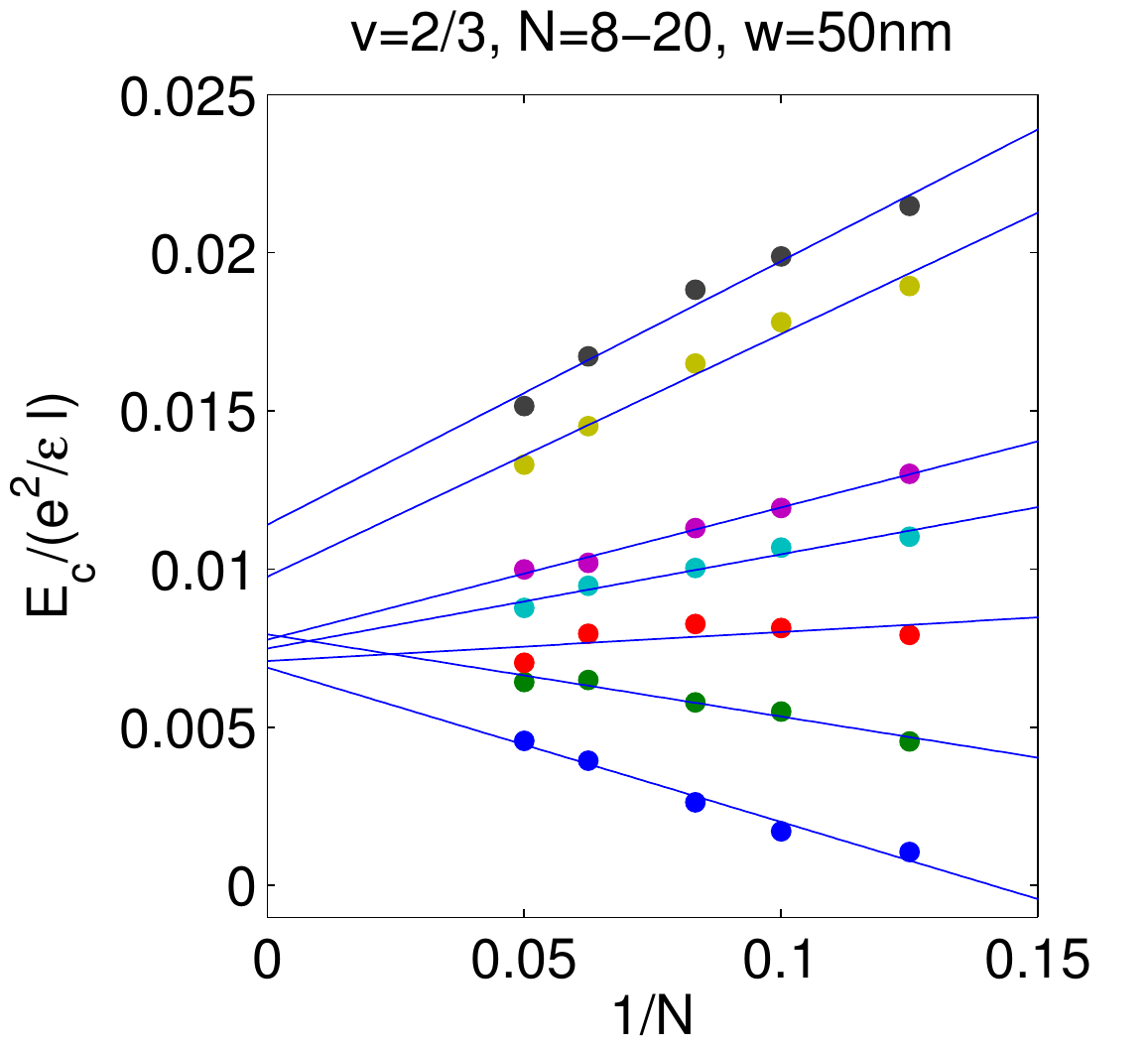}}\\
\resizebox{0.32\textwidth}{!}{\includegraphics{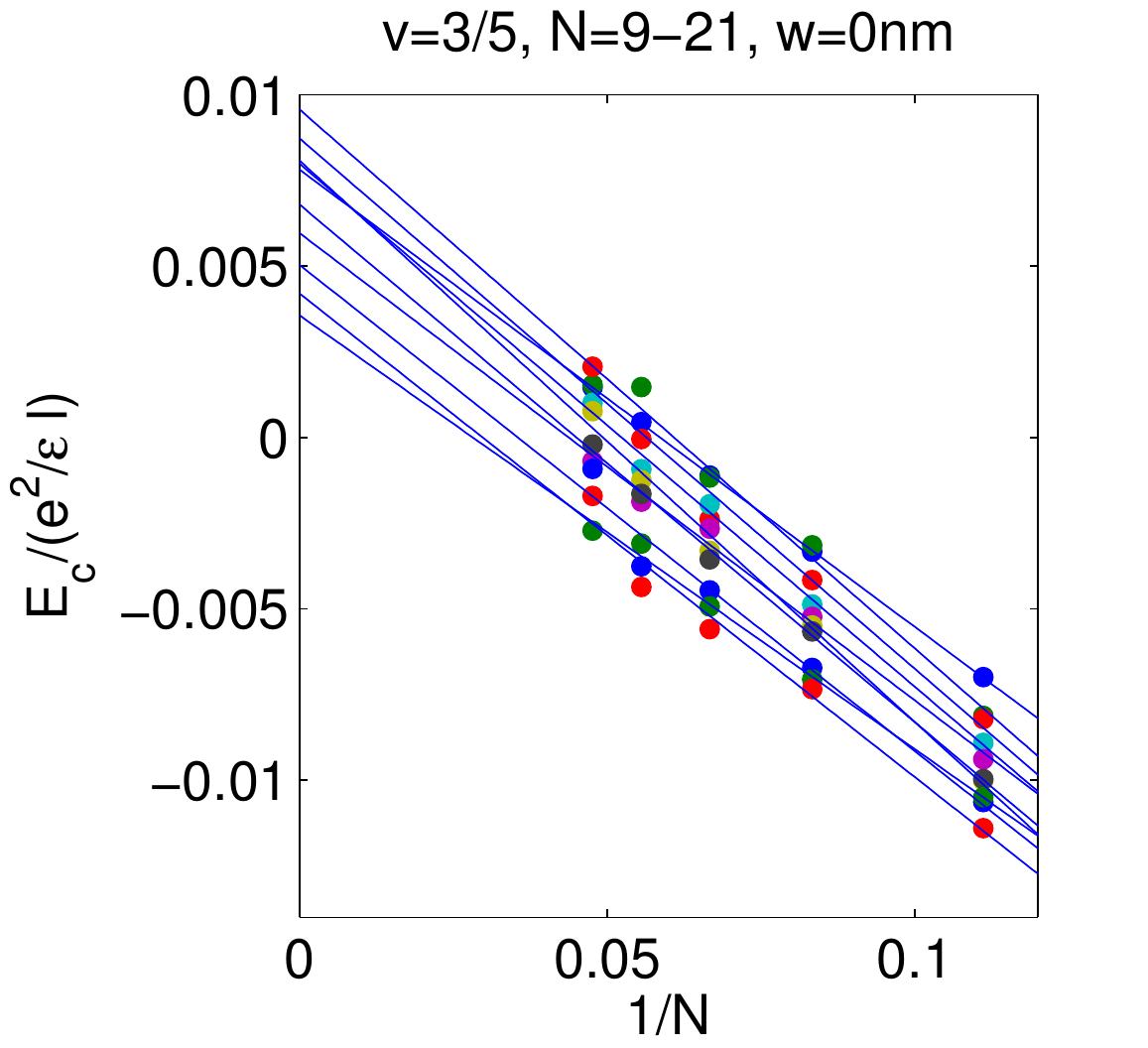}}
\hspace{-3mm}
\resizebox{0.32\textwidth}{!}{\includegraphics{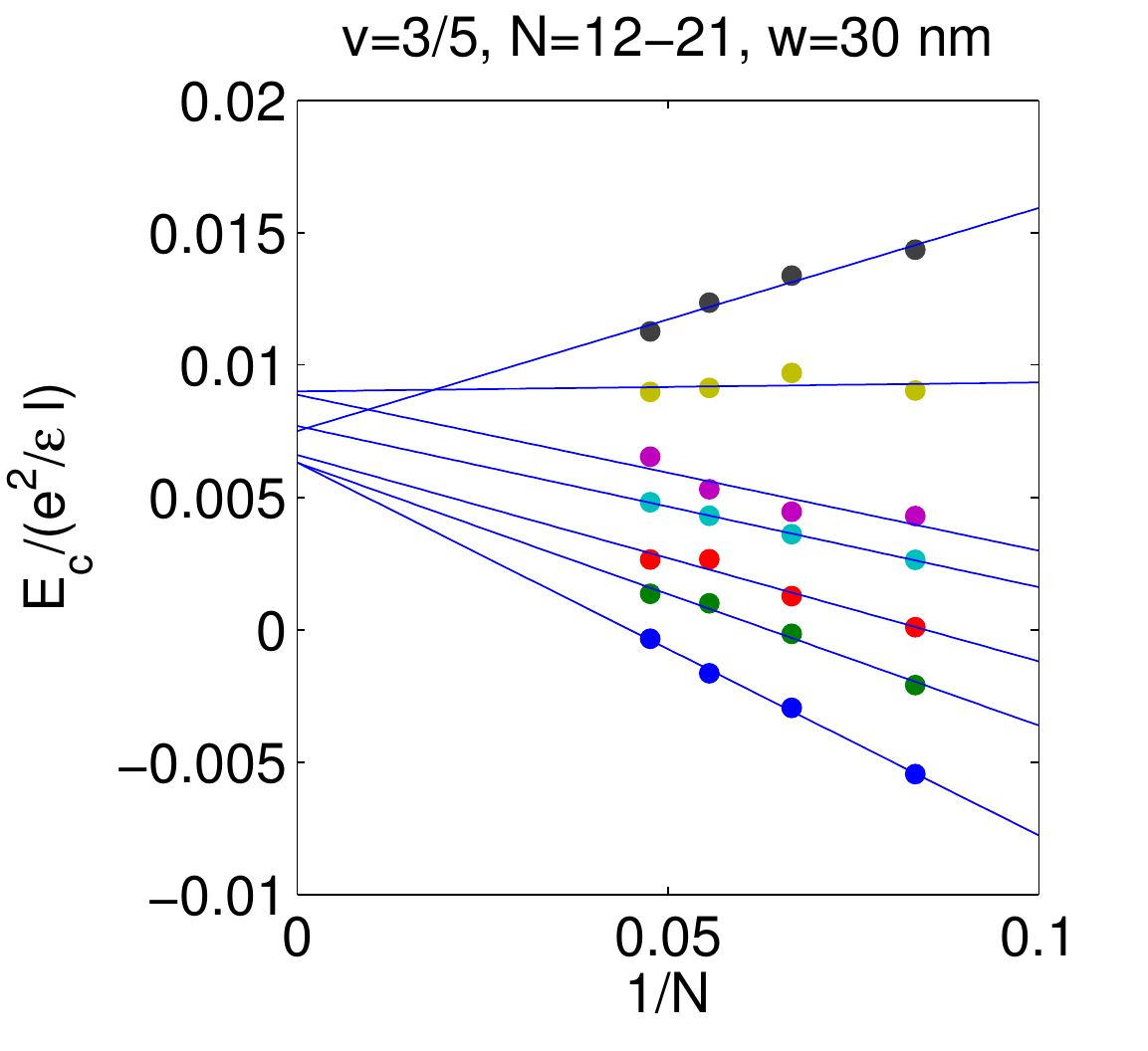}}
\hspace{-3mm}
\resizebox{0.32\textwidth}{!}{\includegraphics{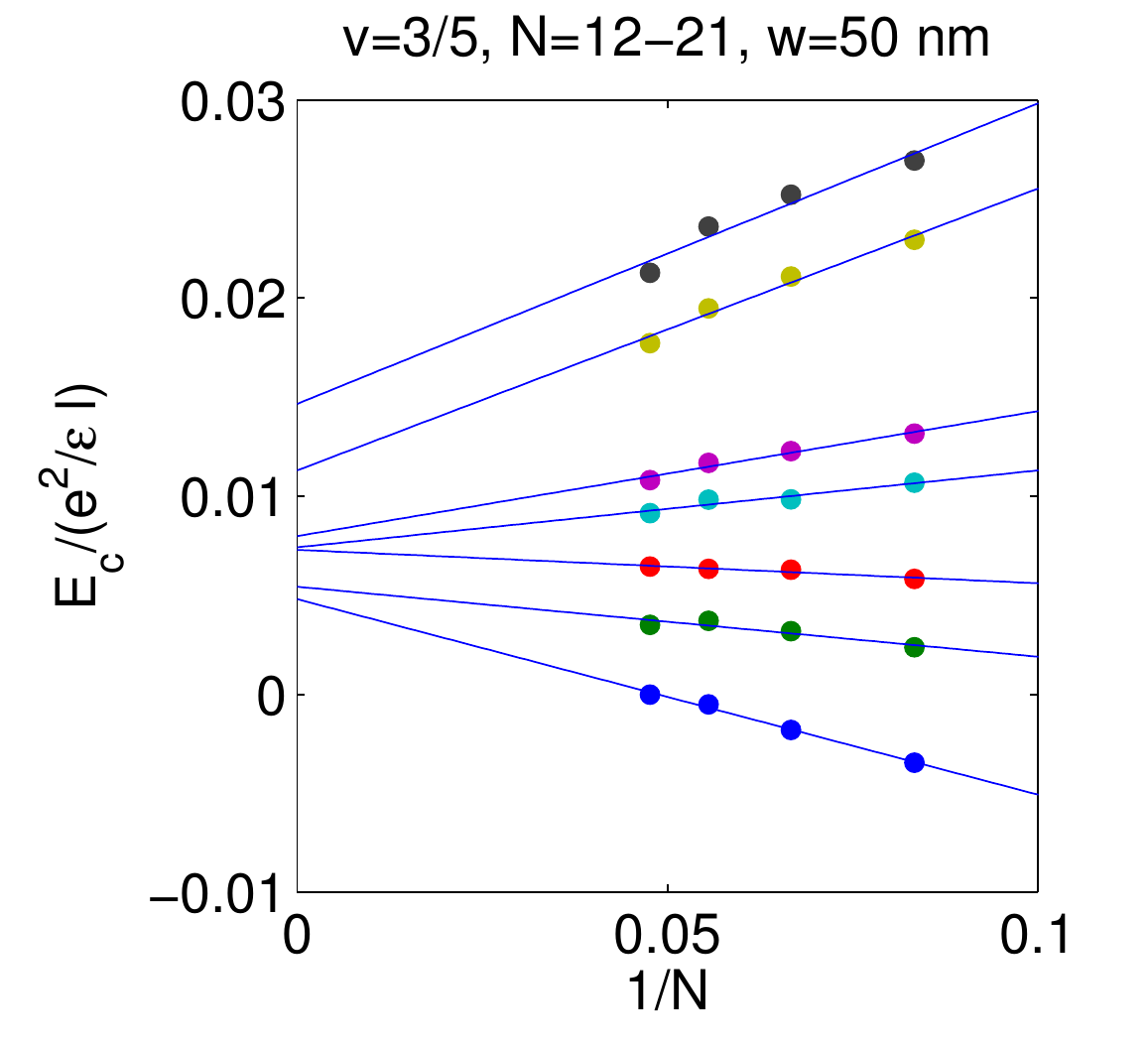}}
\vspace{-2mm}
\caption{(Color online). Extrapolation of the critical Zeeman energies, using the Method I, at $\nu=2/3$ and $3/5$ at different widths. The reverse-flux-attached wave functions of Eq.~2 are used as $\psi_T$. See the legends in Fig. \ref{fit1} for the values of $\kappa$'s (when $w=0$) and $\rho$'s (when $w\neq0$).  
}
\label{fit4}
\end{figure*}

Fig. \ref{fit4} shows finite-size extrapolation for DMC energies obtained by using the reverse-flux-attached Jain wave function as $\psi_{T}$ at $\nu=2/3$ and $3/5$. Here Method I is used to obtain the critical Zeeman energies.

\section{IV. Perturbative study of LL mixing}
An alternative approach for treating LL mixing involves using an effective interaction within an isolated LL that simulates the effect of LL mixing. (This further modifies the effective interaction for finite width.) This interaction involves a two-body term, a three-body term and, in principle, higher body terms. The two-body term conserves particle-hole symmetry within each LL, but the three- and higher body terms break the particle-hole symmetry and are responsible for different values of 
$\alpha_{\rm Z}^{\rm crit}$ for $\nu$ and $2-\nu$. The interaction is parametrized by the so-called pseudopotentials, which are the energies of pairs, triplets, or n-tuplets with well defined relative angular momentum and spin quantum numbers. 

The effective interaction can be included in two ways: (i) One can calculate the expectation value of the effective interaction for ``unperturbed" ground state wave function, i.e. the wave function in the absence of LL mixing. This can be done most conveniently by determining the amplitude of each pair or triplet in the wave function. Because these amplitudes are different for differently spin polarized states at a given filling factor, the resulting $\alpha_{\rm Z}^{\rm crit}$ values depend on LL mixing. Because the corrections to the various pseudopotentials are linear in $\kappa$, so is the correction to $\alpha_{\rm Z}^{\rm crit}$. This approach assumes that the wave function itself is not significantly modified by LL mixing, which should be the case for small LL mixing where the correction to the interaction is small compared to the excitation gap of the state. 
(ii) Alternatively, one can diagonalize the total effective Hamiltonian to obtain the ground state and its energy.
The diagonalization must be carried out independently at each value of $\kappa$.
As this approach admits the effect of LL mixing on the wave functions, the corrections 
to $\alpha_{\rm Z}^{\rm crit}$ will no longer be strictly linear in $\kappa$, but only the 
part that is linear in $\kappa$ is meaningful within the perturbative approach. 
We have tested both of these approaches and found that they produce the same 
correction to $\alpha_{\rm Z}^{\rm crit}$ to linear order in $\kappa$. 
The magnitude of the resulting nonlinearities in $\alpha_{\rm Z}^{\rm crit}(\kappa)$ in the second approach 
presumably serves as a guide for the upper limit on $\kappa$ for which the 
effective pseudopotentials can be used; we find 
noticeable nonlinearities in $\Delta\alpha_{\rm Z}^{\rm crit}$ only above $\kappa\approx1$. We will be using the approach (i) in what follows below.

We have used exact numerical diagonalization in the configuration interaction basis to compute series of Coulomb ground states on a sphere, labeled by the electron number $N$, magnetic flux $2Q$, spin polarization, and the 
quasi-2D layer width $w$ (expressed in the units of magnetic length $\ell$). The standard combinations of $(N,2Q)$ were chosen to represent fractional quantum Hall states with relevant filling factors such as $\nu=2/5$, $3/7$, $3/5$, or $2/3$. For each $\nu$ we have computed as many systems as possible, limited only by the dimension of the Hilbert space.
Naturally, inclusion of the spin degree of freedom enlarges space and thus reduces the maximum computable $N$. For example, for the fully polarized (FP) state at $\nu=2/5$ the largest system is $(N,2Q)=(18,41)$ with space dimension exceeding $3.5\times10^9$, while the largest system for the spin-singlet (SS) state at the same fraction $\nu=2/5$ is $(N,2Q)=
(12,27)$ with space dimension exceeding $2.2\times10^9$.

For each of these systems we have determined short-range pair and triplet amplitudes, 
$P^{(2)}(S,m)$ and $P^{(3)}(S,m)$, where $S$ is the total spin and $m$ is the total relative angular momentum.
They were computed as expectation values of appropriate model pseudopotentials with
only one nonvanishing coefficient.
Specifically, for pair amplitudes we used $m\le5$, i.e., $(S,m)=(0,0)$, $(1,1)$, $(0,2)$, 
$(1,3)$, $(0,4)$, and $(1,5)$, while for triplet amplitudes $m\le3$, i.e., $(S,m)=(1/2,1)$, $(1/2,2)$, 
$(1/2,3)$, and $(3/2,3)$. The pseudopotentials for these pairs and triplets 
have been calculated perturbatively by Peterson and Nayak in the limit of small $\kappa$ \cite{Peterson13}, and we use their values. The LL mixing corrections to higher order pseutopotentials are not included in our analysis.

Ref.~\onlinecite{Peterson13} also gives pseudopotentials for finite thickness, assuming a $\cos z\pi/w$ wave function in direction $z$ perpendicular to the plane. We give below perturbative results also for finite thickness, but stress that these may not be directly compared to the finite width results in our DMC calculation, which uses a more sophisticated self-consistent LDA treatment of the finite width effects. 

\begin{figure}
\includegraphics[width=0.45\textwidth]{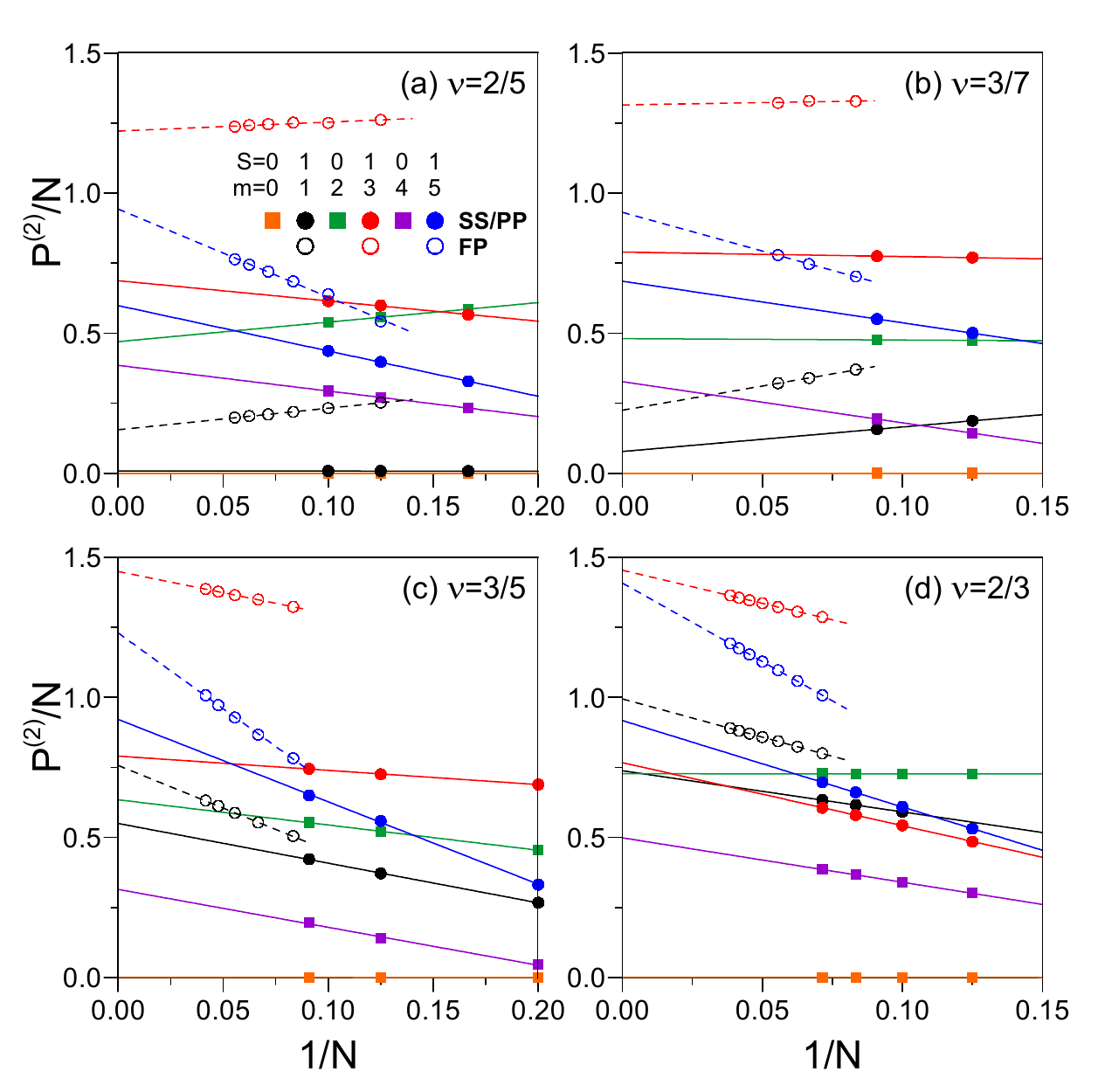}
\caption{(color online) Pair amplitudes per particle, $P^{(2)}(S,m)/N$, as a function
of inverse system size, $1/N$, for Coulomb ground states of different spin polarization
at different filling fractions $\nu=2/5$ (a), $3/7$ (b), $3/5$ (c), and $2/3$ (d), in strictly
2D layers (i.e., layer thickness $w=0$).}
\label{fig-arek-1}
\end{figure}

In Fig.~\ref{fig-arek-1} we plot pair amplitudes per particle, $P^{(2)}(S,m)/N$, as a function of inverse system size, $1/N$, for several filling factors of interest. For each $\nu$, we show data for two relevant spin polarizations: FP and either SS (for $\nu=2/5$ and $2/3$) or PP (for $\nu=3/5$ and $3/7$). Of course, the amplitudes involving unpolarized pairs, $P^{(2)}(S=0,m)$, vanish for the FP states and have not been shown. The regular size dependence of all amplitudes ensures reliable extrapolation to the limit of infinite system, $1/N\rightarrow0$, even in those cases where relatively few
data points are available.

\begin{figure}
\includegraphics[width=0.45\textwidth]{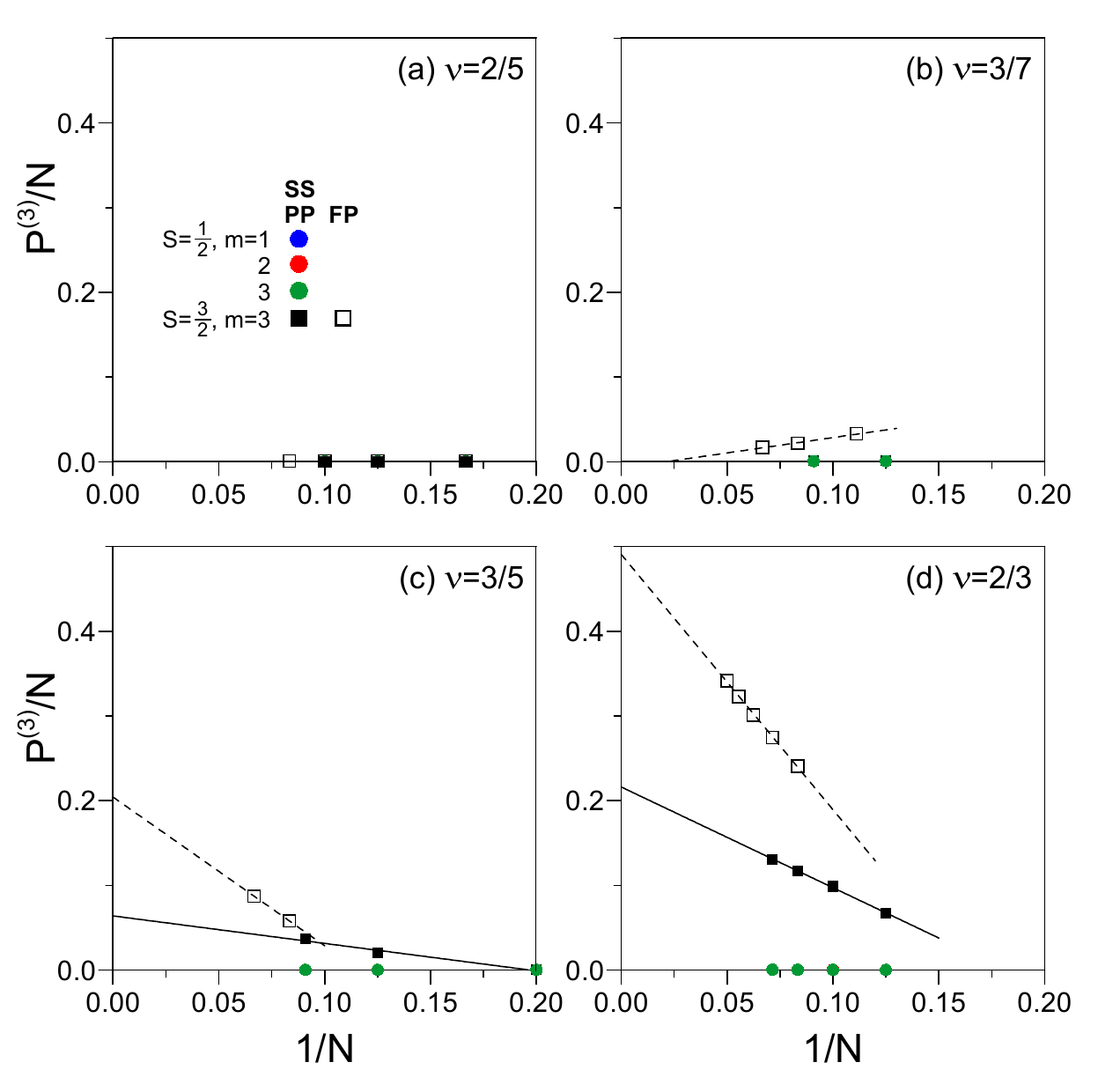}\\
\vspace{-2mm}
\caption{(color online) Similar as Fig.~\ref{fig-arek-1}, but for triplet amplitudes per particle, 
$P^{(3)}(S,m)/N$.}
\label{fig-arek-2}
\end{figure}

In Fig.~\ref{fig-arek-2} we show an analogous plot of triplet amplitudes per particle, 
$P^{(3)}(S,m)/N$ in the same states, with the amplitudes involving unpolarized triplets, 
$P^{(3)}(S=1/2,m)$, only shown for the SS and PP states where they may not vanish.
Here also the apparently quite regular size dependence of all amplitudes enables reliable 
extrapolation, which is fortunate because for the triplet amplitudes we were
unable to use the largest systems due to the fact that three-body hamiltonians produce denser matrices in the configuration space.

\begin{figure}
\includegraphics[width=0.45\textwidth]{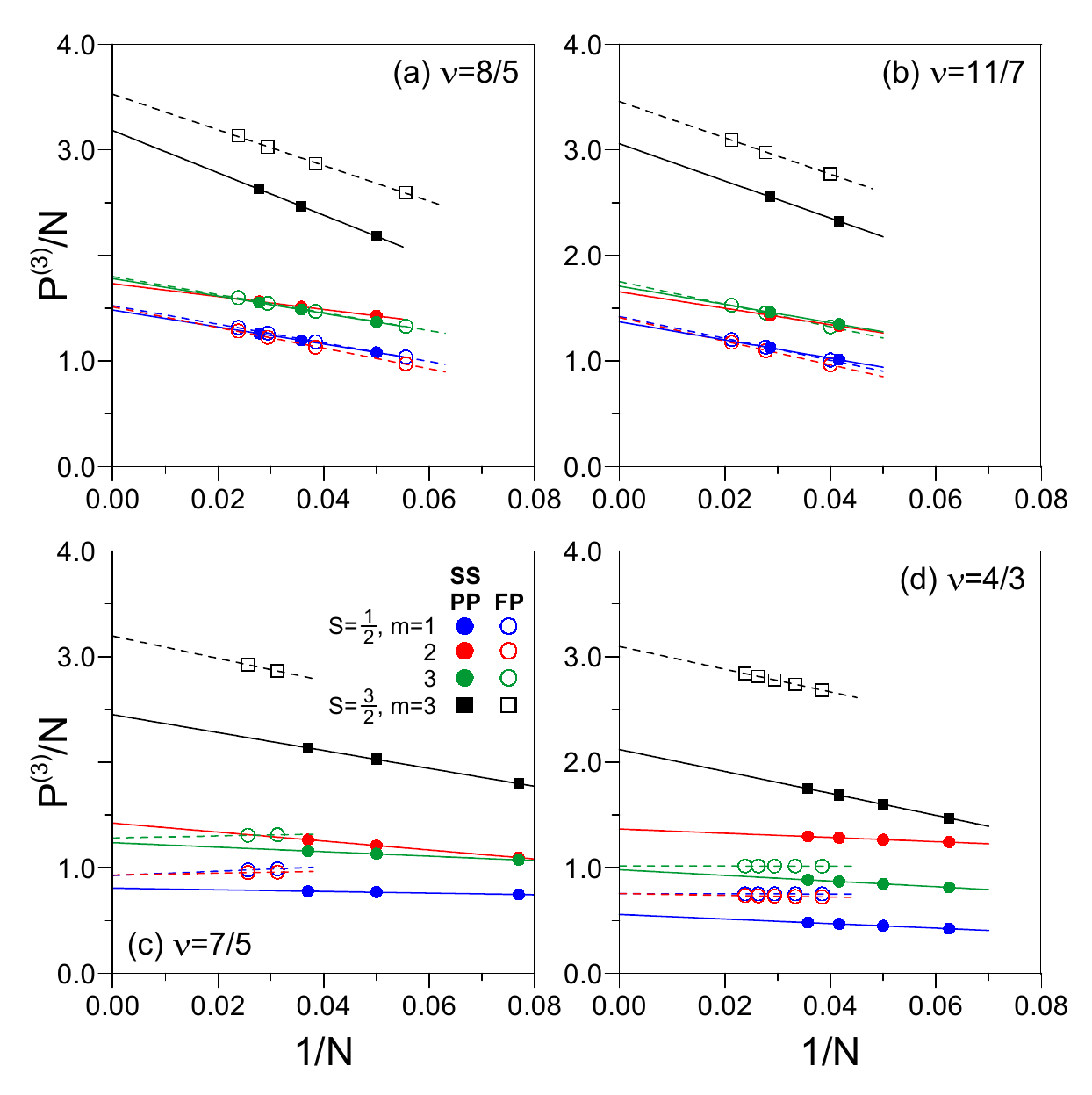}
\vspace{-2mm}
\caption{(color online) Similar as Fig.~\ref{fig-arek-2}, but at conjugate filling factors
$\nu=8/5$ (a), $11/7$ (b), $7/5$ (c), and $4/3$ (d).}
\label{fig-arek-3}
\end{figure}

Since the three-body interaction breaks particle-hole symmetry, 
we must calculate its effect on the ground state energy separately at the conjugate 
filling factors $\nu$ and $2-\nu$.
In Fig.~\ref{fig-arek-3} we show the plots of $P^{(3)}(S,m)/N$ vs.\ $1/N$ at the
conjugate fractions $\nu=8/5$, $11/7$, $7/5$, and $4/3$.
The amplitudes at $\nu$ and $2-\nu$ are indeed related, but we show
both plots of $P^{(3)}$ to make it clear that the amplitudes are generally larger at 
$\nu>1$ than at $\nu<1$, and that at $\nu>1$ even the PP states (obtained by particle-hole conjugation of the corresponding FP states at $\nu<1$) involve electrons with both
spins, and hence their unpolarized amplitudes $P^{(3)}(S=1/2,m)$ do not vanish.

\begin{figure}
\includegraphics[width=0.45\textwidth]{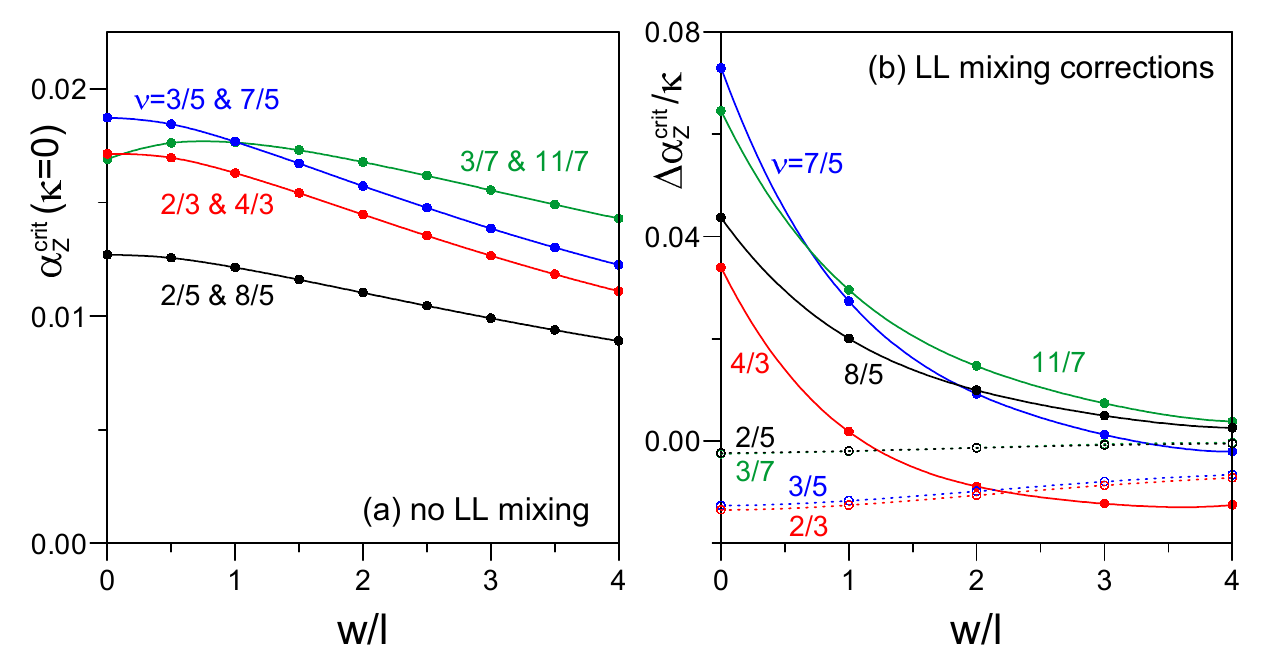}
\vspace{-2mm}
\caption{(color online) (a) Dimensionless critical Zeeman energies $\alpha_{\rm Z}^{\rm crit}$
for spin transitions between differently polarized fractional quantum Hall states at the
indicated filling factors $\nu$, in the absence of LL mixing (which ensures 
equal values of $\alpha_{\rm Z}^{\rm crit}$ at he conjugate fractions $\nu$ and $2-\nu$).
(b) LL mixing corrections to $\alpha_{\rm Z}^{\rm crit}$, calculated from the convolution
of pair and triplet amplitudes in Figs.~\ref{fig-arek-1}--\ref{fig-arek-3} with effective
LL mixing pseudopotentials of Ref.~\onlinecite{Peterson13}.
Both $\alpha_{\rm Z}^{\rm crit}$ (a) and $\Delta\alpha_{\rm Z}^{\rm crit}/\kappa$ are 
plotted as a function of the thickness $w$ of the quasi-2D layer, expressed in the units of magnetic length $\ell$.}
\label{fig-arek-4}
\end{figure}

The convolution of the thermodynamic values of the two- and three-body amplitudes 
with the effective LL mixing pseudopotentials of Peterson and Nayak~\cite{Peterson13} produces the 
LL mixing correction to the ground state energies per particle.
The difference between these energy corrections for the differently polarized ground 
states at the same filling factor $\nu$ yields the LL mixing correction to the critical 
Zeeman energy for the spin transition between these states, 
$\Delta\alpha_{\rm Z}^{\rm crit}$.
These corrections are proportional to $\kappa$, so in Fig.~\ref{fig-arek-4}(b) we
show $\Delta\alpha_{\rm Z}^{\rm crit}/\kappa=d\alpha_{\rm Z}^{\rm crit}/d\kappa$, as a function of thickness $w$
of the quasi-2D layer expressed in the units of magnetic length $\ell$.
For completeness, in Fig.~\ref{fig-arek-4}(a) we have plotted the thickness dependence of $\alpha_{\rm Z}^{\rm crit}$ in the absence of LL mixing, to which $\Delta\alpha_{\rm Z}^{\rm crit}$ of Fig.~\ref{fig-arek-4}(b) is the LL mixing correction. Note that the corrections due to finite width modeled through a $\cos z\pi/w$ wave function are much smaller than those seen in Fig.~1. 

\section{V. Decrease in energy due to LL mixing}

\begin{figure}
\includegraphics[width=0.47\textwidth]{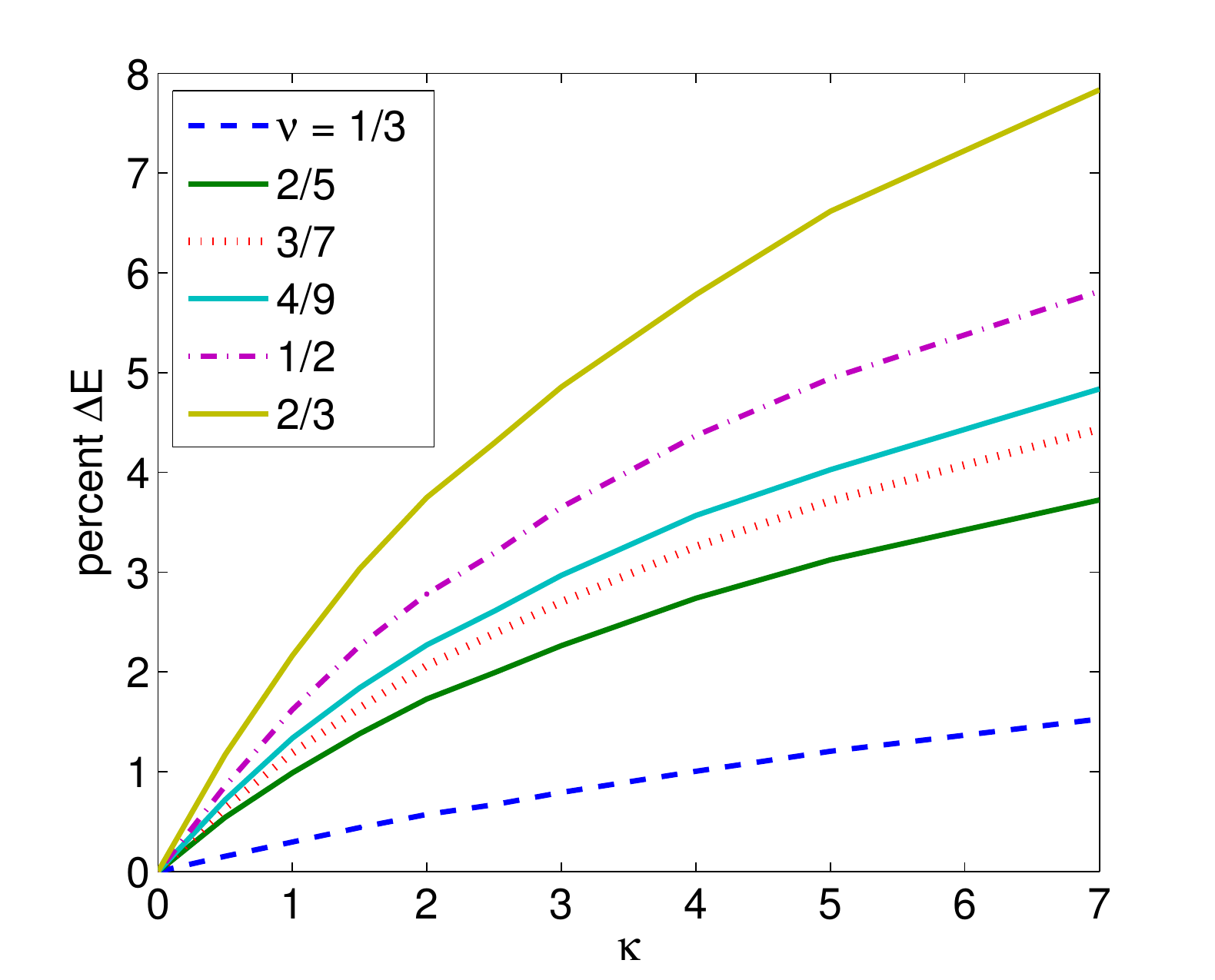}
\vspace{-2mm}
\caption{(color online) The percent decrease in energy, quoted in units of $e^2/\epsilon \ell$, as a function of $\kappa$ for several filling factors. All results represent thermodynamic limits. Zero thickness is assumed. The $1/2$ filling represents the CF Fermi Sea.}
\label{dE}
\end{figure}

In the above, we have discussed the relative change in the energies of differently spin polarized states, which are relevant for the spin phase transitions. In Fig.~\ref{dE} we plot the percent decrease in the energies (quoted in units of $e^2/\epsilon \ell$) of fully spin polarized states at filling factors 1/3, 2/5, 3/7, 4/9, 1/2 and 2/3, for a sample with zero thickness. For low $\kappa$, the change in energy is approximately 0.3-2\% per unit of $\kappa$. The correction due to LL mixing is seen to grow with increasing filling factor, as expected from the observation that at small fillings electrons are able to avoid one another effectively even without LL mixing.

\end{document}